\documentclass[longauth]{aa}
\usepackage{caption}
\usepackage{natbib}
\usepackage{scalerel}
\usepackage{cancel}
\usepackage[table]{xcolor}
\usepackage{soul}

\usepackage{subcaption}
\usepackage[utf8]{inputenc}
\usepackage[table]{xcolor}
\usepackage{amsmath}

\makeatletter

\newcommand{\prob}{\mathbb{P}}

\newcommand{\dd}{\mathrm{d}}
\newcommand{\btheta}{\boldsymbol{\theta}}

\makeatother

\usepackage[utf8]{inputenc}

\usepackage{txfonts}
\usepackage[pdfencoding=auto,psdextra]{hyperref}
\hypersetup{
    colorlinks=true,
    linkcolor=blue,
    filecolor=magenta,      
    urlcolor=blue,
    citecolor=blue
}
\definecolor{cadmiumgreen}{rgb}{0.0, 0.42, 0.24}
\usepackage{graphicx}

\makeatletter
\renewcommand*\aa@pageof{, page \thepage{} of \pageref*{LastPage}}

\makeatother

\usepackage[utf8]{inputenc}

\usepackage[switch, modulo]{lineno}

\usepackage{euclid}

\begin{document}

   \title{\texttt{Owl-z}: a Bayesian tool to select $z \gtrsim 7$ quasars}

\newcommand{\orcid}[1]{} 
\author{\normalsize M.~Ezziati$^{1}$\thanks{Corresponding author, \email{meriam.ezziati@lam.fr}},  R.~Pello$^{1}$, J.-G.~Cuby$^{1,4}$, P.~Pudlo$^{2}$, F.-X.~Dupé$^{3}$,J.-C.~Lambert$^{1}$, J.-C.~Cuillandre$^{5}$, 
 O.~Ilbert$^{1}$,  S.~de la Torre$^{1}$, S.~Arnouts$^{1}$, E.~Jullo$^{1}$,D.~Yang$^{6}$}

\institute{$^{1}$ Aix Marseille Univ, CNRS, CNES, LAM, Marseille, France\\
$^{2}$ Aix Marseille Univ, CNRS, I2M, Marseille, France\\
$^{3}$ Aix Marseille Univ, CNRS, LIS, Marseille, France\\
$^{4}$ Canada-France-Hawai'i Telescope, Waimea, Hawai'i\\
$^{5}$ Université Paris-Saclay, Université Paris Cité, CEA, CNRS, AIM, 91191, Gif-sur-Yvette, France\\
$^{6}$Leiden Observatory, Leiden University, P.O. Box 9513, 2300 RA Leiden, The Netherlands
}

\abstract{This paper presents \texttt{Owl-z}, a Bayesian code aiming at identifying $z\geq 7$ quasars in wide field optical and near-infrared surveys. By construction, the code can also be used to select objects that contaminate the high-$z$ quasar population, i.e. brown dwarfs and early-type galaxies at intermediate redshifts. The code can be adapted for the selection of high-$z$ galaxies, and although it has been tuned to the Euclid Wide Survey, it can be easily adapted to other photometric surveys. 
The code input data are the object's photometric data and its galactic longitude and latitude, and the code output data are the probabilities of the modelled populations of high-$z$ quasars, brown dwarfs and early-type galaxies at intermediate redshift.
As part of the validation, \texttt{Owl-z} could re-identify all spectroscopically confirmed quasars at $ z\geq 7$, demonstrating the code's versatility in applying to different photometric catalogues. 
The performance of \texttt{Owl-z}, based on a metric combining completeness and purity called $F$-measure, is analysed in the case of Euclid using simulated data in a wide range of redshifts ( $7 \leq z \leq 12$) and H-band Euclid magnitudes ($ 18 \leq \HE \leq 24.5 $).
The results show that \texttt{Owl-z} reaches full performance for bright sources ($\HE \lessapprox 22$), somewhat independently of the redshift.
We show that the probability threshold used to select promising quasar candidates can be adjusted after processing to fine-tune the $F$-measure value of candidates depending on their magnitude and redshift estimates. We show that for objects brighter than about two magnitudes above the survey detection limit, \texttt{Owl-z} provides a good classification that will facilitate the optimisation of photometric and spectroscopic confirmation campaigns.
In conclusion, \texttt{Owl-z} is a powerful public tool to help select high-$z$ quasars, brown dwarfs or early-type galaxies at intermediate redshifts in Euclid or other wide-field surveys.
}

\keywords{
Methods: statistical; Surveys; Cosmology: observations; reionization; quasars: general
}

   \titlerunning{Euclid\/: \texttt{Owl-z}: Bayesian selection tool for $z \gtrsim 7$ quasars}
   \authorrunning{M. Ezziati et al.}
   
   \maketitle

\section{\label{sc:Introduction}Introduction}

High-redshift quasars are significant in illuminating the early Universe, shedding light on the genesis of initial galaxies and black holes. The luminosity of these cosmic beacons allows for unprecedented insights into the evolution and patchiness of neutral hydrogen and its re-ionization, particularly during the epoch of reionization ~\citep[EoR; see e.g.][]{Fan2006, Becker2015, Bosman2022}. Beyond tracing the EoR, the identification of supermassive black holes (SMBHs) using quasars at high redshifts challenges conventional models, prompting investigations into alternative scenarios ~\citep{Bennett2024}. In particular, the identification and study of $z>7$ quasars are challenging due to their ambiguous nature and the limited observations/identifications of such objects in the early Universe
~\citep[see e.g.][]{Banados2018,Fan2023}.
Also, the recent discoveries of young quasars with small Lyman-alpha proximity zones \citep{hubler2024,Maiolino2024} add complexity, opening avenues for refining our understanding of early cosmic phenomena.
Spectroscopic observations with the James Webb Space Telescope (JWST) have uncovered Active Galactic Nuclei (AGN) signatures in the spectra of the most distant galaxies known today ~\citep{Maiolino2024b}, providing further insights on their environment and formation history ~\citep{Scholtz2024}. 

Despite their high intrinsic luminosity, high-$z$ quasars appear faint at $z>7$, and their photometric selection on wide cosmological surveys is particularly challenging. Indeed, their red colours, often used to identify them with the appropriate colour-cuts or photometric redshift estimates, are subject to severe contamination by brown dwarfs and intermediate-redshift galaxies. 
With the arrival of massive cosmological surveys, such as Euclid ~\citep{Laureijs2011, Mellier2024}, the Nancy Grace Roman Space Telescope ~\citep{Akeson2019} and the Large Synoptic Survey Telescope ~\citep[LSST;][]{LSST_2019}, it becomes possible to identify significant samples of quasars at high redshifts, which will greatly improve our understanding of galaxy and SMBH formation mechanisms. However, for this, it is essential to build efficient and optimised methods and tools. This is precisely the purpose of this work. 

The method introduced here, and its associated code \texttt{Owl-z}, is a Bayesian comparison model aiming at optimising the statistical analysis in such a way that the probability of detecting high-$z$ quasars is enhanced with respect to the classical colour-cuts method. 
Our approach is directly inspired by the Bayesian method developed by  \citealt{Mortlock_2012_bayes} and \citealt{Barnett2019} for the search of quasars at high redshifts.

Our method is one of the methods based on Bayesian analysis, along with \citealt{Barnett2019}, however, there are other probabilistic methods, such as the one presented in \citealt{Nanni2023,Kang2024}, which uses a probabilistic classification approach using density estimation in flux ratio space, specifically using the extreme deconvolution (XD) technique to accurately model the density distribution of high redshift quasar and contaminant populations  \citep{Bovy2009}. There are also other efficient colour-cutting methods paired with radio detections, such as those used in \citealt{Belladitta2020}, \citealt{Banados2014} and \citealt{Banados2016}, which proved to be very efficient. Last, but not least, there are many recent methods that use Random Forests, a supervised machine learning approach, to effectively select high redshift quasars using photometric data from surveys such as Pan-STARRS and WISE \citep{Wenzl2021}.

\texttt{Owl-z} also offers a flexible, fast, and efficient tool for analysing large photometric data sets, looking for $z>7$ quasars with high performance in terms of completeness and purity of the extracted samples. 

From a technical point of view, \texttt{Owl-z} is an open-source code developed specifically for the selection of high-$z$ quasars in wide field near-infrared (NIR) surveys such as the Euclid Wide  Survey  ~\citep[hereafter EWS;][]{Scaramella2022}, but it is adaptable to tackle various other surveys, as also shown in this work.

This paper is organized as follows. In Sect.~\ref{sec:proba}, we present the probabilistic method used for the identification of high-$z$ quasars, the primary objective being the Bayesian selection and classification of each source as either a high-$z$ quasar, an intermediate-redshift contaminant galaxy, or a dwarf star. The detailed modelling performed for each one of these populations is also presented in this section. A brief technical description of \texttt{Owl-z} is presented in Sect.~\ref{Technical}. Sect.~\ref{Validation} is devoted to the validation of \texttt{Owl-z} using two different approaches: the successful re-identification of known and spectroscopically-confirmed quasars at $z>7$ selected from near-IR surveys and simulations of the EWS allowing us to quantify the performance of the code based on the measurement of the completeness and purity of the extracted samples. In Sect.~\ref{Discussion}, we discuss the influence of several important parameters of the method on the performance of \texttt{Owl-z}, such as the threshold used for the selection of high-$z$ quasars, and we provide some guidelines to optimise their selection and follow-up photometric or spectroscopic campaigns. The summary of conclusions is presented in  
Sect.~\ref{Conclusions}. 

Throughout this paper, the following cosmological parameters have been adopted: $H _0=70$ km s$^{-1}$ Mpc$^{-1}$, $\Omega_m$ = 0.3 and $\Omega_{\lambda}$ = 0.7. All magnitudes are given in the AB system \citep{Oke1983}.

\section{Probabilistic selection of sources at high redshift}
\label{sec:proba}
This section presents the probabilistic selection method for identifying high-$z$ quasars in the EWS. The method can be adapted to other wide-field surveys and to the selection of other high-$z$ astronomical sources. The objective of this method is to identify and select a distinct set of potentially high-$z$ sources, referred to as candidates, by effectively distinguishing them from low-redshift sources designated as contaminants. Our approach is based on the Bayesian method developed by  \citealt{Mortlock_2012_bayes} and \citealt{Barnett2019} for the search of quasars at high redshift.

The primary spectral signature of objects with a redshift exceeding 7 ($z > 7$) is the quasi-absence of flux blue-ward the Lyman alpha line, which is a consequence of the combined effects of the Lyman forest and the Gunn-Peterson trough. It is thus possible to identify objects with a redshift of more than 7 by the lack of signal in the optical domain below  $\approx 1 \mu$m (0.97 $\mu$m). This allows us to restrict the modelling of high-$z$ quasars and their contaminants to objects detected in near-infrared bands and pre-selected on the basis of the absence or near-absence of detection in optical bands. This work can be compared with others, such as \citealt{Barnett2019} and \citealt{Pipien2018}. For details, please see Sect.~\ref{comparison_results} 

\begin{figure}[ht!]
        \centering
        \includegraphics[angle=0,width=0.9\hsize]{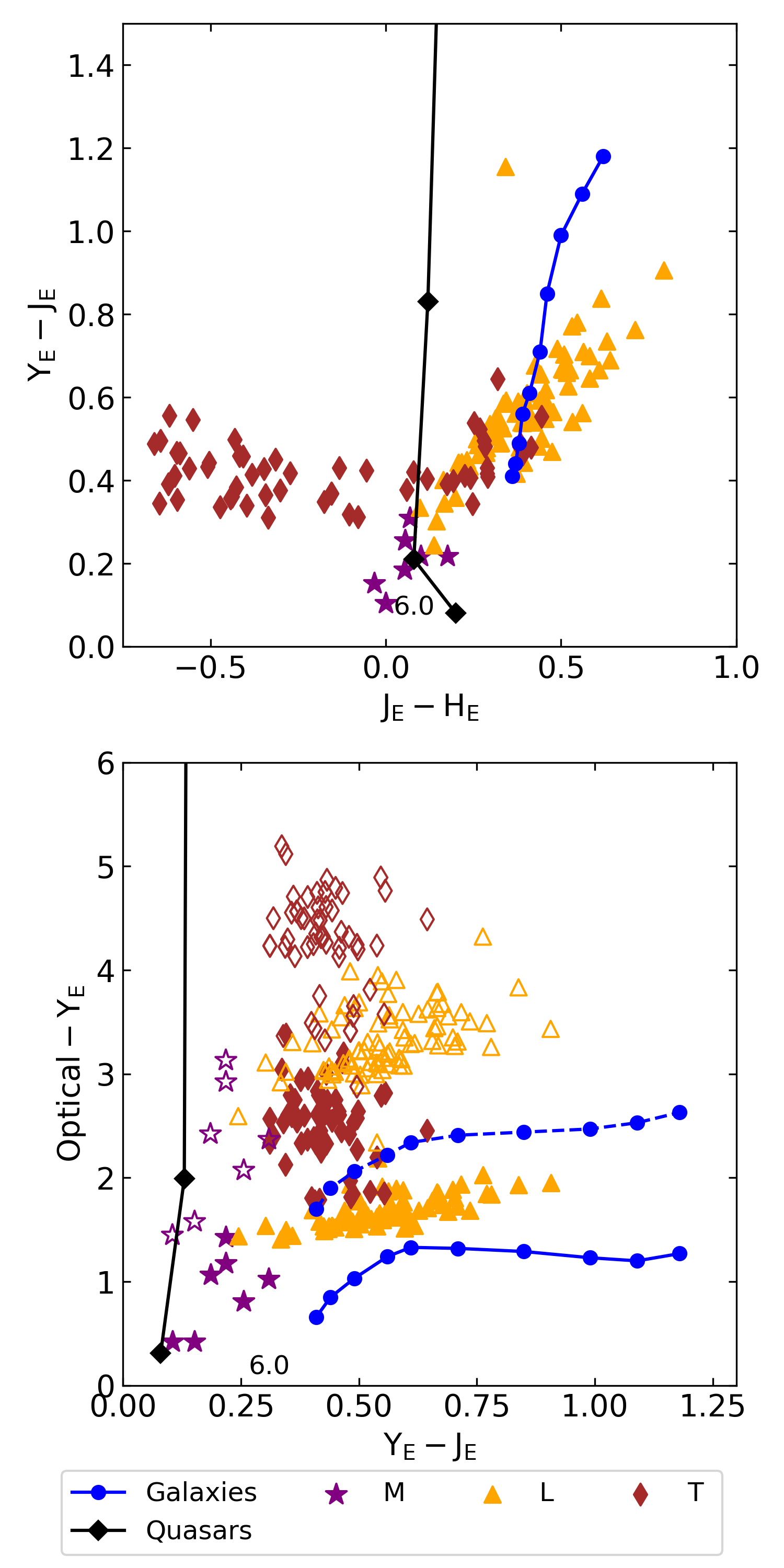}

    \caption{NIR (top) and Optical-NIR (bottom) colours for the three classes of objects considered in this work, using Euclid filters. Black solid lines display quasars in the redshift domains captured by the filters, with redshifts indicated directly on the lines. Blue lines display galaxies at $(1\leq z \leq 2)$, susceptible to contaminating the quasar samples. MLT: stars M, triangles L, and  diamonds T. For the M, L and T, filled: $z-\YE$ and hollow: $\IE-\YE$}
    \label{Color}
\end{figure}

We employ Bayesian selection methodology to ascertain the posterior probability that a specific observed astronomical source, initially identified as non-detected in optical bands, is classified as a high-$z$ quasar or as a contaminant. This is achieved by considering the prior surface density of the quasars and contaminants. Contaminant populations of high-$z$ galaxies and quasars in near-infrared and mid-infrared bands are well known \citep[see e.g.][]{Mortlock_2012_bayes, Barnett2019} and consist of early-type galaxies at intermediate redshift \citep[$1\lesssim z \lesssim 2 $, see e.g.][]{vanMierlo2022} and low mass dwarf stars of spectral types late-M, L or T \citep[MLT hereafter;][]{Stern2007,Caballero2008,Wilkins2014,Hainline2024_2}.
 The difficulty in distinguishing high-$z$ quasars from the above-mentioned contaminants is illustrated in Fig.~\ref{Color}, showing the comparison of NIR  and optical-NIR colours for the three populations considered in this work using Euclid filters \citep{Schirmer2022}. Analysis of this figure highlights the distinct colour characteristics of these populations in the $\YE-\JE$ colour space, which is key to the quasar selection process. The top panel
 of the figure illustrates that at higher redshifts $(z>8)$, quasars are significantly redder than the contaminating populations, allowing for effective separation based on their $\YE-\JE$ colours. However, the analysis also notes that at lower redshifts $(7<z<8)$, the NIR broadband colours of quasars and contaminants overlap more, making it difficult to distinguish between them without deep complementary data, especially in the LSST z-band (lower panel).

\subsection{Principles}
In this section, we present the Bayesian formalism in a general context and then apply it in more detail to each of the populations.
Bayesian model selection \citep{robert2007bayesian} is a relevant method to establish the nature of a source in a survey given photometric data \citep[see][and the references therein]{Mortlock_2012_bayes}.
The posterior probability that a given source is of a type $t\in\mathcal{T}$, given the data $\mathbf{D}$, is driven by two terms: the prior probability $\prob(t)$ of the source being of type $t$, and the integrated likelihood or the evidence of the data given the type of source, $\prob(\mathbf{D}|t)$. 
The latter is the marginal likelihood of the data, which is the integral of the likelihood of the data given the parameters of the model, $\prob(\mathbf{D}|\btheta_t,t)$, with respect to the prior distribution of the parameters, $\prob(\btheta_t|t)$. 
The marginal likelihood is the key quantity in Bayesian model selection: Unlike methods based on best-fit parameters and type, it accounts for the diversity of photometric data of a given type. The posterior probability of a source being of type $t$ is then given by Bayes' theorem:
\begin{equation}
\prob(t|\mathbf{D}) = \frac{W(t|\mathbf{D})}{\sum_{t'\in\mathcal{T}}W(t'|\mathbf{D})}\text{,}
\label{bayes}
\end{equation}
where $W(t|\mathbf{D}) = \prob(t)\prob(\mathbf{D}|t)$ is the weighted evidence of the model $t$ given the data $\mathbf{D}$.
The weighted evidence of type $t$ is given by
\begin{align}
W(t|\mathbf{D}) &= \prob(t)\prob(\mathbf{D}|t)=\prob(t)\int \prob(\btheta_t|t)\prob(\mathbf{D}|\btheta_t,t)\dd\btheta_t \notag
\\
& = \int \rho_t(\btheta_t)\prob(\mathbf{D}|\btheta_t,t)\dd\btheta_t\text{,} \label{eq:weight}
\end{align}
where $\btheta_t$ is the vector of parameters of $t$-type object and $\rho_t(\btheta_t)$ is the prior value of surface density of detected objects of type $t$ and parameters $\btheta_t$ in the sky. 
A photometric dataset is a set $\widehat{\mathbf F}=(\widehat F_1,\ldots, \widehat F_N)$ of $N$ measurements of the flux of a source in $N$ photometric bands, and their standard error $\widehat{\boldsymbol\sigma}=(\widehat\sigma_1,\ldots, \widehat\sigma_N)$. The statistical model of each type of source is built on a set of spectral energy densities (SEDs) of $n_t$ templates. Given the $i$-th SED of type $t$, we can compute expected fluxes in each band, $\mathbf F_{t,i}(\btheta_t)=(F_{t,i,1}(\btheta_t),\ldots,F_{t,i,N}(\btheta_t))$ at value $\btheta_t$ of the parameter. The likelihood of the data given $\btheta_t$ and $t$ is set as the mixture of multivariate densities is given by
\begin{equation}
\prob\big[\widehat{\mathbf{F}}\big|\btheta_t,t\big] \sim \frac{1}{n_t}\sum_{i=1}^{n_t} p\Big(\widehat{\mathbf F}\Big|\mathbf F_{t,i}(\btheta_t), \widehat{\boldsymbol\sigma}\Big)\text{,}\label{eq:likelihood1}
\end{equation}
centred at $\mathbf F_{t,i}(\btheta_t)$ and with dispersion given by $\widehat{\boldsymbol\sigma}$.  

The above mixture sets a uniform prior distribution on the SED templates modelling each population.
If the photometric bands do not overlap, we can neglect the correlation between the fluxes in different bands. Thus, we rely on a product of Gaussian distributions truncated to be non-negative, given by

\begin{align}
    p\big(\widehat{\mathbf F}\big|\mathbf F_{t,i}(\btheta_t), \widehat{\boldsymbol\sigma}\big) =\prod_{j=1}^N \mathcal{L}_j \text{,}
\end{align}
where each truncated Gaussian distribution is given by
\begin{align}
\mathcal{L}_j = 
\frac{\displaystyle 
    \exp\left(-\frac{1}{2}\left(
        \frac{\widehat F_j - F_{t,i,j}(\btheta_t)}{\widehat\sigma_j}
    \right)^2\right)
    \mathbb{I}_{\widehat F_j\ge 0}
}
{\displaystyle 
    \widehat\sigma_j\sqrt{2\pi} 
    \left(
        1-\mathcal N\left(
            -\frac{F_{t,i,j}(\btheta_t)}{\widehat\sigma_j}
        \right)
    \right)
} \text{,}\label{eq:likelihood2}
\end{align}
where $\mathbb{I}_{\widehat F_j\ge0}$ is the indicator function that the flux is non-negative and $\mathcal N$ is the cumulative distribution function of the standard normal distribution given by
\[
\mathcal N(z)= \frac 12 + \frac 12 \operatorname{erf}\left(\frac{z}{\sqrt{2}}\right).
\]
Note that the $\chi^2$-criterion that is used to produce the best-fit parameters in the frequentist approach appears in the exponent of the likelihood given in Eq.~\eqref{eq:likelihood2}. 

In a general context,  faint sources are measured with a low signal-to-noise ratio (SNR) in one or more bands. 
Forced photometry is a practice that is applied in many surveys, including Euclid, using detected images from which aperture fluxes are measured for all undetected sources. The likelihood is then calculated using Eq.~\eqref{eq:likelihood2} even when the forced photometry flux has a negative or null value. However, when this is impossible i.e.there is no estimation of the flux in a certain band, we re-write the likelihood with an unknown measured flux. This allows us to calculate the probability that a source 
is observed with a measured flux below the stated detection limit. This probability can be expressed as follows:

\begin{align}
\mathcal{L}_j = 
\frac{\displaystyle 
     \mathcal{N}\left(
        \frac{\widehat F_{lim} - F_{t,i,j}(\btheta_t)}{\sqrt{\widehat\sigma_j}}
    \right)
}
{\displaystyle 
        1-\mathcal N\left(
            -\frac{F_{t,i,j}(\btheta_t)}{\widehat\sigma_j}
        \right) }\text{.}
        \label{nondet}
\end{align}

In \texttt{Owl-z}, Eq.~\eqref{eq:weight}  is calculated in two separate steps: initially to build a model and subsequently during comparison of the model with the sources' photometry. 

 For each population, \texttt{Owl-z} calculates the colours in all bands to build a model by generating maps of the values of colour for each population for each parameter value. The SED and the desired ranges of the parameters are used. Subsequently, the colours are calculated through the integration of magnitudes, as outlined by \citealt{hogg}, across all SEDs employed. The model comparison step is located within the central portion of the code, where \texttt{Owl-z} examines each candidate for which comprehensive photometry is available (including associated errors), and contrasted it with the models. For each parameter set, it compares the photometry to all model fluxes. The model fluxes are calculated using the colours and model reference magnitude that we choose to be the $H$ band in this work. However, any reference magnitude could be selected, provided that a detection is made in this band.
 It should be noted that \texttt{Owl-z} does not calculate the probability from the best value of the weighted evidence but rather from the integration of all values. 
 Subsequently, the probability for each population is calculated using Bayes probability in Eq.~\eqref{bayes}. In addition, \texttt{Owl-z} incorporates snippets of code that facilitate the recovery of parameters enabling the calculation of the maximum a posteriori, thus providing the best-fit parameters for a given candidate (see Sec. \ref{output_owl}).

\subsection{\label{sc:high-$z$_QSO} The high-$z$ quasars population}

The statistical model of the quasar (QSO) population is based on a collection $n_\text{QSO}$ of SEDs (also referred to as spectral templates or templates) and a set of parameters $\btheta_\text{QSO}$ describing the quasars' properties. The spectral templates used in this paper are smoothed versions of the quasar composites presented in \citealt{Banados2016_template} and parametric SED models that accurately reproduce the observed optical and NIR colours of luminous Type 1 quasars over a wide range of redshifts and luminosities from \citealt{Temple}. The parameters are the redshift $z$ and the apparent magnitude in a reference band, which we choose to be the Euclid $H$ band, denoted \HE. This allows us to compute the expected fluxes in each band, $\mathbf F_{\text{QSO},i}(\btheta_\text{QSO})$ at the $\btheta_\text{QSO}=(\HE,z)$ value of the parameter set given the $i$-th SED model. The QSO likelihood of the data given $\btheta_\text{QSO}$ is defined as the mixture of multivariate densities in Eq. ~\eqref{eq:likelihood1}. For the purposes of our paper, we let the redshift vary in the range $7 \leq z \leq 12$. We aim for a maximum redshift of 12 as the Euclid bands will cover up to the \HE\ band, and the Lyman-alpha emission line stays in the \JE\ band until $z\approx12$.

\begin{figure}[t!]
    \centering
    \includegraphics[angle=0,width=1.0\hsize]{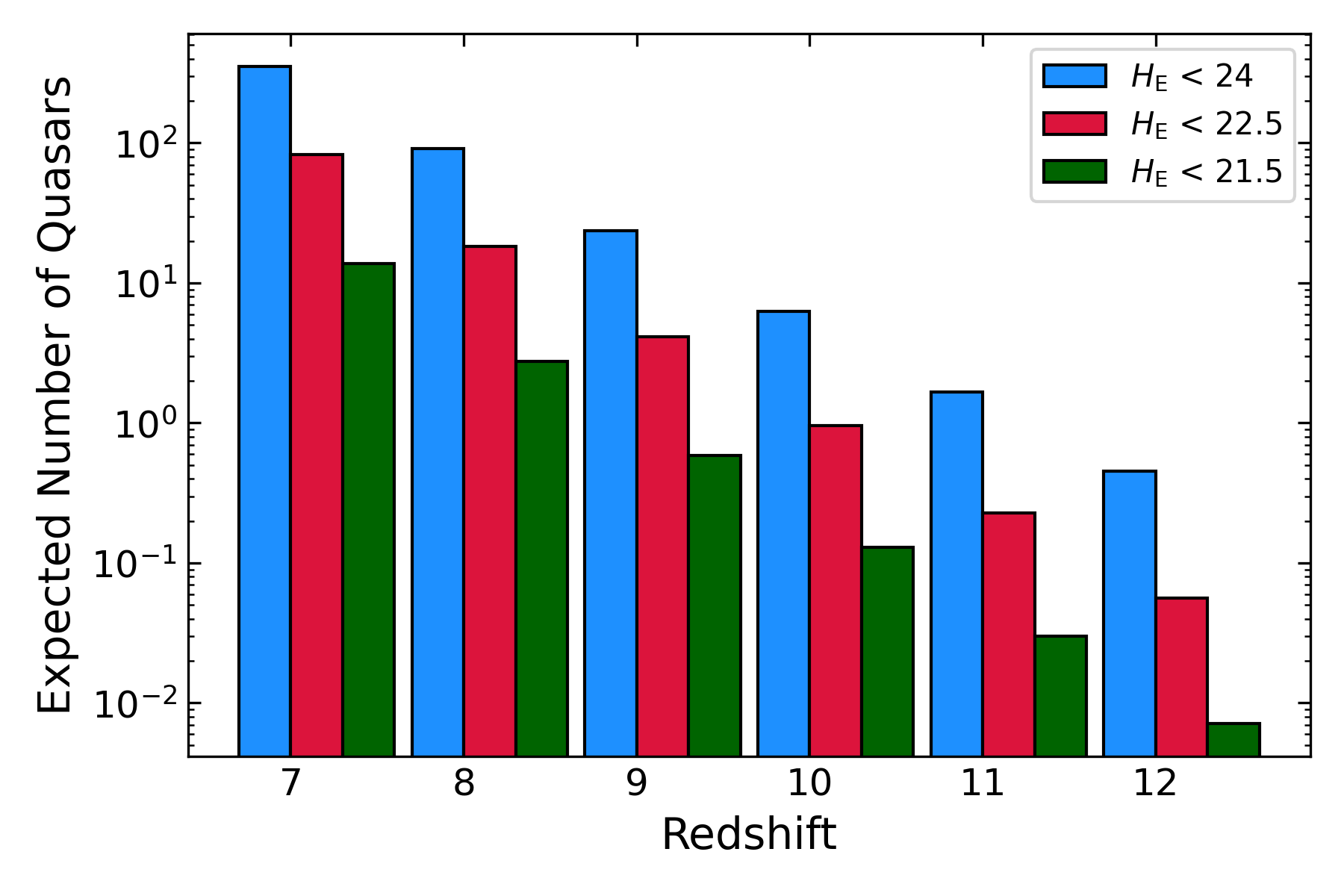}
    \caption{The expected number of quasars per redshift bin in the 15,000 deg$^2$ EWS  given the high-$z$ quasar Luminosity Function adopted in this paper. In blue are shown the numbers of quasars expected up to $\HE<24$, in crimson with $\HE<22.5$ and in green with $\HE<21.5$}
    \label{prior_qso_plot}
 \end{figure}

To characterise the prior distribution of high-$z$ quasars, we adopt the double power law parametric  Luminosity Function (LF) from \cite{Willott2010} and adjust the parameters fit to \citealt{Matsuoka2023}:

\begin{align}
    \Phi(M_{1450}, z)= \dfrac{  10^{k(z-7)} \Phi^*              }
    {10^{0.4(\alpha+1)(M_{1450}-M_{1450}^*)} +10^{0.4(\beta+1)(M_{1450}-M_{1450}^*)}   },
\end{align}
where $M_{1450}$ is the quasar rest-frame absolute magnitude at $\lambda=1450\text{\AA}$  and $\Phi(M_{1450}, z)$ is the number of quasars per magnitude bin, and unit volume at redshift $z$.
The parameters are the normalisation $\Phi^*$, the break magnitude $M_{1450}^*$, the bright end
slope $\beta$ and the faint end slope $\alpha$, their values can be found in Table~\ref{fit_LF}. 
Using the apparent magnitude in the \HE\ band, we can write the surface density of quasars in the sky in mag$^{-1} \times $deg$^{-2} \times$ d$z^{-1}$ units as:
\begin{align}
    \rho_\text{QSO}(H_E,z)=\dfrac{1}{4\pi}\times \dfrac{dV_c}{dz}\times \Phi \left[ H_{E} -\mu -K_{corr}(z),z  \right],
\label{eq:qso_surface_density}
  \end{align}
where \begin{itemize}
    \item $\mu$ represents the distance modulus, defined as a function of the luminosity distance $D_L$ by 
    \begin{align}
       \mu = \dfrac{5\log_{10}(D_L/10 \ \text{pc})}{\text{pc}},
    \end{align}
    \item $K_{\text{corr}}(z)$ denotes the K-correction, which converts the absolute magnitude at rest-frame $\lambda=1450\text{\AA}$ into the observed H-band magnitude, and
    \item $\dfrac{1}{4\pi} \times \dfrac{dV_c}{dz}$ is the co-moving volume element per steradian and per redshift interval $dz$.
\end{itemize}

Fig.~\ref{prior_qso_plot} illustrates the number of quasars per redshift bin and up to three different values of \HE\ across the entire 15,000 deg$^2$ of the EWS, as derived from Eq.~\eqref{eq:qso_surface_density}. This demonstrates that Euclid will detect tens of quasars brighter than \HE\ $\approx$ 22.5 between redshifts 7 and 8, and potentially a few up to z $\approx$ 9. Up to this \HE\ magnitude, the selection with \texttt{Owl-z} will be robust while becoming increasingly more hazardous beyond it.
\begin{table}[htbp!]
\caption{Quasar Luminosity function parameters}
\centering
\begin{tabular}{ccccc}
\hline
\cr $\alpha$ &$ \beta$&k&$M^*_{1450}$ & $\phi^* 10^{-8}$ $(\mathrm{Mpc}^{-3} \mathrm{mag}^{-1})$\\
 \hline
\cr $ -1.2$ &$-3.34$& $-0.47$& $-24.38$ &$0.475$ \\
\hline
\end{tabular}
\label{fit_LF}
\end{table}

\subsection{\label{sc:MLT-subsection}The MLT stars population}

\begin{figure}[t!]
    \centering
    \includegraphics[angle=0,width=1.0\hsize]{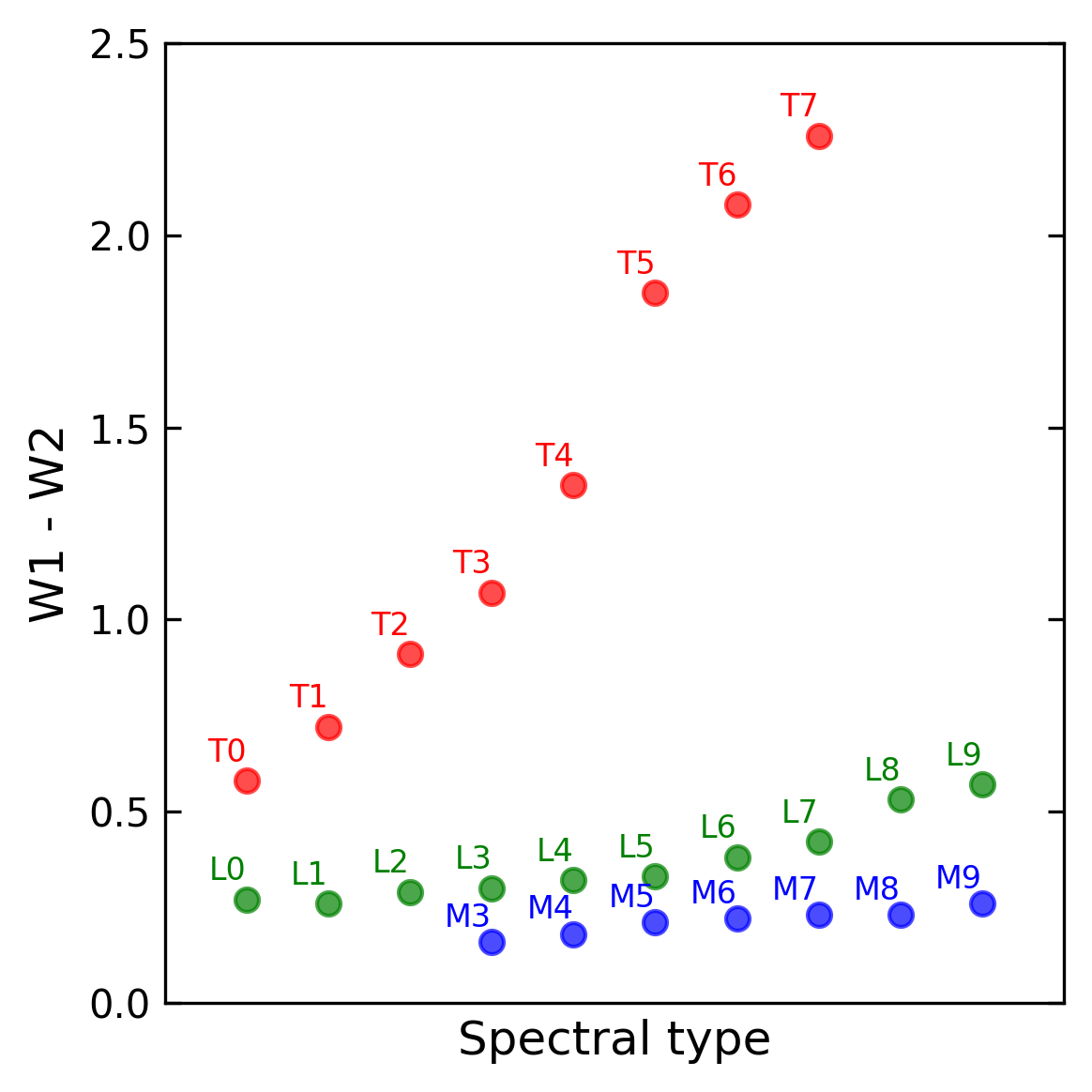}
    \caption{WISE $W1-W2$ colours of MLT dwarfs from \citealt{Best2017} shown in blue, green, and red, respectively, with the values indicating their spectral type.}
    \label{colorwise}
\end{figure}

For the purpose of our Bayesian analysis, we need to model the spatial distribution of MLT dwarfs, their luminosity, and spectral energy distributions from the optical to the near-infrared. For spectral energy distributions, we use the stellar dwarf Luyten Half-Second (LHS) library \citep{Bakos2002} for M3-M5 brown dwarfs and the brown dwarf spectral library SpeX prism \citep{Burgasser2014,Burgasser2017} for other spectral types. The use of spectral data enables magnitudes and colours to be correctly determined and modelled in the Euclid filters or any other filters as long as they overlap the spectral range of the spectra.
Moreover, there are between 5 and $\approx 50$ spectra available for each spectral type. 
Our models also include  WISE photometric data for the W1 and W2 bands, anchored to the average JHK photometry of brown dwarfs as tabulated in \citealt{Best2017}. This allows us to include in our model the WISE photometry that cannot be calculated from the MLT spectra that do not extend beyond the K band. Fig.~\ref{colorwise} shows the $W1-W2$ colours for all MLT types as indicated in \citealt{Best2017}, showing that T dwarfs, especially late types, have very strong $W1 - W2$ colours that can easily be discriminated from the colours of high-$z$ quasars.

To estimate the probability of brown dwarf contamination when searching for high redshift objects in wide-field imaging datasets, the spatial distribution of stellar dwarfs in the Galaxy is usually modelled and restricted to the thin disk of the Milky Way \citep{Caballero2008, Barnett2019, Pipien2018}. Detailed modelling of the thin disk is beyond the scope of this paper, and we therefore follow \citealt{Caballero2008} by assuming a similar thin-disk scale height for LT dwarfs to its value determined from earlier stellar dwarf types (GKM) and a simplified exponential vertical and horizontal distribution. For detailed theoretical modelling of thin disk parameters as a function of stellar age and metallicity, calibrated against Gaia and APOGEE data, see, for example,  \citealt{Sysoliatina2022}.

At the magnitudes corresponding to the depth of the EWS ($\sim 24$), the distance to T dwarfs varies from $\sim 150$ pc for the coldest to $\sim 450$ pc for the hottest. With a scale height of the thin disk of the order of 300 pc \citep{Recio-Blanco2023, Vieira2022}, we infer that the sole thin disk approximation will reasonably well represent the population of T dwarfs detectable in the EWS. However, the same will not be true for M dwarfs. An M6 dwarf at an apparent magnitude $\HE \approx 24$ is at a distance of $\sim$ 4000 pc, extending well beyond the thin disk into the thick disk or the halo. As a matter of fact, the \JWST has revealed in extragalactic fields a significant number of extremely faint and low-temperature brown dwarfs extending well into the thick disk and possibly into the halo, at distances of up to 2 kpc \citep{Hainline2024_1, Hainline2024_2, Burgasser2024}. An analysis of the thick disk contamination is therefore warranted.

For this analysis, we follow a similar approach to that used in \citealt{Ryan2016} and compare the distribution of stellar dwarfs in thin and thick disks as a function of magnitude. We adopt a value of 330 pc for the scale height of the thin disk \citep{Caballero2008}, 800 pc for the scale height of the thick disk \citep{Vieira2023}, 2250 pc  \citep{Caballero2008} for the value of the scale length of both disks and a normalisation factor between the thick and thin disks local densities of 10\%.  Using Eq. (12) of \citealt{Pipien2018}, we compare brown dwarf densities for different spectral types in the thin and thick disks for two lines of sight: $b = 90^{\circ}$ and ($l$, $b$) = ($90^{\circ}$, $30^{\circ}$) (the lowest Galactic latitude of the EWS is $b = 23^{\circ}$). For local volume densities in the Galactic plane and absolute Euclid magnitudes by spectral type, we refer to Table 2 of \citealt{Barnett2019} and the references therein \citep{Dupuy2012,Skrzypek2016,Bochanski2010}. Late-type M dwarfs can potentially contaminate searches for galaxies or quasars at high redshift in some surveys, whereas, as we will see later, only L and T-type brown dwarfs contaminate the search for $z \ge 7$ with Euclid. Consequently, for the purposes of this paper, we restrict our analysis to the brown dwarf population within the M dwarf population, that is, dwarfs of a type later than approximately M6. 

\begin{figure}[t!]
    \centering
    \includegraphics[angle=0,width=1.0\hsize]{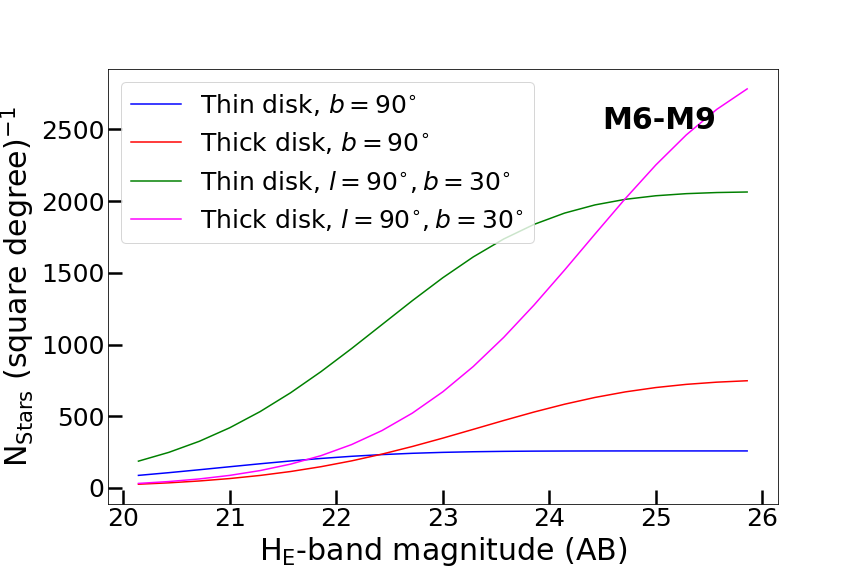}
    \includegraphics[angle=0,width=1.0\hsize]{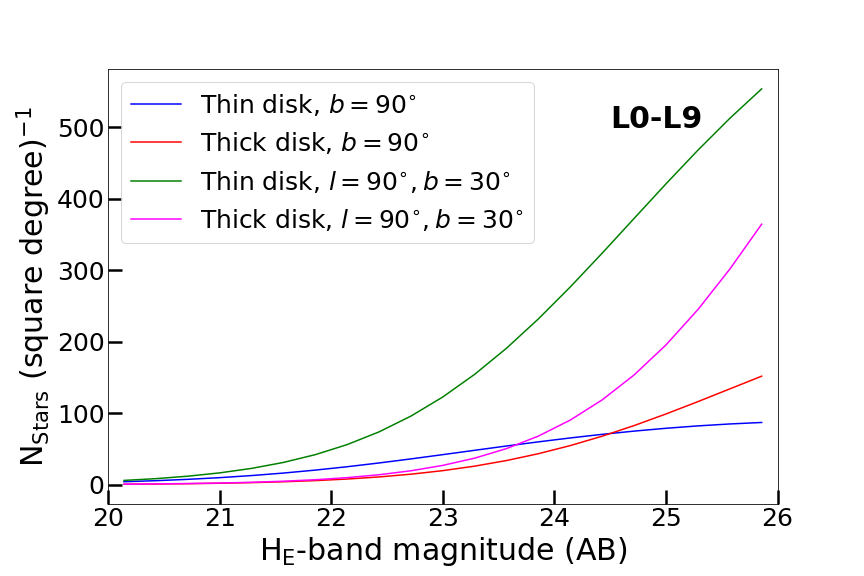}
    \includegraphics[angle=0,width=1.0\hsize]{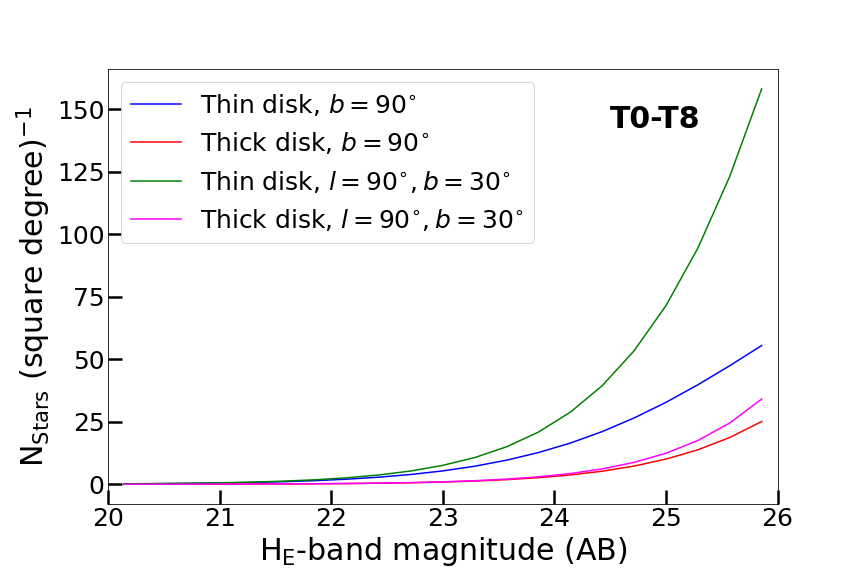}
    \caption{Brown dwarf number counts as a function of H-band magnitude for the thin and thick discs in two different field locations ($b=90^{\circ}$ and ($l$,$b$)=(90$^{\circ}$,30$^{\circ}$)). Top: M6 to M9 spectral types, middle: L0 to L9 and bottom: T0 to T8.}
    \label{fig:thin-thick}
\end{figure}

The analysis (see Fig.~\ref{fig:thin-thick}) indicates that up to Euclid magnitude $\HE \sim 24$, the thin disk dominates the number of L and T dwarfs at low and high Galactic latitudes. For M6 to M9 dwarfs, the thick disk is the main contributor above $\mathrm{H} \sim 22.5$ magnitudes at high Galactic latitudes, whereas at low Galactic latitudes, the thin disk dominates up to  $\HE \sim 24.5$ and beyond. For the purposes of our paper, Galactic latitude is, therefore, the dominant parameter affecting the population of contaminating stars. We explore the effect of Galactic latitude on the performance of our code in Sect.~\ref{Validation}.

This simple analysis assumes that stellar dwarf populations are the same in both thin and thick disks. This oversimplification ignores the differences in age and metallicity between the constituents of the thin and thick disks and, therefore, the significant differences that may exist between the stellar dwarf populations within them. Early M-type dwarfs (M0 to M5) have masses above the hydrogen-burning minimum mass, and their luminosity evolves little on time scales up to 10 billion years, whereas brown dwarfs below this limit cool on much faster time scales \citep{Reid2013}. As a result, late M-type dwarfs evolve into later spectral types, and their luminosities drop rapidly over timescales of a Gyr or less. The thick disc, which is older than the thin disc, should, therefore, be depleted of late-type brown dwarfs (M6 to M9 and L), and T dwarfs should be significantly less luminous in the thick disk than in the thin disk (see also the discussion in \citealt{Caballero2008}). This suggests that the stellar densities reported in Fig.~\ref{fig:thin-thick} are likely to be seriously overestimated in the thick disk at a given magnitude for spectral types later than M5. 

As a conclusion to this analysis, we restrict the modelling of the Galaxy's brown dwarf population to the thin disk and analyse the impact of stellar density through its dependence on Galactic latitude. In the coming years, Euclid and \JWST data will enable a more detailed understanding of the relative populations of brown dwarfs in the thin and thick disks.

We now return to the formulation of our Bayesian model for the description of the stellar dwarf population. The parameters of this population are magnitude (or heliocentric distance), and spectral type $\{\HE,spt\}$. We use the Euclid \HE\ band as the reference band for calculating magnitudes and the spectral types from M6 to T9. Using the notation described in Sect.~\ref{sec:proba}, we write the weighted evidence of this population as follows:

\begin{equation}
     W_{s}=\sum_{M6}^{T9}\int_{-\infty}^{+\infty} \rho_\mathrm{s}^{(l,b)}(H_{mod},spt)\prob(D|H_{mod},spt)\mathrm{d}H_{mod}\text{,}
\end{equation}

\noindent where $\rho_\mathrm{s}(H_{mod},spt)$ is the surface density of the contaminating brown dwarf population of Euclid magnitude $H_{mod}$ and spectral type $spt$, at a given Galactic longitude $l$ and latitude $b$. We model the spatial distribution across the Galaxy  in $\mathrm{mag}^{-1} \times \mathrm{deg}^{-2} \times \mathrm{spt}^{-1}$  units as a function of the heliocentric distance to the star $d$, assumed to be far smaller than the solar galactocentric distance $R_{\odot}$ \citep{Caballero2008}:

\begin{equation}
\label{eqapprox2}
\rho_\mathrm{s}^{(l,b)}(H_{mod},spt) \approx \rho_{0,spt}\, \exp \left({\frac{\mp Z_{\odot}}{h_{Z}}}\right)\mathcal{R}(d_{H_{mod}},l,b)
\mathrm{,}
\end{equation}
where,
\begin{equation}
    \mathcal{R}(d_{H_{mod}},l,b)=\exp \left[{-d_{H_{mod}}\left(-\frac{\cos(b)\cos(l)}{h_{R}}\pm \frac{\sin(b)}{h_{Z}}\right)}\right]
\end{equation}
\noindent where $\rho_{0,spt}\, \exp \left({\frac{\mp Z_{\odot}}{h_{Z}}}\right)$ is the local volume density of brown dwarfs of spectral type $spt$, $Z_{\odot}$ the height of the Sun above the Galactic plane (assumed to be 27 pc), \(h_Z\) and \(h_R\) the scale height and length of the thin disk as mentioned above, and the sign convention indicates whether the source is above or below the Galactic plane \citep[see][]{Caballero2008}. Finally, $H_{mod}$ and $d_{H_{mod}}$ are related through:

\begin{align}
    H_{mod}-M_{H_{mod}}=5\log (d_{H_{mod}})-4+3.09~E(B-V )\text{,}
\end{align}

\noindent where $H_{mod}$ and $M_{H_{mod}}$ are the apparent and absolute magnitudes in the reference band \HE\ and $E(B-V)$ the interstellar reddening.

\begin{figure}[t!]
    \centering
    \includegraphics[angle=0,width=1.0\hsize]{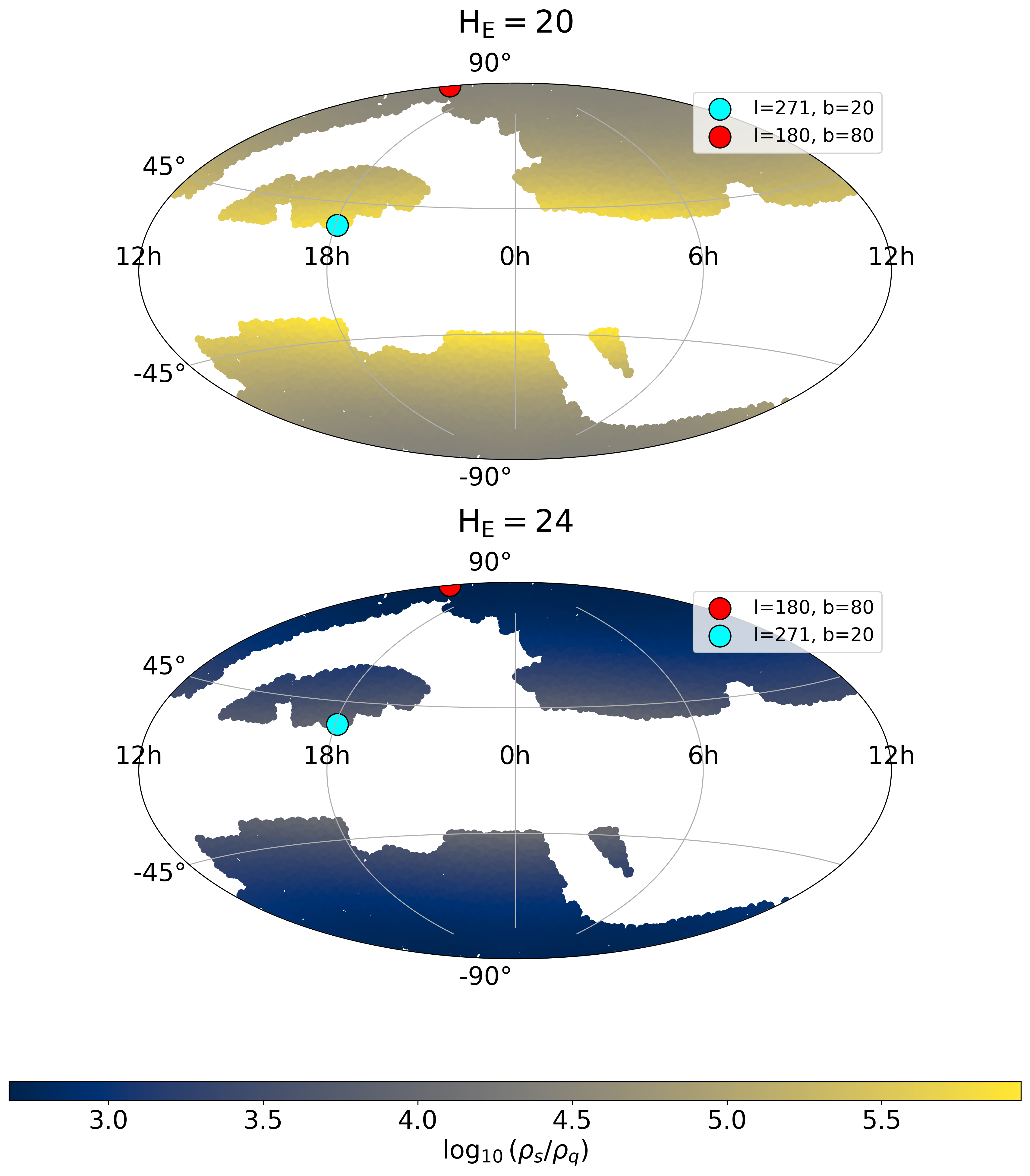}
    \caption{Ratio of the surface density of brown dwarfs of spectral type M6 to T8 to the surface density of $z>7$ quasars over the Euclid footprint shown in Galactic coordinates and galactic projection. Top: for magnitude \HE = 20, bottom for magnitude \HE = 24.}
    \label{prior_MLT_ciel}
\end{figure}
\begin{figure}[t!]
    \centering
    \includegraphics[angle=0,width=1.0\hsize]{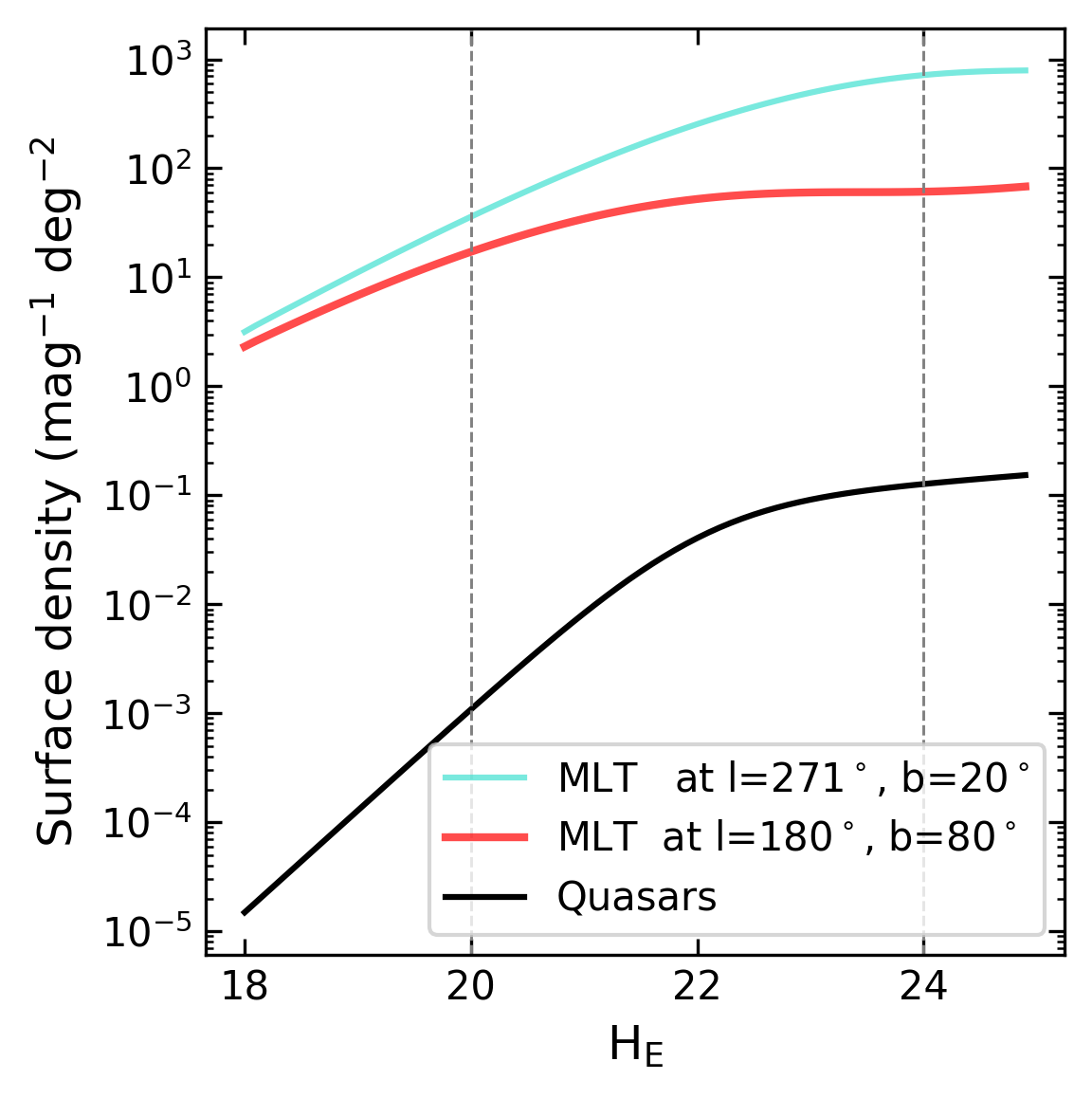}
    \caption{
    Surface number densities of MLT stars compared to $z>7$ quasars, as a function of apparent magnitude, in two different locations of the Euclid footprint. }
\label{prior_MLT_20}
\end{figure}

Using the modelling of the high-$z$ quasar population (Sect.~\ref{sc:high-$z$_QSO}) and the MLT population described in this section (M6 to T9), we derive the ratio of the brown dwarf to high-$z$ QSO density on the Euclid footprint. This is shown in Fig.~\ref{prior_MLT_ciel} for two different apparent magnitudes.

Fig.~\ref{prior_MLT_20}  shows the densities of quasars and M6 to T9 brown dwarfs for two positions with different Galactic coordinates in the Euclid footprint.

\subsection{The contaminant galaxy model}\label{sc:galaxy_model}
There are two different types of galaxies susceptible to contaminating the quasar samples because they display red colours that mimic those of  high-$z$ quasars, in particular at low SNR: early-type galaxies at intermediate redshift ($1<z<2$) on the one hand, and dusty star-burst galaxies on the other hand. The former exhibit an important 4000$\text{\AA}$ break because they are dominated by an old stellar population, whereas the presence of dust reddens the latter. To assess the extent of this contamination by intermediate- $z$ galaxies, we have carried out two different tests. The first one consists of precisely determining the contaminant population by using straightforward simulations. The second one is to quantify the contamination in the real Universe based on the COSMOS2022 field  \citep{Weaver_2022} by computing the Luminosity Function (LF) of the contaminant population. The LF will be used later to model the abundance of contaminant galaxies.  

To precisely identify the contaminant population, we run \texttt{Owl-z} on a simulated catalogue containing 100,000 galaxies uniformly sorted in redshift ( $0\le z\le 6$), spectro-photometic templates, and extinction values (between $0\le A_V\le 3$ magnitudes), using the Calzetti extinction law \citep{Calzetti}. The simulated catalogue was generated using \texttt{make\_catalogue}, a tool from the HyperZ tool \citep{Bolzonella2000}. The template library includes two evolutionary synthesis models: a delta burst -Single Stellar Population (SSP)- from the Bruzual \& Charlot code \citep{BC2003},
with Chabrier IMF \citep{Chabrier2003} and solar metallicity, and 
10 Starbursts99 templates \citep{STARBURST99}, including emission lines, for single bursts and constant star formation rate models, each one spanning five metallicities ($Z=$0.04, 0.02, 0.008, 0.004 and 0.001) and 37 ages for the stellar population (between 0 and 1 Gyr). Apparent magnitudes and associated errors have been computed using these templates, sorted to uniformly sample the range of magnitudes where galaxies are expected to be the dominant population in the EWS, in the reference \HE-band, that is $22\le \HE \le 25$. Photometric error bars and noise are scaled to apparent magnitudes, assuming a Gaussian distribution, according to the expected SNR in the different bands used in the EWS. The probability threshold for the selection as a quasar is set to $P_q=0.1$ for this analysis, following \citealt{Mortlock_2012_bayes} and \citealt{Barnett2019}.

The first result of this experiment is that young starbursts, represented by the Starburst99 models, only contaminate the sample at low SNR values for extremely high values of $A_V \ge1.5$, spanning a relatively broad domain in redshift ($1\le z\le 5$). It is important to acknowledge the significant influence of SNR values in the reference magnitude on our selection process.  Indeed, a low SNR in the reference filter $H$ induces a noisy determination of the $J-H$ colour,
which in turn leads to a higher degree of confusion in the probability calculation and a greater likelihood of misidentifying high-$z$ quasars.  This is a general remark affecting all contaminant galaxies but is particularly important for dusty starbursts. The main contaminants among this population are galaxies at $z\gtrsim 4$ with $A_V \ge1.5$, and galaxies at $1\le z\le 2$ with $A_V \ge3$. In the two cases, such extremely reddened galaxies are relatively rare in the real Universe at these redshifts. 

Our analysis shows that the main contamination comes from early-type galaxies, represented by SSP models, with ages above 1 Gyr, at intermediate redshifts, $1<z<2$.  Indeed, the spectral energy distribution of early-type galaxies, such as elliptical and lenticular galaxies, is characterised by an old stellar population, well represented by an initial burst of star formation followed by a rapid decline \citep[see e.g.][]{Sadman}. Contrary to dusty starbursts, these galaxies are relatively abundant at $1<z<2$. For this reason, in the following, we consider that the main contamination comes from early-type galaxies. In this regard, our results are consistent with the assumptions of previous works \citep[e.g.][]{Barnett2019}. 
The abundance of these galaxies has been well studied at $z\sim1.5$, for instance, by   \citealt{Zucca2006}, who studied the evolution of the LF for different filters and spectrophotometric types of galaxies. We can use these previous findings to guide our modelling in the sensitive magnitude domain, as shown below.

After identifying the nature of the contaminant galaxies, we determine the LF of this population. For this need, we use the Euclidised COSMOS2022 catalogue \citep{Weaver_2022}, named E-COSMOS hereafter. This catalogue is described in more detail in Sect.~\ref{sc:Purity}. Note that for galaxies detected in the \JE-band at intermediate redshift, the relevant filter for determining the LF is the rest-frame B band. 

To represent the LF, we use the classical parameterisation provided by the  Schechter function \citep{LF} in terms of magnitude given by:
\begin{equation}
    n(M)=(0.4\ \ln 10)\ \phi ^{*}\ [10^{0.4(M^{*}-M)}]^{\alpha +1}\exp[-10^{0.4(M^{*}-M)}].
    \label{Paul}
\end{equation}

Here, $n(M)~dM$ represents the number of galaxies per co-moving Mpc$^3$ with magnitudes between $M$ and $M + dM$. The parameters are as follows:
\begin{itemize}
    \item $\phi^{*}$: The normalisation factor, representing the overall volume density of galaxies in Mpc$^{-3}\times$ mag$^{-1}$.
    \item $M^{*}$: The characteristic magnitude represents the cut-off between the (bright) luminosity regime dominated by the exponential function and the (faint) regime dominated by the power-law.  
    \item $\alpha$: The faint-end slope of the luminosity function, describing the distribution of the faintest galaxies.
\end{itemize}

 For the sake of consistency, before selecting the target population of contaminant galaxies in E-COSMOS, we run \texttt{Owl-z} on the E-COSMOS catalogue, with only one contamination source: MLT stars, and the same threshold as before for the selection of high-$z$ quasars. We check the spectrophotometric type and redshift of the 16 galaxies that have been selected as quasars. They are all early-type galaxies at intermediate redshift ($z\sim1.5$), as expected from previous simulations. 

In order to model the LF for this population of contaminants, we extract from the E-COSMOS catalogue the subset of early-type galaxies within the sensitive redshift domain $1<z<2$.
To compute absolute magnitudes, we model the spectral energy distribution with a template well suited to represent this population, namely a short, exponentially decaying model with characteristic star-formation time $\tau = 0.1$ Gyr, age $= 10~\text{Gyr}$ and 
solar metallicity ($Z = 0.02$). It is worth mentioning that the precise choice of this model does not change the results. The LF points and their associated error bars are then computed; error bars in the LF  include Poisson noise and field-to-field variance. 
Field-to-field variance is derived using the Trenti \& Stiavelli method and calculator \citep{Trenti-Stiavelli08}. 
We then fit the data points with the Schechter function in Eq. ~\eqref{Paul} using a $\chi^2$ minimisation. 

\begin{figure}[t!]
    \centering
    \includegraphics[angle=0,width=1.0\hsize]{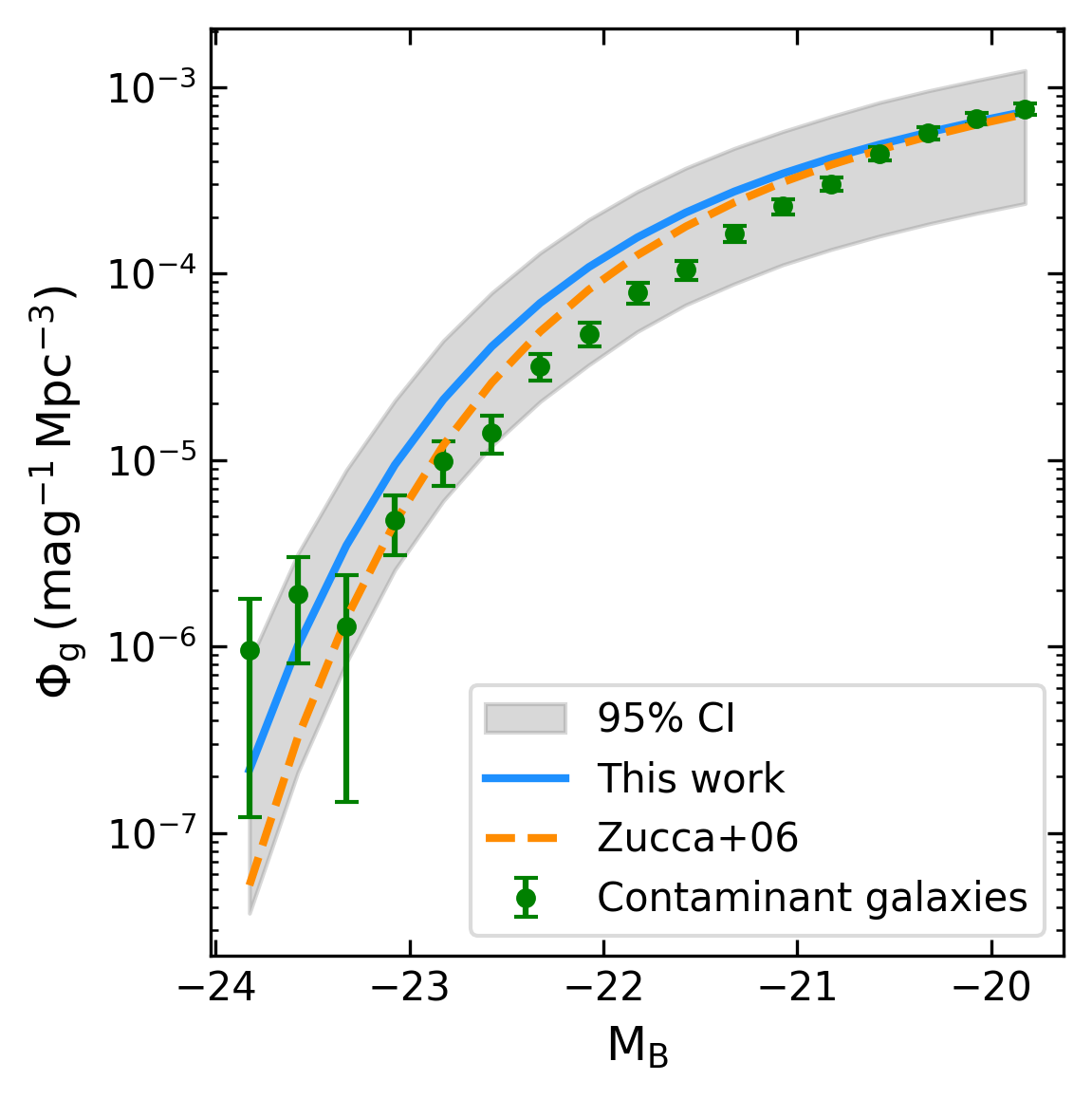}
    \caption{Luminosity function fit in the B-band (blue) in addition to its 95\% confidence interval (grey) for the population of early-type galaxies at $1<z<2$ in the  E-COSMOS catalogue data (green),  compared to the early-type galaxies (dashed orange) luminosity function fit by \citealt{Zucca2006}. }
    \label{LF_plot}
\end{figure}

\begin{figure}[t!]
    \centering
    \includegraphics[angle=0,width=1.0\hsize]{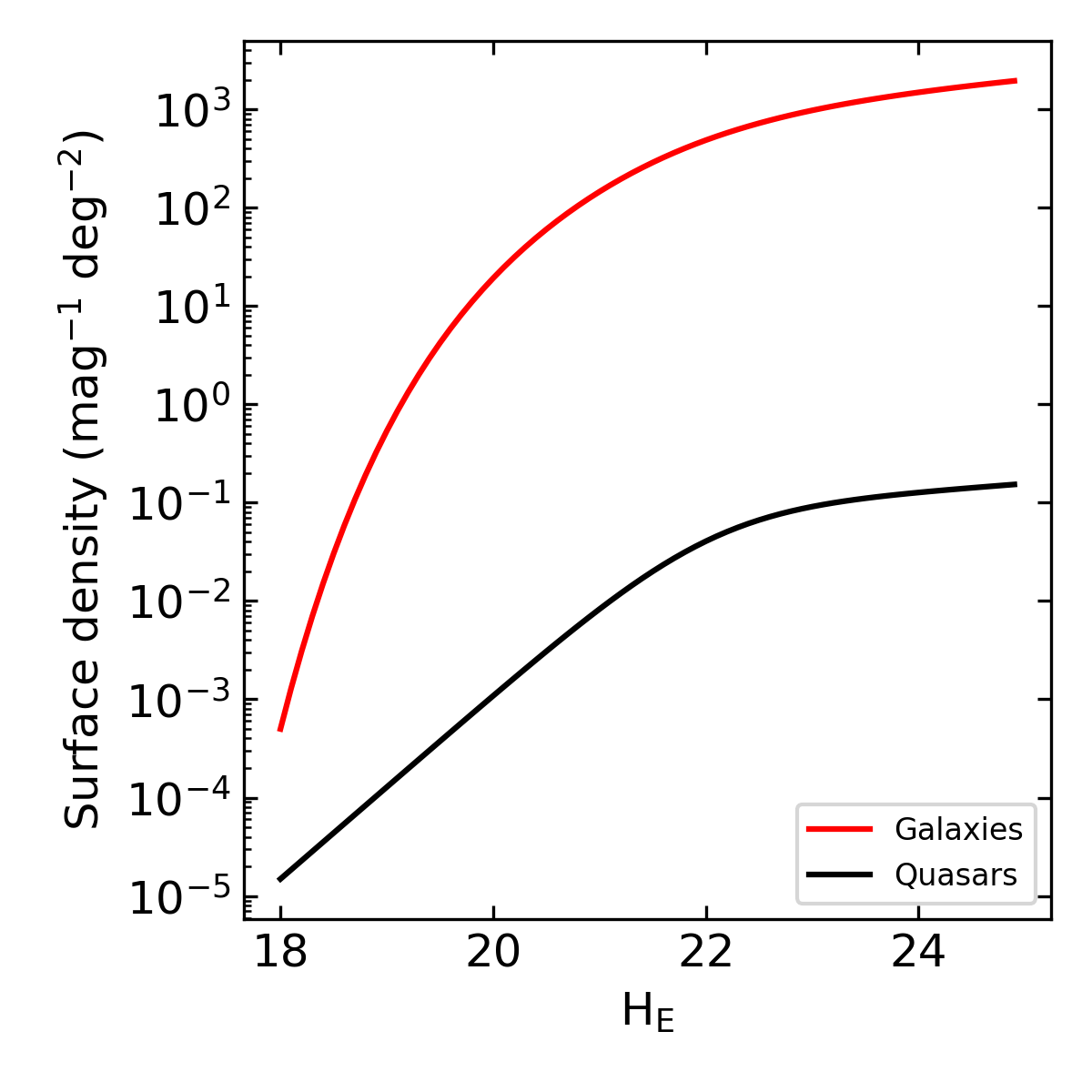}
    \caption{
     The surface number density of early-type galaxies integrated in the redshift interval  $1<z<2$ compared to the quasars surface number density integrated over the redshift interval $7<z<12$ quasars, as a function of apparent magnitude.
    }
\label{Prior_QSO_GAL}
\end{figure}

\begin{table}[htbp!]
\caption{Comparison of the Schechter function parameters.}

\centering
\begin{tabular}{lccc}
\hline
 & $\alpha$ & $M^*$ & $\phi^*$\\
 &   &   &  $\times 10^{-4}$ Mpc$^{-3}$\\
 \hline
\cite{Zucca2006} & $-1.23 $& $-21.5^{+0.48}_{-0.57} $ & $3.0^{+1.4}_{-1.2}$\\
\hline
This work &$ -1.23$ & $-21.7^{+ 0.17}_{- 0.17} $ &$ 2.9^{+0.6}_{-0.6}$ \\
\hline
\end{tabular}
\label{fit_LF_table}
\tablefoot{Comparison with the results by \citet{Zucca2006} for early-type galaxies at $1<z<1.2$. The value of the slope has been fixed to $\alpha=-1.23$ in our fit.}
\end{table}

Towards the faint end, we applied a magnitude cut to $M_B< - 19.7$ because beyond this limit, our sample suffers from incompleteness. 
The parameters were chosen to enable easy comparison with the work of \citealt{Zucca2006}. 
This motivated us to fix the same value for the parameter $\alpha$. For comparison, a value of $\alpha=1.0$ gives a similar result to the function adopted by \citealt{Barnett2019} to represent the same population. Note that the precise choice of $\alpha$ is irrelevant here because, on the one hand, only the faintest part of the LF is dominated by the power law and, on the other hand, the population of contaminant galaxies lies preferentially in the bright end. 

Fig.~\ref{LF_plot} shows that our LF points are in good agreement with \citealt{Zucca2006}. Our fit parameters are shown in Table \ref{fit_LF_table}.
The LF obtained above is then used to model the prior distribution of early-type galaxies $\rho_{g_i}(H_{\text{mod}}, z)$ given in mag$^{-1} \times$ deg$^{-2} \times$ d$z^{-1}$  units by:
\begin{align}
    \rho_g(H, z_g) = \dfrac{1}{4\pi} \times \dfrac{dV_c}{dz_g} \times \Phi_g \left[ H_{\text{mod}} - \mu - K_{\text{corr}}(z_g), z_g \right]\text{,}
\end{align}
where $z_g$ is the redshift range of the contaminant galaxies $1 < z < 2$ and $\Phi_g$ is the luminosity function fit obtained in this work.

For a given set of parameters $\{z_g, H_{\text{mod}}\}$ and an early-type galaxy model $g_i$, the weighted evidence can be quantified as:
\begin{align}
    W_{g_i} = & \int_{-\infty}^{+\infty} \int_{-\infty}^{+\infty} \rho_{g_i}(H_{\text{mod}}, z_g) 
 P(\text{bands}, H_{\text{obs}} | H_{\text{obs}}, g_i) dH_{\text{mod}} dz_g.
\end{align}

 Fig.~\ref{Prior_QSO_GAL} displays a comparison between the surface number density of early-type galaxies in the $1<z<2$ interval and $z>7$ quasars as a function of apparent magnitude. The population of contaminant galaxies clearly dominates the number counts.

\section{\label{Technical}Technical description of \texttt{Owl-z}}
The \texttt{Owl-z} code calculates the probability of each preselected candidate belonging to one of the modelled populations, namely high-$z$ quasars, early-type galaxies at intermediate redshifts, and MLT dwarfs. The code is self-contained and written in the Python programming language and its various packages, such as  NumPy, SciPy, and Astropy. It is capable of functioning without any dependencies on external software. 

\subsection{Inputs}
 
The \texttt{Owl-z} code relies on the computation of colours from the SEDs for all three populations in addition to the K-corrections for quasars and galaxies. To do so, the model requires the filter transmission curve data for each band, as well as a designated reference magnitude.
Additional parameters are the magnitude range and the redshift range to explore, for each population. 
In the case of brown dwarfs, for each spectral type, the range of absolute magnitudes and the local densities should be provided. Templates for each population should be provided, 
and the code distribution already includes a large number of them for the three populations considered.  
It is possible to add as many templates as required, provided that they are in the correct format. 
\subsection{Outputs}\label{output_owl}

The program provides a list of probabilities of belonging to each class for each source . The user then interprets the data and makes a decision by comparing these probabilities with a threshold $\zeta$. Furthermore, it provides a list of useful output parameters, including the optimal redshift for galaxies and quasars and the optimal spectral type for the best-fit stellar model. It also provides the SEDs of the galaxies and quasars that provide the maximum posterior condition. Note that the "best-fit" output of a Bayesian code is  the "maximum posterior" (MAP), calculated by maximising the posterior, that is $\max(w_q)$, where $w_q=(w_{q_1},w_{q_2}, ...,w_{q_{n_i}})$ are given by 

\begin{align}
    w_{q_i}=\rho_{q_i}(H_{\text{mod}}, z)
    \times P(\text{D}, H_{\text{E}} | H_{\text{E}}, q_i)\text{.}
\end{align}

In the following sections, we use the terms best-fit or MAP, but we always refer to the same Bayesian definition above.

\subsection{Efficiency}
\texttt{Owl-z}  has several  efficiency indicators:
\begin{itemize}
\item \texttt{Owl-z}  requires the user to fill a configuration file which is straightforward to modify, requiring only the input of the user to enable its operation. 

\item The code is self-sustained, whereby the models and K-corrections are calculated concurrently with the code and the input parameters.
\item The code is flexible regarding the numbers and nature of bands used, provided that they do not overlap.
\item  The code is written with high-performance computing methods in mind and is partly written in Cython for the purpose of enhancing numerical speed.
 \item In addition to providing a probability value, the model also provides additional output parameters for each source, thus facilitating a more accurate interpretation of the results.
\end{itemize}
\subsection{Limitations}
A natural limitation of the code is that the filters correspond to the spectral range of the templates used, for brown dwarfs on the one hand and for galaxies and quasars in the redshift ranges to be explored for each of them.

In the released version of the code corresponding to the one described in this paper, the brown dwarf templates are observed spectra and cover up to 2500 nm, thus imposing that the filters that can be provided are limited to the K-band. The galaxy and quasar templates have a sufficient spectral range for the redshift ranges explored in this paper and do not impose additional limits.

As will be seen later in this paper, in Sect.~\ref{sec:reidentification}, the colours in the WISE filters have been used to further model the SED beyond the K-band, but this is exclusively possible with the tabulated data provided in \citealt{Best2017} and used in the code and cannot be generalised to other datasets extending beyond K-band. Extending the filters that can be used beyond the K-band would require brown dwarf templates with extended spectral ranges that are not readily available.

\section{\label{Validation}Validation and Performance}

\subsection{\label{sec:methodology}Methodology}
This section is dedicated to the validation of \texttt{Owl-z} using two different approaches. The first one consists of re-identifying all the  $z>7$ spectroscopically confirmed quasars available in the literature. To achieve this, \texttt{Owl-z} was configured on a case-by-case basis with the photometric data from which the individual quasars were identified.
The second approach relies on simulating different flavours of the EWS catalogues, allowing us to quantitatively evaluate the performance expected for \texttt{Owl-z} based on the measurement of the completeness, a measurement of how many (if any) quasars we misidentify with this process and purity, a measurement enables us to ascertain whether we are still contaminated among the objects identified as quasars by \texttt{Owl-z}. A global performance metric is also introduced, combining both completeness and purity. Performance estimators are important for optimising the selection of high-$z$ quasar candidates with \texttt{Owl-z}, minimising the demand for spectroscopic follow-up and maximising the confirmation rate with a reasonable observational effort.

\subsection{\label{sec:reidentification}Re-identifying known  $z>7$ quasars}

\begin{figure*}[t!]
    \centering
    \includegraphics[width=0.9\textwidth]{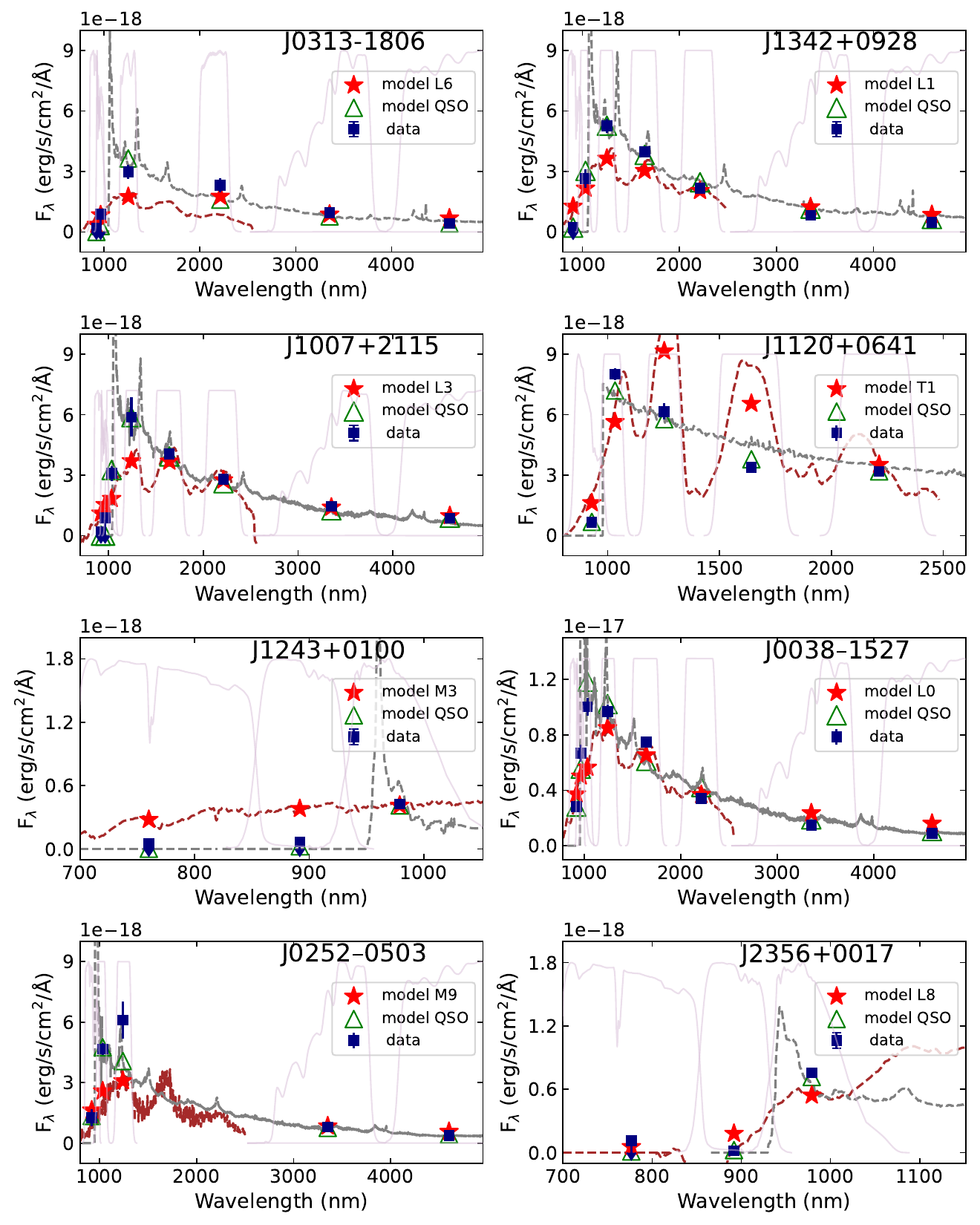}
    \caption{Photometric data for the eight spectroscopically confirmed quasars at $z > 7$ and the \texttt{Owl-z} best-fit solutions for quasars and MLT dwarfs. The original photometric data points are represented by blue squares. The best-fit quasar model is in dashed grey, the best-fit model for an MLT dwarf is in dashed brown, and the filter transmission curves are shown in solid light purple. Model photometry for best-fit solutions is represented by green triangles for quasars and red stars for MLT dwarfs.}
    \label{plot8}
\end{figure*}

We analyse the performance of our code in recovering known and spectroscopically-confirmed quasars at  $z>7$ and selected from near-IR surveys. The versatility of \texttt{Owl-z} allows us to apply it to the different optical and NIR  data sets used in the discovery of these quasars. The objective here is to determine if \texttt{Owl-z} is able to identify them as quasars and to compare their spectroscopic redshift with the redshift obtained by the code. 

Table~\ref{table8} lists the quasars used for this analysis, the photometric bands used for their discovery, their measured spectroscopic redshifts, and their reference magnitude.

 The photometric data, different for each quasar, include $z$-band data ($z_{DECam}$) from DECaLS \citep{DECALS}, $z$- and $y$-band data ($z_{PS1}$ and $y_{PS1}$) from The Pan-STARRS1 Surveys \citep{PS1}, $g$-, $r$-, $i$-, $z$- and $y$-band data ($g_{HSC},r_{HSC},i_{HSC},z_{HSC},y_{HSC}$) from Subaru/HSC and $Y$-, $J$-, $H$- and $K$-band data from UKIDSS \citep {UKIDSS}. Data from WISE \citep{WISE} were also used. Given the shallow depth and non-constraining photometry in the W3 and W4 bands, we only use photometric data from the W1 and W2 bands in the following.
\begin{table*}[ht!]

\caption{Known quasars at $z>7$}
\label{table8}
\centering

\begin{tabular}{lccccccc}
\hline
\hline
\textbf{QSO name }& \textbf{Photometry}&$\mathbf{z_{spec}}$  & $\mathbf{z_{out}}$& $\mathbf{P_q}$ &$\mathbf{P_{q_2}}$&\textbf{mag} & \textbf{Reference}\\
\hline

J0313–1806 &   $z_{PS1}$,$y_{PS1}$ ,$J$,$K $, $W1$,$W2$ &7.64 &7.7&1 & 0.25&$y_{PS1}$ >23.13 & \citealt{Wang2021}\\

J1342+0928  & $z_{DECam}$, $Y$,$J$,$H$,$K$, $W1$,$W2$ &7.54  & 7.8 &1&0.9& Y 21.47$\pm 0.19$ &\citealt{Banados2018}\\

J1007+2115  & $z_{DECam}$,$y_{PS1}$, $Y$,$J$,$H$,$K$, $W1$,$W2$& 7.51& 7.65&1&1& $y_{PS1}$ 21.3$\pm0.13$ & \citealt{Yang2020}\\

J1120+0641 & $z_{DECam}$, $Y$,$J$,$H$,$K  $& 7.08 &7.05&1&-&  Y 19.63$\pm0.04$& \citealt{Mortlock2011}\\

J1243+0100& $g_{HSC},r_{HSC},i_{HSC},z_{HSC},y_{HSC}$&  7.07&6.9&1& -& $y_{HSC}$ 23.57$\pm0.08$   & \citealt{Matsuoka2019J124353.93+010038.5}\\

J0038–1527&   $z_{DECam}$,$y_{PS1}$, $Y$,$J$,$H$,$K$, $W1$,$W2$&7.02 &6.95&1 &0.9&$y_{PS1}$ 20.61$\pm0.1$  & \citealt{Wang2018}\\

J0252–0503  & $z_{DECam}$, $Y,J, W1,W2$ & 7.02&6.9&1& 10$^{-5}$& Y 20.85$\pm0.07 $ & \citealt{Yang2019}\\

J2356+0017 & $ i_{HSC},z_{HSC},y_{HSC}$&7.01 & 6.75&1&- & $y_{HSC}$  22.94$\pm0.05$  & \citealt{Matsuoka2019}\\
\hline
\hline

\end{tabular}
\tablefoot{Spectroscopically confirmed quasars at $z>7$ with photometric data used for discovery and $z_\mathrm{spec}$. Columns $z_\mathrm{out}$ and $P_q$ are from \texttt{Owl-z}; $P_{q_2}$ is given for objects with WISE data.}
\end{table*}

The pre-selection methods used to select these quasars were mainly based on colour selection and the identification of strong Lyman breaks ($z-J>4$) around $\sim 1 \mu$m. An additional criterion was used for the five cases out of eight where WISE photometry was available, consisting of a $W1 - W2 <0.7$ colour selection enabling the rejection contamination by T-dwarfs (see Fig.~\ref{colorwise}).

The results of \texttt{Owl-z} are also presented in Table~\ref{table8}, in particular, the output redshifts and $P_q$ values returned by the code. As further discussed in Sect.~\ref{photoz}, the \texttt{Owl-z} output redshifts are in good agreement with the actual redshift of the objects. The primary potential contaminants expected for the QSOs in this study are L brown dwarfs. \texttt{Owl-z} yields probability $P_q>0.99$ for all of the QSOs above, showing the high performance of the code in re-identifying the whole sample. Fig.~\ref{plot8} shows the photometric data, including error bars, for all quasars and the best-fit quasar and MLT dwarf solutions returned by the code.

Some results deserve a specific comment. \texttt{Owl-z} has provided an excellent result for all quasars, including those detected in a single band with non-detection constraints in the others (J2356+0017 and J1243+0100) because the photometric bands in HSC are very deep, and the  Ly$\alpha$ break much easier to detect. 
These quasars have broad Ly$\alpha$ emission and are low-luminosity quasars, matching well the properties 
 of the quasar population model (Sect.~\ref{sc:high-$z$_QSO}), unlike the SEDs of the MLT dwarfs (Sect.~\ref{sc:MLT-subsection}). 
As expected, the best-fit MLT solutions for quasars with WISE photometry are late M or L dwarfs, given the aforementioned colour selection criterion ($W1 - W2 <0.7$) that excludes T dwarfs from photometric selection (Fig.~\ref{colorwise}). 

To assess the impact of WISE data on the performance of \texttt{Owl-z}, we conducted an experiment in which WISE data were removed from the selection process. The results indicate that WISE data were crucial only in identifying one quasar, J0252$-$0503. Without WISE data, J0252$-$0503 would have been misclassified as a T star. For the remaining candidates, the exclusion of WISE data primarily led to a decrease in selection probability for the quasar J0313-1806, while the probability for candidates J0038–1527, J1342+0928 and J1007+2115 remained very high  (see Table~\ref{table8}).

\subsection{Expected performance on EWS simulated data}\label{sec:performance}
The performance of \texttt{Owl-z} on the EWS is estimated in terms of completeness and purity in order to ascertain the sensitivity to confusion and contamination by MLT dwarfs and early-type galaxies. To this end, a series of catalogues is simulated in the following sections using Euclid filters (optical \IE\ and NIR \YE\JE\HE) and sensitivities detailed in Table~\ref{table_euclid_filters}.

\begin{table}[htbp!]
\caption{Used 
 bands characteristics.}
\centering
\begin{tabular}{lccc}
\hline
 band & $\lambda_{eff}$& median SNR& median depth\\
 \hline
\IE & 6869.74& 15.9 &26.2\\
z&  8678.90   & 5& 25.9\\
\YE &10688.60 &6.5 &24.3 \\
\JE & 13422.62&7.8 & 24.5\\
\HE &17410.63 & 7.2&24.4 \\
W1&33526.00& 5& 19.8\\
W2&46028.00&5&18.9\\
\hline
\end{tabular}
\label{table_euclid_filters}
\tablefoot{ The Euclid bands characteristics are a duplicate of Table 7 in \citealt{Scaramella2022}. The 
SNR statistics and depths ( magnitudes) in the region of interest. All the statistics have been computed for each channel: optical \IE\  and LSST z-band and NIR \YE, \JE, \HE,  and WISE channels W1 and W2. The
median depth here is evaluated for $5\sigma$ point-like source}
\end{table}

\subsubsection{\label{completeness}Performance: Completeness}

\begin{figure}[ht!]
    \centering
    \includegraphics[angle=0,width=1.0\hsize]{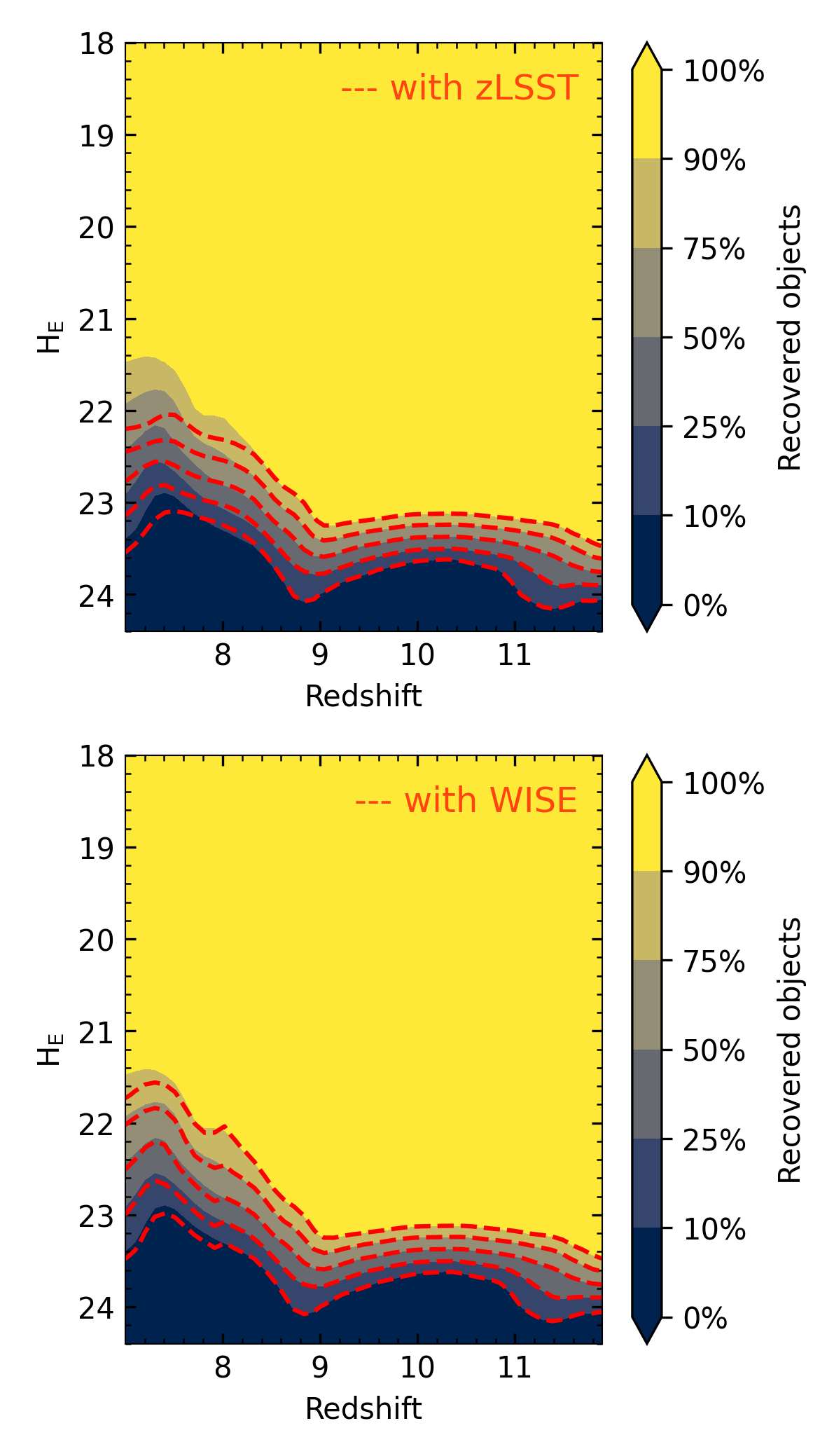}
    \caption{ The selection functions of high redshift quasars determined using \texttt{Owl-z}  for Euclid \IE\YE\JE\HE\ data. The selection function designs the level of completeness per redshift bin; the completeness of a bin is calculated by counting the percentage of quasars in the bin with probability is $ P_q > 0.1$. Several cases are shown:\textit{ (top panel)} In red, the optical band O of Euclid is replaced by the z-band from LSST. \textit{(bottom panel)} Euclid \IE\YE\JE\HE\ data  and in red Euclid \IE\YE\JE\HE\ in addition to WISE (W1 and W2 bands).}
    \label{comp}
\end{figure}

\begin{figure*}[ht!]
    \centering
    \includegraphics[angle=0,width=1.0\hsize]{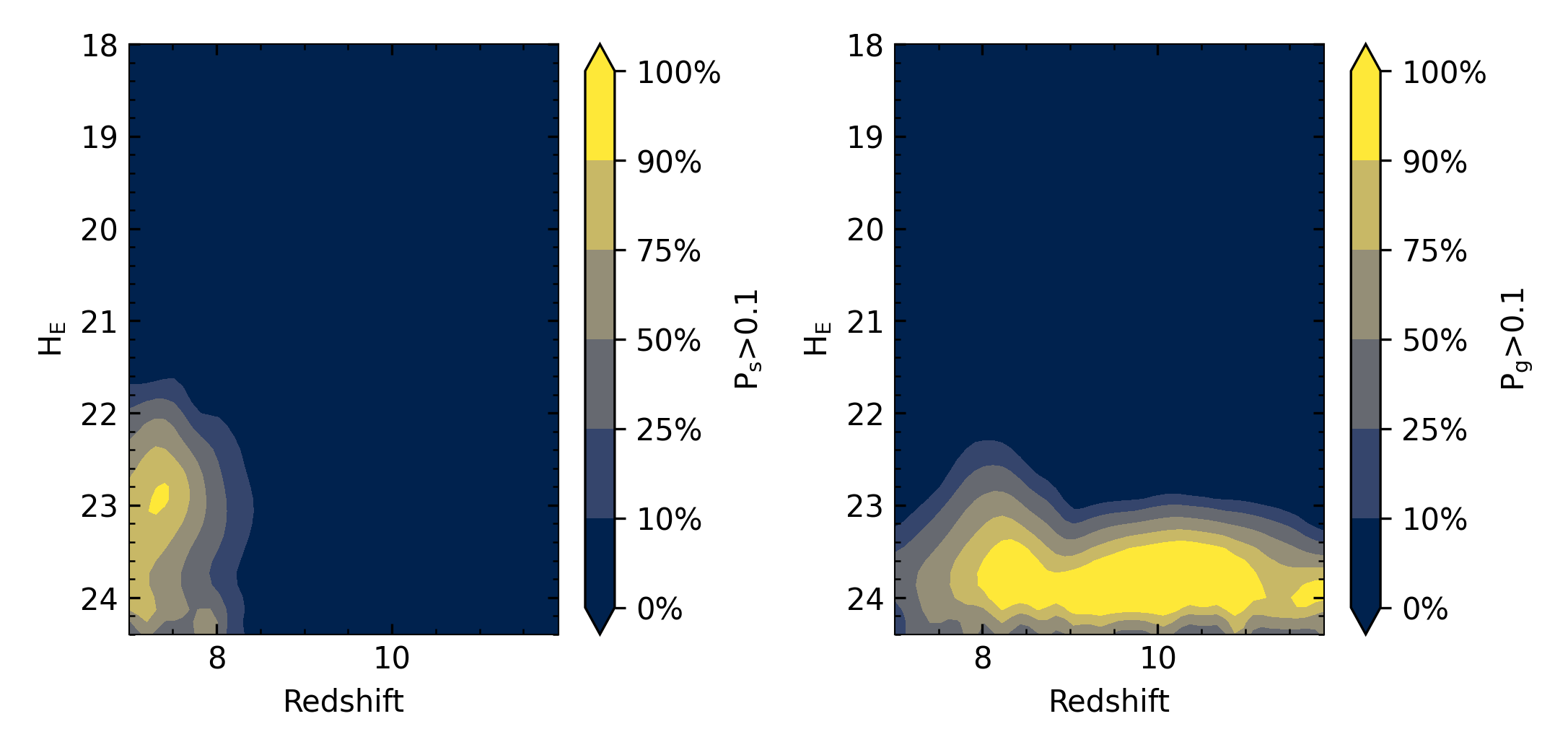}
    \caption{Incompleteness maps of high-$z$ quasar selection using \texttt{Owl-z}, shown as a function of redshift and \HE-band magnitude.  Each panel displays the fraction of quasars with $P_q$<0.1, i.e. quasars misclassified as MLT stars (left panel) or intermediate redshift galaxies (right panel), colour-coded by the percentage of quasars lost in the bin. These maps illustrate the regions of the parameter space where the selection is least complete and help characterise the dominant sources of contamination.}
    \label{incomp}
\end{figure*}
To estimate the completeness of the \texttt{Owl-z} selection method, we create a mock catalogue of 500,000 quasars. In order to gain a deeper understanding of the impact of using only Euclid data and of replacing or supplementing the existing data with other bands, three different scenarios have been developed. The first scenario involves using only Euclid data. In the second scenario, the optical band \IE\ is replaced by the optical z-band of LSST. In the third scenario, in addition to the Euclid optical and 
 NIR data, the photometry is extended to the WISE coverage in the W1 and W2 bands (see Table~\ref{table_completude}). The sensitivities of these additional bands can also be found in Table~\ref{table_euclid_filters}. 
The completeness is calculated on the EWS DR6 \citep{Scaramella2022} footprint, which covers 15,000 deg$^2$. Quasars are randomly selected with a flat distribution in redshift in the range $7 \le z \le 12 $, a flat distribution in absolute magnitude $M_{1450}$ in the range $-29 \le M_{1450} \le -22$, a flat distribution over all the SEDs in our library, and finally, a flat spatial distribution within the 15,000 deg$^2$ of the EWS footprint.
For each object in the catalogue, we calculate apparent magnitudes in the Euclid bands, in the LSST z-band, and in the WISE $W_1$ and $W_2$ bands.
For photometric errors, we adopt those corresponding to the DR1 release of the LSST  data \citep{LSST_2019} and those for the WISE data; we use the W1 and W2 bands and their corresponding errors \textbf{from the survey description}\footnote{\url{wise2.ipac.caltech.edu}}. For Euclid photometric errors, we use the signal-to-noise ratio maps across the EWS footprint as described in \citealt{Scaramella2022}; all of this information can be found in Table~\ref{table_euclid_filters}.

\begin{table}[ht!]
\caption{Scenarios explored in the simulations for the completeness estimation.}
\label{table_completude}
\centering

\begin{tabular}{lcc}
\hline
\hline
 \textbf{Survey }& \textbf{NIR bands}&  \textbf{optical bands}\\
 \hline
Euclid & \YE\JE\HE & \IE \\
 Euclid +LSST&\YE\JE\HE &z\\
Euclid +WISE &\YE\JE\HE W1 W2  &\IE\\
\hline
\hline
\end{tabular}
\end{table}

We define the completeness $C$ of the samples selected by
\texttt{Owl-z} as follows:
\begin{align}
    C=\frac{\text{TP}}{\text{TP}+\text{FN}}\text{,}
    \label{comp_eq}
\end{align}
where $\text{TP}$ (True Positive) is the number of quasars that have been successfully classified as high-$z$ quasars, and FN  (False Negatives) is the number of quasars that have been incorrectly classified as contaminants. For the sake of consistency with other work, we define a successful classification when $P_q>0.1$, and we will examine in Sect.~\ref{Discussion} how a different definition affects the results.

For all the objects in our mock quasar catalogue, we calculate the probability $P_q$, $P_s$ and $P_g$ that the quasar is identified as a quasar, star and galaxy, respectively, as defined in Eq.~\eqref{bayes}. Additionally, the SEDs for each category that allow for a maximum posterior are retrieved.

Results are presented as a function of redshift and  \HE\ magnitude in Fig.~\ref{comp}, for a selection from Euclid's \IE\YE\JE\HE\ data. The colours correspond to different increments of the completeness value. We also evaluate how completeness varies when the LSST z-band (top-hand panel in Fig.~\ref{comp}) and the WISE $W1$ and $W2$ bands (bottom-hand panel in Fig.~\ref{comp}) are added to the photometric dataset. This is represented by the dotted red contour lines corresponding to the completeness increments. 

In addition to the global completeness maps, we show in Fig. \ref{incomp} the  fraction of quasars with $P_q<0.1$, i.e. quasars misclassified as MLT stars (left panel) or intermediate redshift galaxies (right panel), colour-coded by the percentage of quasars lost in the bin. These maps illustrate the regions of the parameter space where the selection is least complete and help characterise the dominant sources of contamination as a function of redshift and \HE\ magnitude.  As seen in this figure, at  $z\sim7-8$ misclassification as MLT stars dominates, whereas at higher redshifts ($z\geq 8$), the incompleteness reflects the overlap between quasars and mid-$z$ interlopers, increasing with \HE\ magnitude. 

To achieve maximum completeness, quasar magnitudes need to be 2 to 3 magnitudes brighter than the EWS detection limit, depending on the redshift. The closer to the detection limit, and therefore the lower the SNR, the greater the confusion between quasars and brown dwarfs or galaxies at intermediate redshifts, preventing reliable identification.

Between redshift 7 and 8, completeness is lower than at higher redshift, up to $\HE\ \approx 21$, due to greater confusion with late-type dwarfs. Due to the strong increase in late-type dwarf flux in the reddest part of the optical band and the wide width of the Euclid \IE\  band,  any late-type dwarf will appear significantly fainter in the \IE\ band than in any z-band image of comparable depth (see Fig.~\ref{Color}). As a result, z-band data from LSST or from the UNIONS survey \footnote{\url{https://www.skysurvey.cc/}} provide significantly improved discrimination against L- and T- dwarfs (dashed red curves in the left-hand plot of Fig.~\ref{comp}). 

\begin{figure}[t!]
    \centering
    \includegraphics[angle=0,width=1.0\hsize]{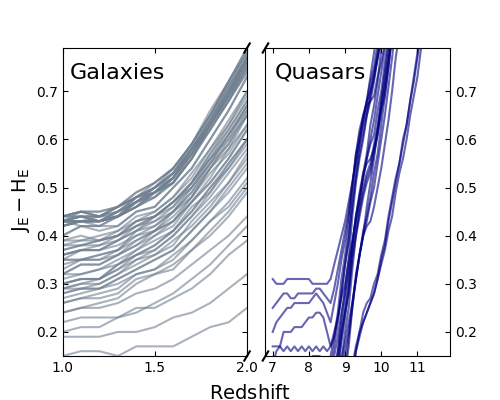}
    \caption{Tracks of \JE-\HE\ colours  as a function of redshift 
 of  Early-Type galaxies (left panel)  and high redshift quasars (right panel). For the early-type galaxies, the lower in colour the track is the younger the galaxy. }
    \label{gal_z_j_h}
\end{figure}

Fig.~\ref{comp} also shows a loss in completeness around redshift 10 and between magnitudes 22 and 23 of \HE, compared with its value at lower redshift values. We attribute this loss in completeness mainly to an increase in confusion with early-type galaxies as their colours are very similar (see Fig.~\ref{gal_z_j_h}). 

Finally, the analysis of completeness, including WISE data (bottom panel in Fig.~\ref{comp}) shows no significant improvement, in contrast to the improvement mentioned in Sect.~\ref{sec:reidentification} for the re-identification of known quasars at high redshift. Only a modest improvement in the decrease of the confusion with late-type dwarfs is perceptible for the brightest quasars at the lowest redshift. This is due to the fact that the depth of the WISE data (AB magnitude $\sim 19$ at 5$\sigma$ in the $W1$ and $W2$ bands) does not match that of the Euclid data, making them unconstrained and allowing no significant improvements over the Euclid data alone, either above or below the WISE detection limit.

\subsubsection{\label{photoz}Redshift estimation}

\begin{figure*}[h!]
\centering
\includegraphics[angle=0,width=0.9\textwidth]{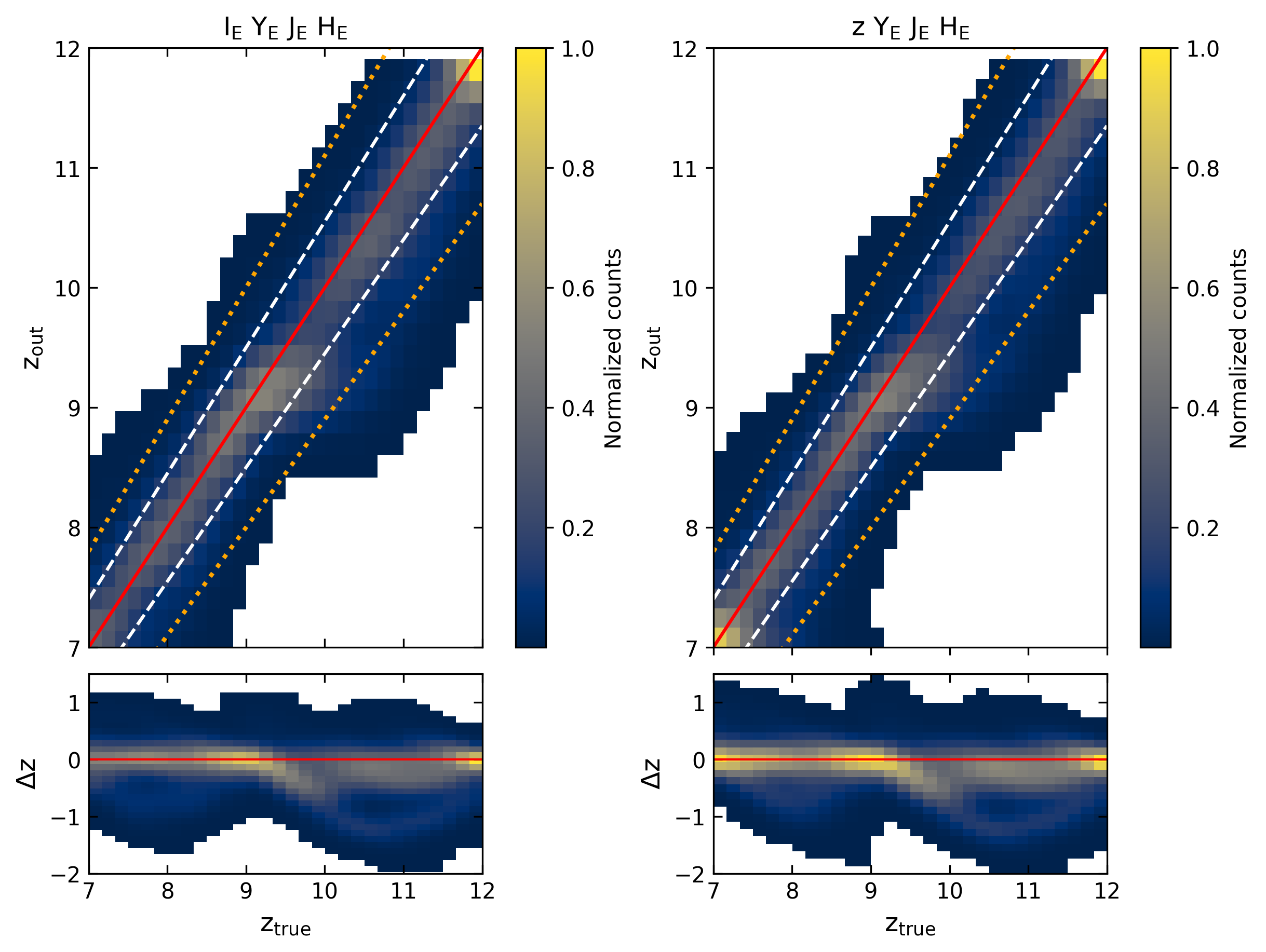}
\caption{Top panel shows the comparison between the true redshift and the redshift estimated by \texttt{Owl-z} with both Euclid \IE\YE\JE\HE\ dataset (left) and  z\YE\JE\HE\ dataset (right). Red solid lines display the 1:1 bisector; the thresholds corresponding to $| \Delta z|>0.1(1+z_{true})$ and $|\Delta z|>0.05(1+z_{true})$ are also displayed to guide the eye as orange dotted and white dashed lines, respectively.
The bottom panel shows the residual $\Delta z=z_{out}-z_{true}$ as a function of redshift, with the red line corresponding to $\Delta z =0$ }
\label{zin_out}
\end{figure*}

Following Sect.~\ref{sec:methodology}, the redshift estimated by \texttt{Owl-z} corresponds to the maximum posterior value of the parameter $z$, that is, the maximum posterior $\int W_q(z_{out},H_E)dH_E$. We refer to it as "photometric redshift" or $z_{out}$ hereafter. 

The accuracy of photometric redshift has been estimated by comparing the true redshift injected in the simulation $z_{true}$ to $z_{out}$. Fig.~\ref{zin_out} displays this comparison for the redshift interval explored in the simulations presented in Sect.~\ref{completeness}. To evaluate the accuracy of the photometric redshift, the sample has been limited to objects brighter than $H_E<24$, identified as quasars using the same criterion as in Sect.~\ref{completeness}, that is $P_q>0.1$. 

As shown in Fig.~\ref{zin_out}, an excellent correlation between $z_{true}$ and $z_{out}$ is found, irrespective of the photometric dataset. For the \IE\YE\JE\HE\ filter set, the Pearson correlation coefficient is found to be 0.97, whereas the Spearman correlation coefficient is 0.98, indicating a strong linear correlation. The same results are found for the z\YE\JE\HE\ dataset. 
The classical quality estimators used for photometric redshift yield the following results, with $\Delta z = z_{out} - z_{true}$:
$\sigma(\Delta z/(1+z))$ = 0.030,
the median of $(\Delta z/(1+z))$ = -0.007, and the normalised median absolute deviation, which is less sensitive to outliers:
$\sigma_{z,MAD}$ = 1.48 $\times$ median $(|\Delta z|/(1+z))$ = 0.021. 
Regarding the fraction of outliers, defined in a conservative way as sources with 
$\Delta z >0.1(1+z_{true})$, it is found to be only 2.45\% (2.23\%) for the \IE\YE\JE\HE(z\YE\JE\HE) datasets. Only 13.7\% (13.3\%) of the sample exhibit $|\Delta z|>0.05(1+z_{true})$ for the \IE\YE\JE\HE(z\YE\JE\HE) datasets.

These results demonstrate the quality and the reliability of the photometric redshift obtained by \texttt{Owl-z} for sources identified as high-$z$ quasars. This is of utmost importance, given the use expected for the output redshift in the selection of samples for spectroscopic follow-up. It is also important in the determination of the purity, as shown in 
Sect.~\ref{sc:Purity}.

\subsubsection{\label{sc:Purity}Purity}
The purity of the sample expresses the expected reliability of \texttt{Owl-z} in extracting true quasars from photometric observations of real fields, which is applied to the EWS in this particular case.
Here, we use the Euclid bands \IE\YE\JE\HE, the photometric depths and the variable SNR maps of Euclid.  

We define the purity as:
\begin{align}
    P=\frac{1+\text{TP}}{1+\text{TP}+\text{FP}}\text{,}
\end{align}
that is, the number of true positive identifications (TP)  divided by the total number of quasar identifications (TP $+$ FP), where FP stands for false positives. With the above definition, a purity of 100\%  can either mean that no object( quasar or contaminant) has been identified as a quasar or all the identified objects are quasars.
To estimate the purity, we need to simulate as accurately as possible the content of the input photometric catalogues, including the contaminant populations of stars and galaxies in a realistic way. Bright high-$z$ quasars are rare, according to the current LF. For this reason, we force the inclusion of such sources in the simulated catalogues, as explained below, to determine how difficult it will be to identify and study these sources if they ever exist. In order to achieve this, the simulations are conducted on a surface that is 10 times larger than that of the EWS. Subsequently, scaling back to the original surface (1000 deg$^2$) is performed, whereby only those sources that the LF predicts are accounted for. Note that the definition of the purity as given above accounts for the possible detection of these unique, bright and rare quasars. 

The COSMOS2020 field \citep{Weaver_2022} represents an ideal base for this exercise, given the extended wavelength coverage and the exceptionally good quality of the photometric redshifts that can be used as spectroscopic redshifts for our needs. All sources in this field have been fitted using LePHARE \citep{LEPHARE} and a library of reference templates, and this provided us with a best-fit template and a redshift. We use these results to compute the  Euclid Optical and NIR photometry via a process we call  "Euclidisation" (additional details will be provided in Euclid preparation, EC \& Garnett et al.). This process has also allowed us to simulate the SNR and associated photometric errors in each filter. This new Euclid-like COSMOS catalogue will be called hereafter E-COSMOS. 

\begin{figure}[t!]
    \centering
    \includegraphics[angle=0,width=1.0\hsize]{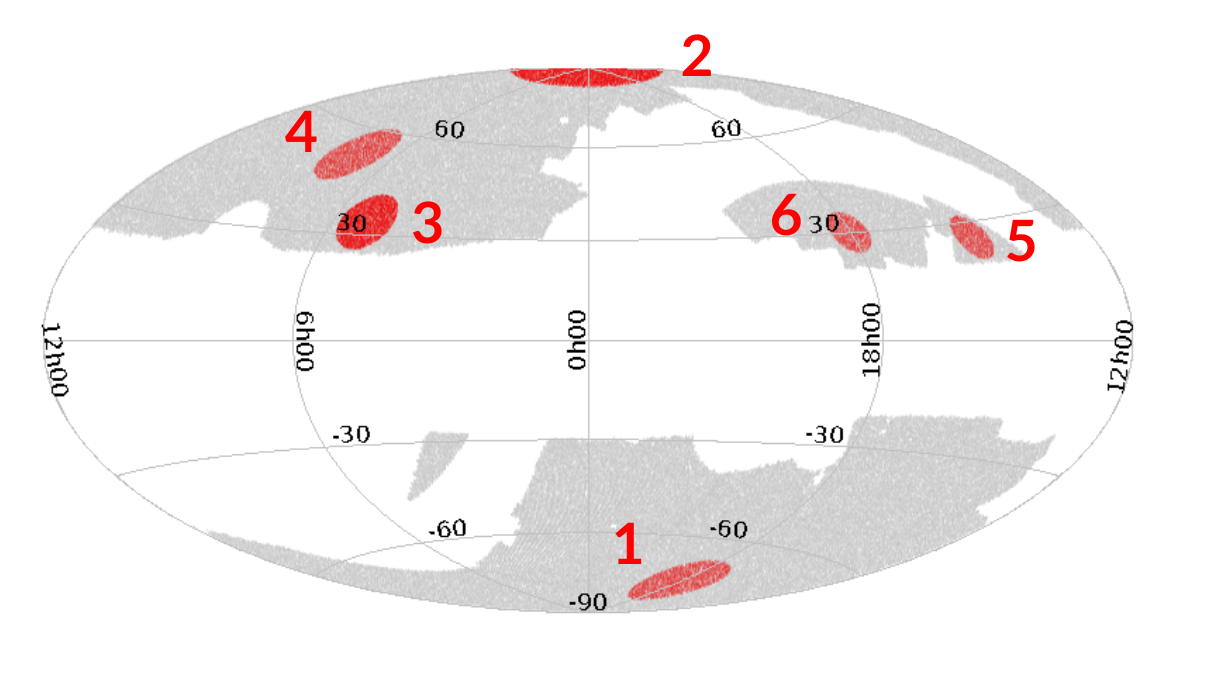}
    \caption{(\textit{red}) The location of the regions of focus used in the computation of the purity how in Galactic coordinates and galactic projection. Details on these surfaces can be found in Table~\ref{table_surfaces}. (\textit{grey} ) the EWS  DR6 footprint.}
    \label{surcafes_1000}
\end{figure}

The original COSMOS catalogue covers 2 deg$^2$. To cover a statistically significant field of view, multiple realisations of the parent E-COSMOS catalogue are needed. This is done by randomly sorting the coordinates of the field centre within the EWS footprint in such a way that a full (not overlapping) large catalogue is created. This catalogue covers 1000 deg$^2$ separated into six different areas. The rationale behind this choice is that it represents the smallest surface area over which we can account for a sufficient number of quasars. Furthermore, this surface has been divided across the region of interest in order to accommodate the various scenarios in galactic latitude/longitude, as well as to account for the inevitable differences in resolution, however slight they may be.
Fig.~\ref{surcafes_1000} and Table~\ref{table_surfaces} display the location of these areas around the Euclid Footprint. It is worth noting that the location of the field is expected to have a direct impact on the performance, given the different distribution of contaminant stars on the one hand and the different SNR achieved across the EWS due to zodiacal light variations on the other hand \citep{Scaramella2022}. These effects are discussed below.
The E-COSMOS catalogue has been cleansed of all stellar objects and is situated at a considerable distance from the Galactic plane.
To properly account for the presence of contaminant stars, we randomly inject additional MLT dwarfs into the catalogues, following the distribution presented in Sect.~\ref{sc:MLT-subsection}, reproducing the number densities expected at the galactic coordinates where the simulated field is located. 
Since E-COSMOS does not contain any spectroscopically identified high-$z$ quasar, this population has been randomly injected in the simulated fields, following the prescriptions presented in Sect.~\ref{completeness} to properly account for the SNR in the EWS. High-$z$ quasars are randomly sorted according to their LF. Obviously, the number of bright  $z>7$ quasars is expected to be very small in the EWS field. In particular, the current LF does not predict any quasar brighter than $M_{1450}<-23$
in the entire field. For these rare bright objects, if they exist, we force the measurement of the purity as explained above. To better capture this population, 10 realisations of the entire Euclid footprint are performed, allowing us to retrieve some quasars towards the brightest luminosities and up to redshifts $z>10$. However, in order to facilitate consistent comparison of the statistical performance estimators $C$ and $P$, it is necessary to normalise all populations to the same surface area  (15,000 deg$^2$).      

\texttt{Owl-z} has been run on these simulated catalogues. The pre-selection criteria and $P_q>0.1$ threshold applied are identical to those used for the completeness in Sect.~\ref{completeness}. We study the variation of the purity as a function of redshift ($z$$_{out}$) and apparent magnitude in the reference filter. The results presented in Fig.~\ref{pq01} provide a summary of the performance of \texttt{Owl-z} in terms of completeness and purity. In each bin, the figure displays the values of $C$ and $P$. Additionally, it illustrates another performance estimator, the F-measure, which will be defined in Sect.~\ref{Performance}. 
As shown in this figure, the code is expected to achieve a high purity for the brightest sources, irrespective of the redshift. The influence of the precise choice for the threshold in $P_q$ is discussed in Sect.~\ref{threshold}. 

\subsubsection{\label{Performance}Global performance}
In order to evaluate the global performance of \texttt{Owl-z} by combining both the completeness and purity measurements, we introduce the $F$-measure as follows:
\begin{align}
    F=2\frac{C P}{C+P}\text{,}
\end{align}
where $P$ is the purity and $C$ is the completeness, according to the definitions in Sect.~\ref{completeness} and Sect.~\ref{sc:Purity} respectively.  
The $F$-measure is a statistical measurement employed to assess the performance of a classifier \citep{fmeasure}.

The classification performance{---}specifically, the code's ability to identify high-$z$ quasars{---}is evaluated using the $F$-measure, which indicates the quality of the classification. A higher $F$-measure indicates a superior classification performance. We use $F$-measure hereafter as a performance metric.  

The results summarising the performance of \texttt{Owl-z} in terms of $F$-measure are also provided in Fig.~\ref{pq01}. As shown in this figure, high values of $F$-measure are obtained for bright (\HE$<21$) sources at all redshifts between $7<z<11$, indicating that the performance in the spectroscopic confirmation of bright rare quasars is expected to be high. On the contrary, the $F$-measure drops significantly towards fainter magnitudes and high redshifts, following the drop in the purity value. Note that satisfying results can still be achieved for $\HE<23$ and $z<8$. Also, the performance depends on the choice of the probability threshold, hereafter referred to as $\zeta$, as discussed in  Sect.~\ref{threshold}.    

To illustrate the effects of contamination, we show in Fig. ~\ref{pq_contamination} the contamination by galaxies (background colour) and indicate the relative fractions $I_g$ of galaxies and $I_s$ of MLT stars among the contaminant population. Contamination is almost entirely dominated by  galaxies, except in one redshift and magnitude bin where MLT stars also contribute. This is explained by the fact that the colours of MLT stars mainly contaminate the colours of quasars in the redshift bin [7-8] (see  Fig.~\ref{Color}), and that the number of contaminants at high magnitudes are dominated by galaxies (see figures \ref{prior_MLT_ciel} and \ref{Prior_QSO_GAL}). 
~~

\section{\label{Discussion} Discussion }

In this section, we study the influence of different parameters on the performance of \texttt{Owl-z}, such as the threshold used to select quasars as a function of redshift and magnitude. We also give some guidelines to optimise the selection of high-$z$ quasars and their spectroscopic follow-up. The discussion in this section is based on the performance expected on the EWS.

\subsection{\label{threshold}Influence of the selection threshold}

In the previous section, we consider an object as a quasar candidate when the posterior probability $\prob$ is above the threshold $\zeta=0.1$.
This parameter clearly impacts the performance because lower values will improve the completeness while inducing worse purity values and vice-versa. There is a trade-off to find to optimise the performance. Here, we discuss the influence of  $\zeta$ on $F$-measure. 

\begin{figure}[t!]

    \centering
    \includegraphics[angle=0,width=1.0\hsize]{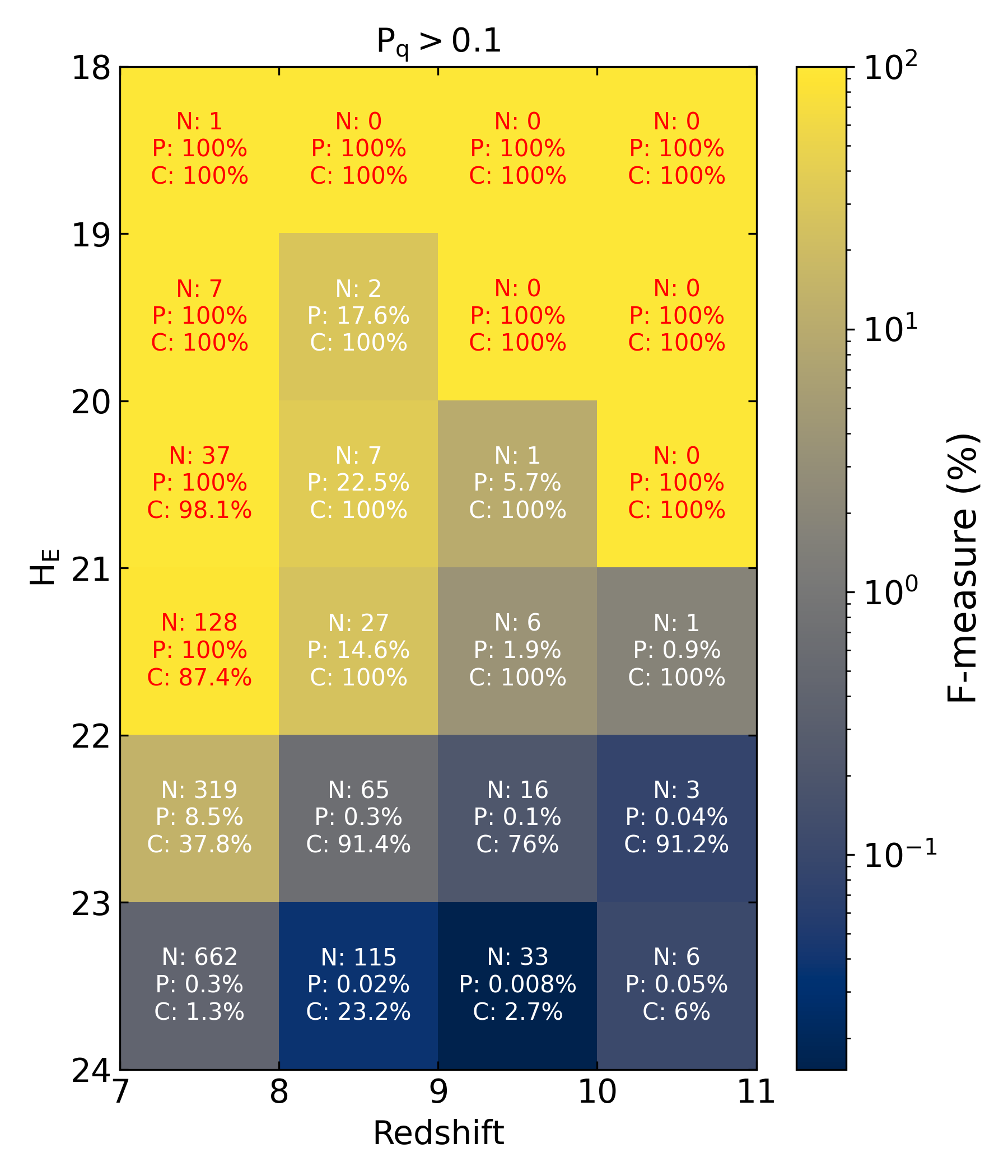}
    \caption{Classification performance of \texttt{Owl-z} determined using the threshold $\zeta=0.1$, in each magnitude and redshift bin, N is the number of quasars in the bin, P is the purity, and C is the completeness. The colour code indicates the performance measurement $F$-measure in each bin }
    \label{pq01}
\end{figure}

\begin{figure}[t!]
    \centering
    \includegraphics[angle=0,width=1.0\hsize]{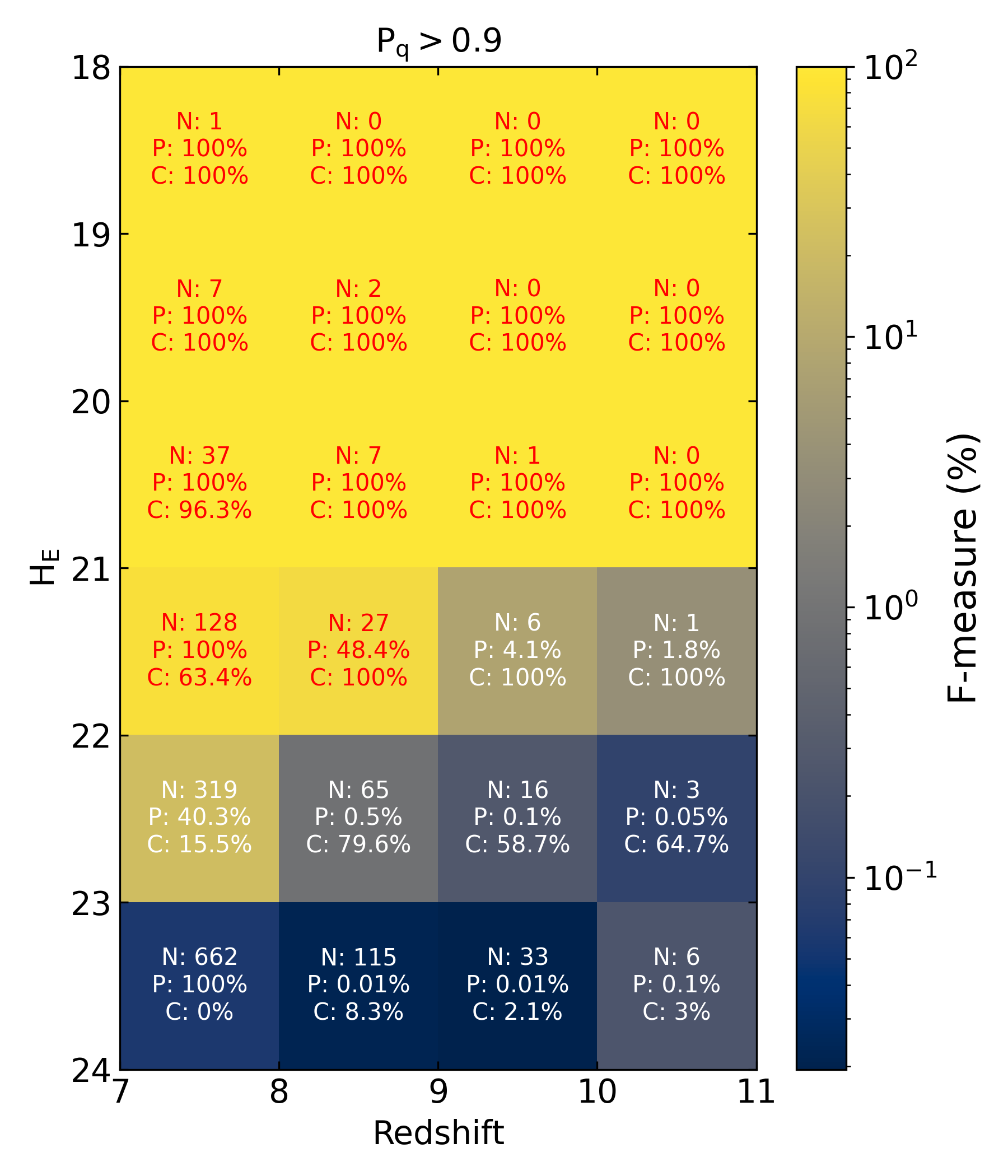}
    \caption{Classification performance of \texttt{Owl-z} determined using the threshold $\zeta=0.9$, in each bin we report N: the number of quasars injected in the bin, P: the purity and C: the completeness. The colour code indicates the performance measurement $F$-measure in each bin.}
    \label{pq09}
\end{figure}

\begin{figure}[t!]
    \centering
    \includegraphics[angle=0,width=1.0\hsize]{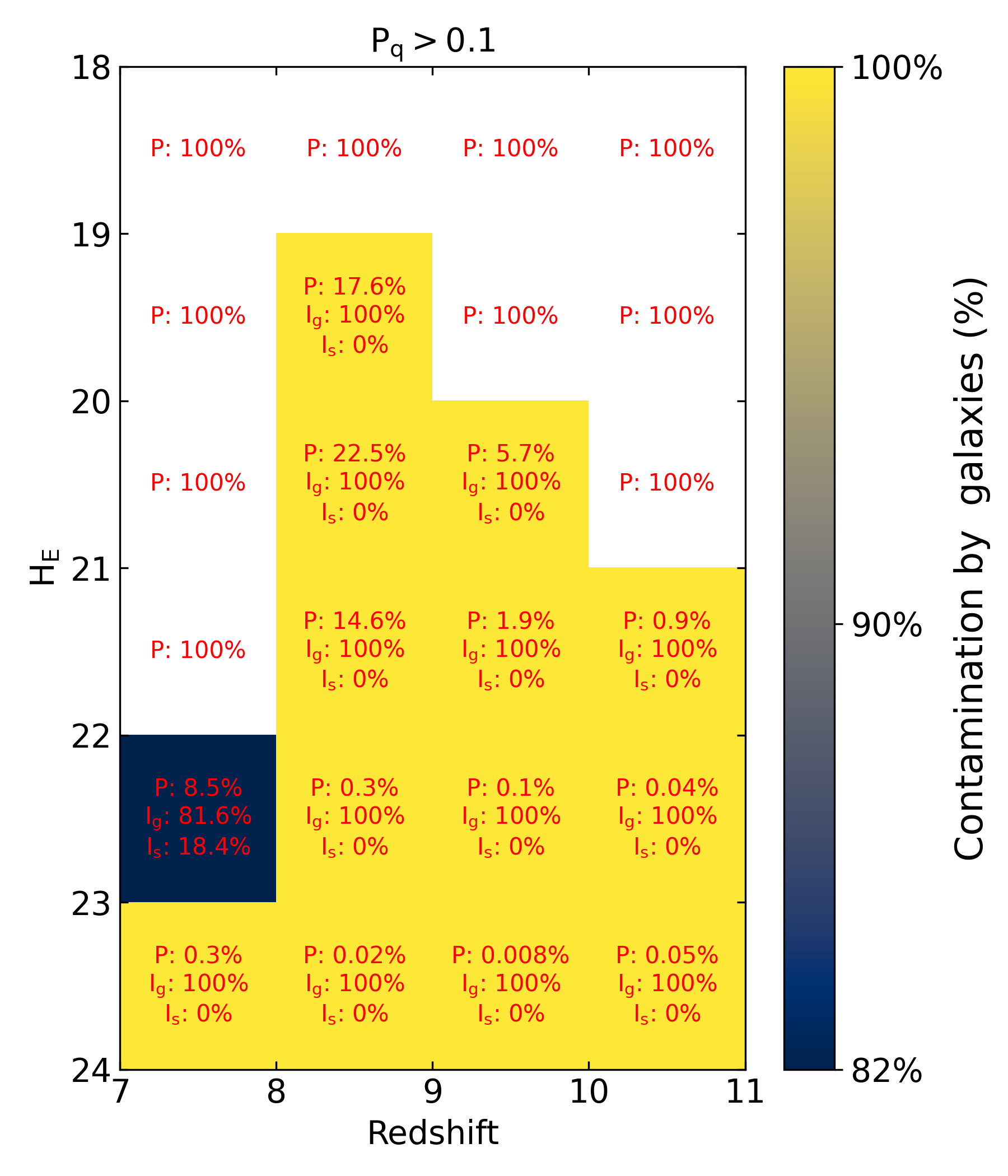}
\caption{Contamination by galaxies indicated by the colour map for the same quasar selection parameters as in Figure~\ref{pq01}. The overlaid text indicates the purity of the quasars as in Figure~\ref{pq01} and the relative fractions of contamination by galaxies ($I_g$) and MLT stars ($I_s$).}
    \label{pq_contamination}
\end{figure}

To this end, we use the data discussed in Sect.~\ref{sc:Purity}, change the value of $\zeta$ to select quasar candidates and recalculate the $F$-measure  in each bin. The results are presented in Fig.~\ref{pq01}  for $\zeta=0.1$ and in Fig.~\ref{pq09} for $\zeta=0.9$.

Two main trends are observed for sources respectively brighter/fainter than \HE$=22$.  

Increasing the threshold improves purity, albeit at the expense of completeness for faint sources ($\HE\ > 22$). For bright sources ($\HE\ < 22$), the completeness is almost unaffected by increasing the classification threshold, thanks to the ability of the code to compute high $P_q$ values for bright quasars.

For faint sources, the performance of \texttt{Owl-z} drops rapidly towards the faintest magnitudes above $\HE\ = 22$, irrespective of the redshift bin and $\zeta$\, due to low SNR values, which contribute to a rapid decrease in purity, even when completeness is still good. The analysis of the results shows that early-type galaxies are the main contributors to contamination at these faint magnitudes ($\HE\ > 22$). Conversely, at bright magnitudes ($\HE< 22$ and brighter), the SNR is high and enables adequate classification between stars, galaxies and quasars with good completeness, and higher values of $\zeta$ improve the purity and the values of the $F$-measure increase.  

In Fig.~\ref{pq01}, the $8 < z < 10$ redshift bins for bright candidates ($19 < \HE\ < 22$) are clearly affected by low purity values. The main sources of contamination in this area are early-type galaxies at redshifts $1 < z < 2$ due to the confusion between the 4000 $\AA$ break at intermediate redshift and the Lyman-$\alpha$ break at high redshift, as explained in Sect. ~\ref{sc:galaxy_model}. There is also a minor contribution from contamination by late L-type dwarfs. Increasing the classification threshold has the immediate effect of reducing this contamination and, therefore, improving the purity. As completeness remains high, this results in a significant improvement in $F-$measure values.

This analysis suggests that an adaptive threshold can be used to optimise the $F-$ measure values in the magnitude/redshift parameter space. 
   
\subsection{\label{optimization}Optimising the identification of $z>7$ quasars}

\begin{figure}
    \centering
    \includegraphics[angle=0,width=1.0\hsize]{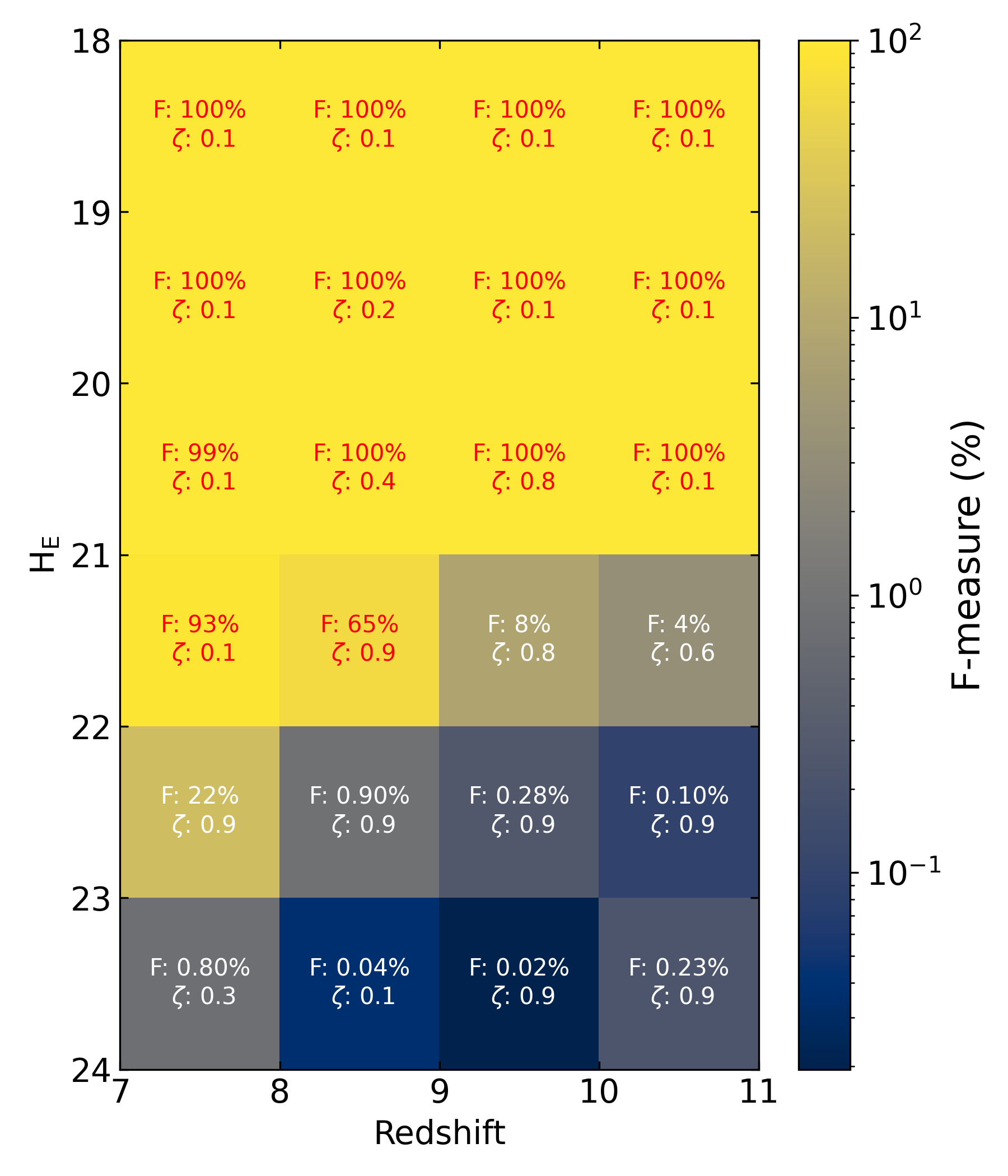}
    \caption{Classification performance of \texttt{Owl-z} on the magnitude \HE\ and redshift $z_{out}$ plane. The threshold value of $\zeta$ that maximises $F$-measure is given for each bin.}
    \label{THvariable}
\end{figure}

We now investigate if the value of the $F$-measure 
 can be further increased by adjusting the selection threshold $\zeta$ in each redshift and magnitude bin. To this end, we use the data discussed in Sect.~\ref{sec:reidentification}, change the value of $\zeta$ in increments of 0.1  and select the value that maximises $F$ in each bin.
The result is summarised in Fig.~\ref{THvariable},~ showing the value of $\zeta$ that maximises  $F$ in each bin of magnitude \HE\ and redshift $z_{out}$. The results show that, for bright magnitudes (\HE$<21$), a low threshold value $\zeta=0.1$ is sufficient to maximise $F$-measure. This is due to the robustness of the selection for high SNR values irrespective of the redshift, as shown previously. On the contrary, for faint magnitudes (\HE$>21$), the value of $\zeta$ needs to be increased to compensate for the decline in purity, requiring $\zeta$ values as high as 0.9 towards the faintest magnitudes.  

In conclusion, adopting a variable threshold enables the optimisation of the $F$-measure  and, consequently, the effectiveness of photometric and spectroscopic follow-up campaigns based on \texttt{Owl-z} selected candidates.  Finding the best threshold value $\zeta$ for a quasar candidate with magnitude \HE\ and a redshift $z_{out}$ from any other survey data can be done by following the same prescription presented in this article in the case of EWS.

\subsection{\label{SNR} Influence of the position in the EWS footprint }

In this section, we evaluate the effect of different locations in the EWS footprint. As described in \citealt{Scaramella2022}, different zodiacal light levels across the footprint modify the SNR for a given magnitude in each filter, and different sky positions and Galactic coordinates introduce varying stellar densities, as described in Sect.~\ref{sc:MLT-subsection}. To this end, we select six different regions in the Euclid footprint, representative of different zodiacal light levels and Galactic coordinates. These regions are shown in Fig.~\ref{surcafes_1000}, and Table~\ref{table_surfaces} lists their Galactic coordinates and SNR values at reference magnitudes in all Euclid filters.

According to the results presented in Sect.~\ref{Validation}, the confusion by MLT dwarfs is only sensible in the redshift interval  $z=[7,8]$, where there is also contamination by early-type galaxies. Therefore, we focus the analysis on this redshift bin because it is more susceptible to being affected by the precise distribution assumed for MLT stars. In addition, this redshift bin is also the most populated by quasars, meaning that the results are expected to be statistically more significant than in the other bins. For each one of these regions, we compute simulated catalogues following the prescriptions in Sect.~\ref{sc:Purity}, and we run \texttt{Owl-z} to derive the value of $F$-measure.   

\begin{figure}
    \centering
    \includegraphics[width=\linewidth]{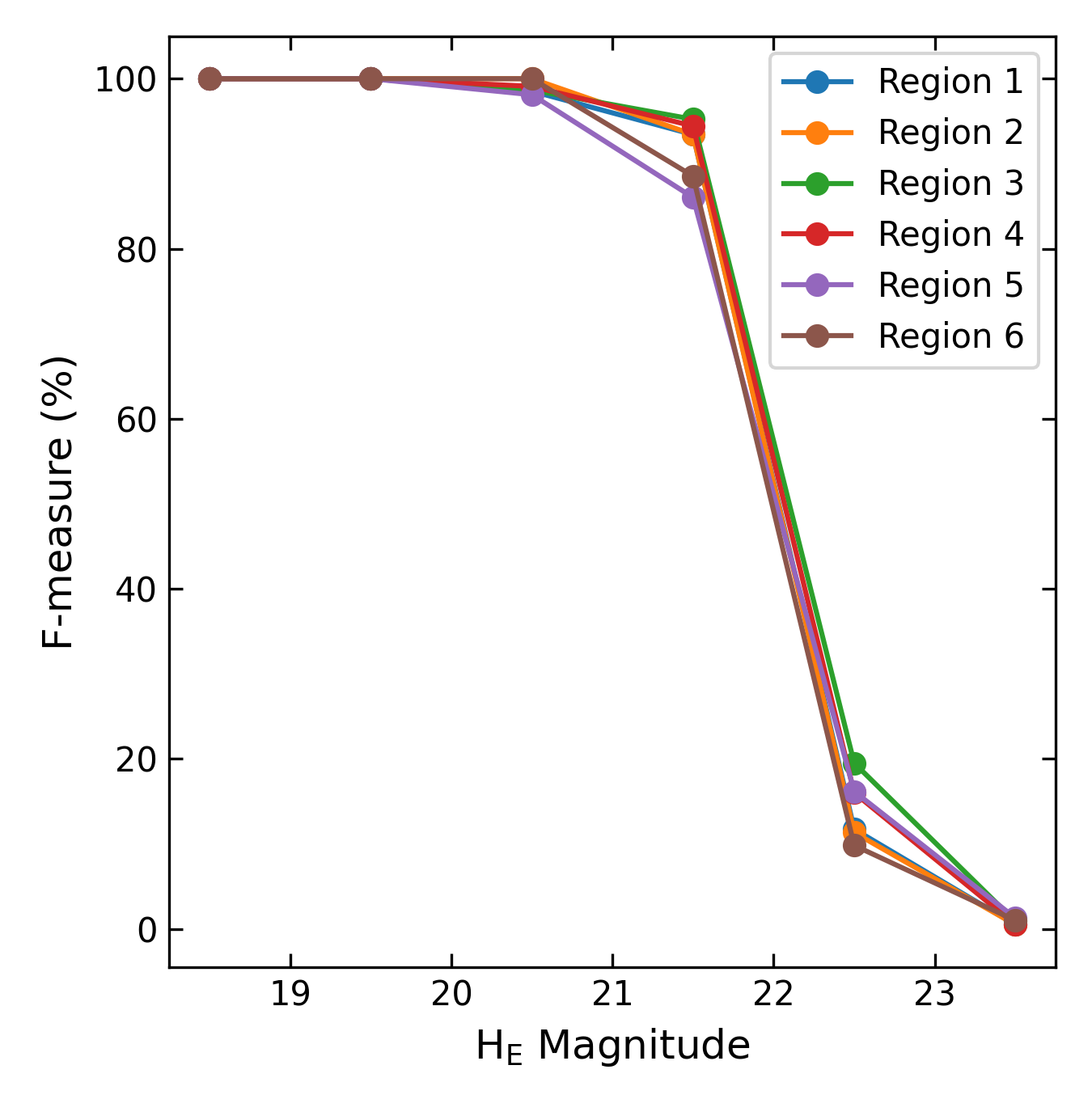}
    \caption{$F$-measure evolution as a function of the \HE\  magnitude in each of the regions shown in Table~\ref{table_surfaces}, computed in the $7<z<8$ interval. }
    \label{areas_all}
\end{figure}
The results are shown in Fig.~\ref {areas_all} and display minor variations in the values of $F$-measure.  Regions 5 and 6 exhibit a slight decline in $F$-measure within the magnitude bin $[21,22]$ compared to the others. This can be attributed either to the fact that these regions possess the lowest Galactic latitudes ($b=26$ and $b=30$) or to the most unfavourable SNR values. According to the results presented in Sect.~\ref{Validation}, the comparison with region 3, which is also at a low Galactic latitude but has higher SNR values, suggests that SNR variation is the dominant factor in variations in $F$-measure. In other words, brown dwarf contamination does not appear to depend significantly on Galactic latitude within the Euclid footprint, which is consistent with  figure~\ref{fig:thin-thick} and with the fact that up to magnitude $\approx 22$ contamination is dominated by late-type L and T dwarfs at distances small compared to the scale height of the Galactic disc.

\subsection{Influence of thick disk MLT}
 We aim to evaluate the robustness of our selection process when adding contamination by thick disk MLTs that are not included in the model. We evaluate the impact on the $F$-measure of adding a contribution of thick-disk brown dwarfs to the input catalogue, following the prescriptions described in section~\ref{sec:performance}, and ignoring differences in brown dwarf populations between the thin and thick disks. 
 To do so, we employ the density function outlined in Sect.~\ref{sec:proba} to generate a population of MLT dwarfs from the thick disc, in addition to the previously added MLT dwarfs from the thin disc, and we analyse the impact on $F$-measure on the six regions introduced previously.  
\begin{figure}[ht!]
    \centering
    \includegraphics[angle=0,width=0.9\hsize]{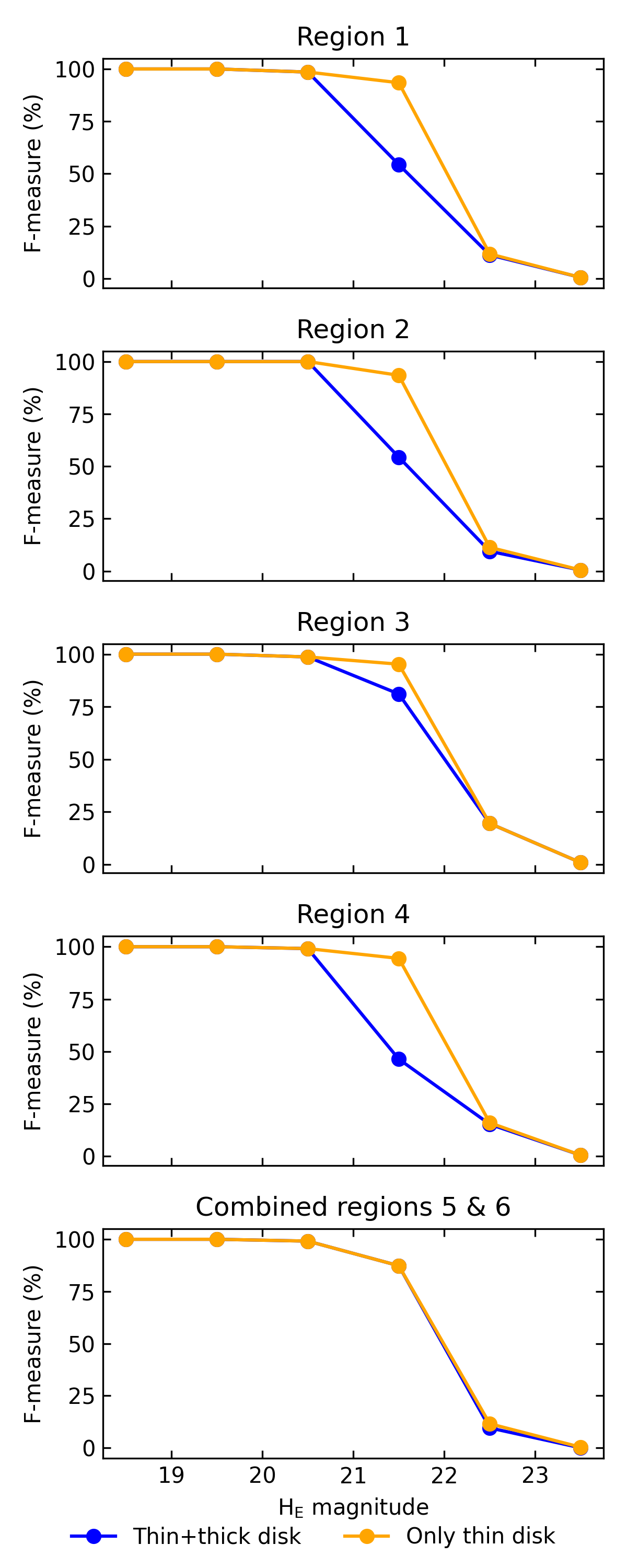}
    \caption{$F$-measure evolution as a function of the \HE\ magnitude in each of the regions shown in Table~\ref{table_surfaces}, computed in the $7<z<8$ interval. Two cases are shown for the contribution of MLT stars. In yellow is presented the evolution when only the thin disk contribution is accounted for, both in the simulation and the Bayesian modelling. In blue, the same result is shown when the contribution of the thick disk is added to the simulations.   
   }
    \label{fig:areas}
\end{figure}
 The results are presented in Fig.~\ref{fig:areas}, comparing two scenarios: one where only thin disk  MLT dwarfs are injected in the input catalogue (shown by yellow lines)  and another where thick disk MLT dwarfs are added (shown by blue lines). 
The trends on these graphs can be qualitatively explained by the relative density of stars between the thin and thick disks as a function of magnitude. At low magnitudes, the thin disk dominates at all Galactic latitudes, and there are no differences between the two scenarios. At magnitudes above 21, however, and at high Galactic latitudes only (see figure~\ref{fig:thin-thick}), the thick disk population begins to contribute to the counts and introduce a difference between the two scenarios (regions 1, 2, 4 and to a lesser extent region 3, which is at a relatively low Galactic latitude but closest to the Galactic centre in longitude).
Overall, this analysis serves as a sanity check in confirming that the precise modelling of the thick disk is unlikely to significantly affect the results for the EWS, but would be warranted for deeper surveys.

 Fig.~\ref{fig:areas} illustrates the drop in F-measure as more contaminants are introduced to the examined catalogues. These added contaminants are MLT stars drawn from the thick disk population, and have been introduced into all catalogues, including Regions 2 and 1, which are the furthest from the Galactic plane. The Bayesian model cannot predict such MLT stars far from the galactic plane, as it only includes thin disk modelling (see section~\ref{sc:MLT-subsection}). In Regions 5 and 6, which are the closest to the Galactic plane, the Bayesian model expects a high number of contamination from MLT. Consequently, \texttt{Owl-z} does not classify MLT resembling QSO as QSO. However, when we move farther from the Galactic plane (Regions 1, 2, and 4), additional contaminants that the model does not expect are encountered. As a result, they are more likely to be misclassified as a QSO, which in turn lowers the accuracy, i.e. the F-measure of the \texttt{Owl-z} classification. 
\begin{table*}[htbp!]
\caption{Properties of simulated areas and catalogues}
\tablefoot{Simulations across the EWS footprint use reference magnitudes of 24 for \YE, \JE, and \HE, and 24.5 for \IE. The SNR values are averaged over each surface.}

\centering
\begin{tabular}{lccccccc}
\hline
 Region & Surface (deg$^2$)& l(deg)&b(deg)&SNR$_{\IE}$&SNR$_{\YE}$&SNR$_{\JE}$&SNR$_{\HE}$\\

 \hline
1& 200&270&$-$75&17.5&7.0&8.45&7.9\\

2& 200 &180&85&16.5&6.6&7.8&7.4\\

3&200&78&35&17.8&7.0&8.5&7.7\\

4&200&107&55&18.6&7.4&8.8&8.22\\

5&100&227&26&13.8&5.6&6.6&6.1\\

6&100&270&30&14.3&5.9&7.1&6.5\\
\hline
\end{tabular}
\label{table_surfaces}
\end{table*}

\subsection{\label{comparison_results}Comparison with previous works}

The results obtained in this article can be compared to previous similar works, at least qualitatively. A proper comparison will need to run the various codes on the same simulated catalogues, which is currently impossible because not all codes are publicly available.  
The Bayesian classification approach is similar to that adopted by ~\citealt{Pipien2018} for the analysis of the CFHQSIR survey ~\citep{CFHQSIR} searching for quasars at $6.5<z<7.5$.
Our results in the lowest redshift bin are comparable to those of ~\citealt{Pipien2018}. Note that ~\citealt{Pipien2018} do not include 
the population of contaminant galaxies, a choice that has no impact on their results because the dominant contamination in this bin comes mainly from LT stars, as shown in Sect.~\ref{Validation}. 

Our work also compares to and complements the study conducted by \citealt{Barnett2019} on the EWS using a similar Bayesian classification method and the same contaminant populations of stars and galaxies.  Although these approaches are comparable, our prior models and definitions of contaminant populations differ. Table~\ref{tab:comparison} highlights the differences between the modelling of quasar and contaminant populations in these different studies.
With respect to \citealt{Barnett2019}, the work with \texttt{Owl-z} extends the study beyond $z>9$, and introduces different performance metrics in addition to the completeness, namely purity and F-measure (combining completeness and purity). Our definition of purity has also been adapted to account for rare quasars beyond $z>9$ that could be found in the EWS if they ever exist. We also use updated prescriptions for the quasar LF and the contaminant populations. It is worth noting that we have used here the EWS SNR maps, making our predictions closer to the real observations.
The performance of \texttt{Owl-z} in terms of completeness in the common redshift range with \citealt{Barnett2019} is found to be better than in this previous work. The reason for this improvement is mainly the fact that the global SNR is expected to be better than assumed by \citealt{Barnett2019}. On the contrary, the effect of the detailed 2D modelling of the SNR in the EWS has little impact on the performance of \texttt{Owl-z} and similar codes, possibly because the depth is not dramatically different across the survey, as discussed in Sect.~\ref{SNR}. 

Possibly the main advantage of \texttt{Owl-z} with respect to other existing codes is its versatility in adapting to different photometric surveys, as shown in Sect.~\ref{sec:reidentification}. The code will be publicly available and open to access.

\section{\label{Conclusions}Conclusions and perspectives}

The conclusions of this work can be summarised as follows: 

\begin{itemize}

\item This work is intended to be the reference article presenting \texttt{Owl-z}, a public Bayesian code aiming at identifying $ z>7$ quasars in massive photometric surveys. \texttt{Owl-z} is based on the classification of sources as high-$z$ quasars or contaminants, the latter being MLT stars or early-type galaxies at intermediate redshifts. The populations of contaminants have been carefully modelled for the needs of the Bayesian model selection. Although the code has been tuned to be used on the EWS, it can be easily adapted to different photometric surveys, as shown in this work. 

\item \texttt{Owl-z} has been able to re-identify all spectroscopically-confirmed quasars at $z>7$. This exercise has also demonstrated the versatility of \texttt{Owl-z} regarding its application to different photometric catalogues. 

\item The performance of \texttt{Owl-z} based on $F$-measure, a metric combining completeness and purity, has been estimated using simulations for the EWS in a wide range of redshifts ($7<z<12$) and reference magnitudes ($18<\HE<24.5$). The results show that \texttt{Owl-z} reaches full performance for bright sources (\HE$<22$) irrespective of the redshift,  meaning that the performance in the spectroscopic confirmation of bright rare quasars selected by \texttt{Owl-z} is expected to be high. The performance drops significantly towards fainter magnitudes and high redshifts.

\item The threshold value for the selection of quasars can be applied in the post-treatment phase, meaning that this value can be tuned to optimise the selection of $z>7$ quasars. For instance, a small/large value can be used for the brightest/faintest sources (\HE$<21$ and $>21$, respectively) to optimise the performance. Given the steep change of the optimum threshold on the magnitude versus redshift plane, a relatively crude cut on this plane could be adopted to optimise the spectroscopic follow-up. 

\item A uniform performance is obtained with \texttt{Owl-z} across the EWS footprint, irrespective of the Galactic coordinates and relative SNR values. 

\item The results of the code remain robust despite inevitable uncertainties in the detailed modelling of the contaminating stellar populations, in particular, the thin and thick disks. 

\end{itemize}

The results obtained with \texttt{Owl-z} are promising and demonstrate its value as a flexible and robust tool for high-$z$ quasar selection. However, several avenues can be explored to enhance its capabilities further and expand its applicability to upcoming large photometric surveys. These perspectives span both methodological improvements and broader scientific applications.

One major development direction involves extending \texttt{Owl-z} to additional photometric datasets, including those from LSST, Roman, and JWST. While the current implementation is tailored to Euclid-like data, its Bayesian framework is inherently adaptable. Nonetheless, brown dwarf spectral templates used in the model become unreliable beyond the K-band, highlighting the need to incorporate theoretical templates or real spectra from complementary instruments to maintain performance across varying wavelength ranges.

Integrating machine learning techniques into the classification pipeline offers another exciting prospect. Deep learning models, such as convolutional neural networks, could be used with \texttt{Owl-z} to refine candidate selection and optimise the balance between completeness and purity. This hybrid approach may prove especially effective at fainter magnitudes, where the performance of \texttt{Owl-z} currently diminishes.

In addition to the photometric selection methods employed in this study, the potential for utilising morphological information in classifying sources presents an intriguing avenue for future research. Incorporating morphological data could differentiate extended sources, such as early-type galaxies, from point sources like high-$z$ quasars. This classification, however, is contingent upon factors such as redshift and magnitude, as higher and elliptical redshift galaxies may appear increasingly compact due to resolution limits. While this work does not include morphological analysis due to its complexity and the technical challenges associated with modelling Euclid detections of extended objects, we acknowledge that such an approach could significantly enhance the quasar selection process. Future studies could explore this integration, potentially refining our ability to mitigate contamination from non-quasar sources and improving detection efficiency in the search for high-$z$ quasars.

Additionally, improved modelling of contaminant populations is vital for maintaining selection accuracy in deeper surveys. Future work should consider a broader range of contaminants, including dusty star-forming galaxies, transients, and thick-disk brown dwarfs. These populations will likely become more prominent as surveys probe deeper into the high-$z$ Universe.

\texttt{Owl-z} would also benefit from being embedded within a broader observational framework, including dedicated spectroscopic follow-up strategies. Systematic spectroscopic validation of candidates would refine the Bayesian thresholds and inform better post-selection filtering.

From a practical standpoint, future developments should also include optimising photometric preselection criteria to better handle systematic uncertainties and noise. This may involve simulation-based studies to understand the code’s response to varying observational conditions and tune selection thresholds accordingly.

Moreover, cross-matching with multi-wavelength surveys (e.g. JWST, ground-based NIR observations) would add confidence to the classification of candidates and help eliminate artefacts or ambiguous detections.

Finally, although \texttt{Owl-z} was not originally designed for precise photometric redshift estimation, its outputs show potential in this area. Coupling \texttt{Owl-z} with dedicated redshift estimation tools in the post-processing stage could provide more accurate redshift predictions, further enriching its scientific utility.

\begin{table*}[h]
    \centering
    \renewcommand{\arraystretch}{1.3}
    \begin{tabular}{|p{3cm}|p{3cm}|p{3cm}|p{3cm}|p{4cm}|}
        \hline
        & \textbf{Models used} & \textbf{Quasar prior} & \textbf{Galaxy prior} & \textbf{MLT stars prior} \\
        \hline
        This work & Quasars, low-z galaxies and MLT stars & \citealt{Matsuoka2023} & LF fitted on real data & \citealt{Caballero2008} \\
        \hline
        \citealt{Pipien2018} & Quasars and LT stars & \citealt{Willott2010} & Not used & \citealt{Caballero2008} \\
        \hline
        \citealt{Mortlock_2012_bayes} & MLT stars, low \textit{z} galaxies and MLT stars & \citealt{Willott2010} & Not used & Power-law number counts in combination with an exponential power-law colour distribution for the reddest stars \\
        \hline
        \citealt{Barnett2019} & MLT stars, low-\textit{z} galaxies and quasars & \citealt{Willott2010} & Parametrised function fit to elliptical galaxies & LT stars based on \citealt{Caballero2008} and M stars based on \citealt{Bochanski2010} \\
        \hline
    \end{tabular}
    \caption{Comparison of models, priors, and references used in various works.}
    \label{tab:comparison}
\end{table*}

\begin{acknowledgements}

This work makes use of publicly available data from the Euclid mission. We are grateful to members of the Euclid Science Working Group Primeval Universe, and in particular the Quasar Work Package, for their informal feedback and discussions during the development of this study. We would like to extend special thanks to Eduardo Ba\~nados and Daniel Mortlock for their insightful input and collaboration. This research has also benefited from the use of quasar templates provided by Paul Hewett. We acknowledge Y\=osuke Matsuoka, Feige Wang, and Jinyi Yang for kindly sharing spectra of quasars at $z > 7$.

This research has made use of the Spanish Virtual Observatory project (\url{https://svo.cab.inta-csic.es}), funded by MCIN/AEI/10.13039/501100011033 through grant PID2020-112949GB-I00, particularly the SpeX Prism Library of brown dwarf spectral templates. We also acknowledge the use of the SVO Filter Profile Service "Carlos Rodrigo", funded by the same grant.

This research has been possible thanks to the computing facilities operated by CeSAM data centre at LAM, Marseille, France.
Finally, we acknowledge the use of several open-source Python packages, including NumPy, SciPy, Astropy, Matplotlib, and Cython, which were instrumental in our data analysis and performance optimisation.

\end{acknowledgements}

\bibstyle{aa}
\bibliography{ref}

@ARTICLE{Fan2006,
       author = {{Fan}, Xiaohui and {Strauss}, Michael A. and {Becker}, Robert H. and {White}, Richard L. and {Gunn}, James E. and {Knapp}, Gillian R. and {Richards}, Gordon T. and {Schneider}, Donald P. and {Brinkmann}, J. and {Fukugita}, Masataka},
        title = "{Constraining the Evolution of the Ionizing Background and the Epoch of Reionization with z\raisebox{-0.5ex}\textasciitilde6 Quasars. II. A Sample of 19 Quasars}",
      journal = {\aj},
         year = 2006,
        month = jul,
       volume = {132},
       number = {1},
        pages = {117-136},
          doi = {10.1086/504836},
archivePrefix = {arXiv},
       eprint = {astro-ph/0512082},
 primaryClass = {astro-ph},
       adsurl = {https://ui.adsabs.harvard.edu/abs/2006AJ....132..117F}
}

@ARTICLE{Becker2015,
       author = {{Becker}, George D. and {Bolton}, James S. and {Madau}, Piero and {Pettini}, Max and {Ryan-Weber}, Emma V. and {Venemans}, Bram P.},
        title = "{Evidence of patchy hydrogen reionization from an extreme Ly{\ensuremath{\alpha}} trough below redshift six}",
      journal = {\mnras},
         year = 2015,
        month = mar,
       volume = {447},
       number = {4},
        pages = {3402-3419},
          doi = {10.1093/mnras/stu2646},
archivePrefix = {arXiv},
       eprint = {1407.4850},
 primaryClass = {astro-ph.CO},
       adsurl = {https://ui.adsabs.harvard.edu/abs/2015MNRAS.447.3402B}
}

@ARTICLE{Bosman2022,
       author = {{Bosman}, Sarah E.~I. and {Davies}, Frederick B. and {Becker}, George D. and {Keating}, Laura C. and {Davies}, Rebecca L. and {Zhu}, Yongda and {Eilers}, Anna-Christina and {D'Odorico}, Valentina and {Bian}, Fuyan and {Bischetti}, Manuela and {Cristiani}, Stefano V. and {Fan}, Xiaohui and {Farina}, Emanuele P. and {Haehnelt}, Martin G. and {Hennawi}, Joseph F. and {Kulkarni}, Girish and {Mesinger}, Andrei and {Meyer}, Romain A. and {Onoue}, Masafusa and {Pallottini}, Andrea and {Qin}, Yuxiang and {Ryan-Weber}, Emma and {Schindler}, Jan-Torge and {Walter}, Fabian and {Wang}, Feige and {Yang}, Jinyi},
        title = "{Hydrogen reionization ends by z = 5.3: Lyman-{\ensuremath{\alpha}} optical depth measured by the XQR-30 sample}",
      journal = {\mnras},
         year = 2022,
        month = jul,
       volume = {514},
       number = {1},
        pages = {55-76},
          doi = {10.1093/mnras/stac1046},
archivePrefix = {arXiv},
       eprint = {2108.03699},
 primaryClass = {astro-ph.CO},
       adsurl = {https://ui.adsabs.harvard.edu/abs/2022MNRAS.514...55B}
}

@ARTICLE{Bennett2024,
       author = {{Bennett}, Jake S. and {Sijacki}, Debora and {Costa}, Tiago and {Laporte}, Nicolas and {Witten}, Callum},
        title = "{The growth of the gargantuan black holes powering high-redshift quasars and their impact on the formation of early galaxies and protoclusters}",
      journal = {\mnras},
         year = 2024,
        month = jan,
       volume = {527},
       number = {1},
        pages = {1033-1054},
          doi = {10.1093/mnras/stad3179},
archivePrefix = {arXiv},
       eprint = {2305.11932},
 primaryClass = {astro-ph.GA},
       adsurl = {https://ui.adsabs.harvard.edu/abs/2024MNRAS.527.1033B}
}

@ARTICLE{Banados2018,
       author = {{Ba{\~n}ados}, Eduardo and {Venemans}, Bram P. and {Mazzucchelli}, Chiara and {Farina}, Emanuele P. and {Walter}, Fabian and {Wang}, Feige and {Decarli}, Roberto and {Stern}, Daniel and {Fan}, Xiaohui and {Davies}, Frederick B. and {Hennawi}, Joseph F. and {Simcoe}, Robert A. and {Turner}, Monica L. and {Rix}, Hans-Walter and {Yang}, Jinyi and {Kelson}, Daniel D. and {Rudie}, Gwen C. and {Winters}, Jan Martin},
        title = "{An 800-million-solar-mass black hole in a significantly neutral Universe at a redshift of 7.5}",
      journal = {\nat},
         year = 2018,
        month = jan,
       volume = {553},
       number = {7689},
        pages = {473-476},
          doi = {10.1038/nature25180},
archivePrefix = {arXiv},
       eprint = {1712.01860},
 primaryClass = {astro-ph.GA},
       adsurl = {https://ui.adsabs.harvard.edu/abs/2018Natur.553..473B}
}

@ARTICLE{Fan2023,
       author = {{Fan}, Xiaohui and {Ba{\~n}ados}, Eduardo and {Simcoe}, Robert A.},
        title = "{Quasars and the Intergalactic Medium at Cosmic Dawn}",
      journal = {\araa},
         year = 2023,
        month = aug,
       volume = {61},
        pages = {373-426},
          doi = {10.1146/annurev-astro-052920-102455},
archivePrefix = {arXiv},
       eprint = {2212.06907},
 primaryClass = {astro-ph.GA},
       adsurl = {https://ui.adsabs.harvard.edu/abs/2023ARA&A..61..373F}
}

@ARTICLE{hubler2024,
       author = {{{\"U}bler}, Hannah and {Maiolino}, Roberto and {P{\'e}rez-Gonz{\'a}lez}, Pablo G. and {D'Eugenio}, Francesco and {Perna}, Michele and {Curti}, Mirko and {Arribas}, Santiago and {Bunker}, Andrew and {Carniani}, Stefano and {Charlot}, St{\'e}phane and {Rodr{\'\i}guez Del Pino}, Bruno and {Baker}, William and {B{\"o}ker}, Torsten and {Cresci}, Giovanni and {Dunlop}, James and {Grogin}, Norman A. and {Jones}, Gareth C. and {Kumari}, Nimisha and {Lamperti}, Isabella and {Laporte}, Nicolas and {Marshall}, Madeline A. and {Mazzolari}, Giovanni and {Parlanti}, Eleonora and {Rawle}, Tim and {Scholtz}, Jan and {Venturi}, Giacomo and {Witstok}, Joris},
        title = "{GA-NIFS: JWST discovers an offset AGN 740 million years after the big bang}",
      journal = {\mnras},
         year = 2024,
        month = jun,
       volume = {531},
       number = {1},
        pages = {355-365},
          doi = {10.1093/mnras/stae943},
archivePrefix = {arXiv},
       eprint = {2312.03589},
 primaryClass = {astro-ph.GA},
       adsurl = {https://ui.adsabs.harvard.edu/abs/2024MNRAS.531..355U}
}

@ARTICLE{Maiolino2024,
       author = {{Maiolino}, Roberto and {Scholtz}, Jan and {Curtis-Lake}, Emma and {Carniani}, Stefano and {Baker}, William and {de Graaff}, Anna and {Tacchella}, Sandro and {{\"U}bler}, Hannah and {D'Eugenio}, Francesco and {Witstok}, Joris and {Curti}, Mirko and {Arribas}, Santiago and {Bunker}, Andrew J. and {Charlot}, St{\'e}phane and {Chevallard}, Jacopo and {Eisenstein}, Daniel J. and {Egami}, Eiichi and {Ji}, Zhiyuan and {Jones}, Gareth C. and {Lyu}, Jianwei and {Rawle}, Tim and {Robertson}, Brant and {Rujopakarn}, Wiphu and {Perna}, Michele and {Sun}, Fengwu and {Venturi}, Giacomo and {Williams}, Christina C. and {Willott}, Chris},
        title = "{JADES: The diverse population of infant black holes at 4 < z < 11: Merging, tiny, poor, but mighty}",
      journal = {\aap},
         year = 2024,
        month = nov,
       volume = {691},
          eid = {A145},
        pages = {A145},
          doi = {10.1051/0004-6361/202347640},
archivePrefix = {arXiv},
       eprint = {2308.01230},
 primaryClass = {astro-ph.GA},
       adsurl = {https://ui.adsabs.harvard.edu/abs/2024A&A...691A.145M}
}

@ARTICLE{Maiolino2024b,
       author = {{Maiolino}, Roberto and {Scholtz}, Jan and {Witstok}, Joris and {Carniani}, Stefano and {D'Eugenio}, Francesco and {de Graaff}, Anna and {{\"U}bler}, Hannah and {Tacchella}, Sandro and {Curtis-Lake}, Emma and {Arribas}, Santiago and {Bunker}, Andrew and {Charlot}, St{\'e}phane and {Chevallard}, Jacopo and {Curti}, Mirko and {Looser}, Tobias J. and {Maseda}, Michael V. and {Rawle}, Timothy D. and {Rodr{\'\i}guez del Pino}, Bruno and {Willott}, Chris J. and {Egami}, Eiichi and {Eisenstein}, Daniel J. and {Hainline}, Kevin N. and {Robertson}, Brant and {Williams}, Christina C. and {Willmer}, Christopher N.~A. and {Baker}, William M. and {Boyett}, Kristan and {DeCoursey}, Christa and {Fabian}, Andrew C. and {Helton}, Jakob M. and {Ji}, Zhiyuan and {Jones}, Gareth C. and {Kumari}, Nimisha and {Laporte}, Nicolas and {Nelson}, Erica J. and {Perna}, Michele and {Sandles}, Lester and {Shivaei}, Irene and {Sun}, Fengwu},
        title = "{A small and vigorous black hole in the early Universe}",
      journal = {\nat},
         year = 2024,
        month = mar,
       volume = {627},
       number = {8002},
        pages = {59-63},
          doi = {10.1038/s41586-024-07052-5},
archivePrefix = {arXiv},
       eprint = {2305.12492},
 primaryClass = {astro-ph.GA},
       adsurl = {https://ui.adsabs.harvard.edu/abs/2024Natur.627...59M}
}

@ARTICLE{Scholtz2024,
       author = {{Scholtz}, Jan and {Witten}, Callum and {Laporte}, Nicolas and {{\"U}bler}, Hannah and {Perna}, Michele and {Maiolino}, Roberto and {Arribas}, Santiago and {Baker}, William M. and {Bennett}, Jake S. and {D'Eugenio}, Francesco and {Simmonds}, Charlotte and {Tacchella}, Sandro and {Witstok}, Joris and {Bunker}, Andrew J. and {Carniani}, Stefano and {Charlot}, St{\'e}phane and {Cresci}, Giovanni and {Curtis-Lake}, Emma and {Eisenstein}, Daniel J. and {Kumari}, Nimisha and {Robertson}, Brant and {Rodr{\'\i}guez Del Pino}, Bruno and {Smit}, Renske and {Venturi}, Giacomo and {Williams}, Christina C. and {Willmer}, Christopher N.~A.},
        title = "{GN-z11: The environment of an active galactic nucleus at z = 10.603. New insights into the most distant Ly{\ensuremath{\alpha}} detection}",
      journal = {\aap},
         year = 2024,
        month = jul,
       volume = {687},
          eid = {A283},
        pages = {A283},
          doi = {10.1051/0004-6361/202347187},
archivePrefix = {arXiv},
       eprint = {2306.09142},
 primaryClass = {astro-ph.GA},
       adsurl = {https://ui.adsabs.harvard.edu/abs/2024A&A...687A.283S}
}

@ARTICLE{Laureijs2011,
       author = {{Laureijs}, R. and {Amiaux}, J. and {Arduini}, S. and {Augu{\`e}res}, J. -L. and {Brinchmann}, J. and {Cole}, R. and {Cropper}, M. and {Dabin}, C. and {Duvet}, L. and {Ealet}, A. and {Garilli}, B. and {Gondoin}, P. and {Guzzo}, L. and {Hoar}, J. and {Hoekstra}, H. and {Holmes}, R. and {Kitching}, T. and {Maciaszek}, T. and {Mellier}, Y. and {Pasian}, F. and {Percival}, W. and {Rhodes}, J. and {Saavedra Criado}, G. and {Sauvage}, M. and {Scaramella}, R. and {Valenziano}, L. and {Warren}, S. and {Bender}, R. and {Castander}, F. and {Cimatti}, A. and {Le F{\`e}vre}, O. and {Kurki-Suonio}, H. and {Levi}, M. and {Lilje}, P. and {Meylan}, G. and {Nichol}, R. and {Pedersen}, K. and {Popa}, V. and {Rebolo Lopez}, R. and {Rix}, H. -W. and {Rottgering}, H. and {Zeilinger}, W. and {Grupp}, F. and {Hudelot}, P. and {Massey}, R. and {Meneghetti}, M. and {Miller}, L. and {Paltani}, S. and {Paulin-Henriksson}, S. and {Pires}, S. and {Saxton}, C. and {Schrabback}, T. and {Seidel}, G. and {Walsh}, J. and {Aghanim}, N. and {Amendola}, L. and {Bartlett}, J. and {Baccigalupi}, C. and {Beaulieu}, J. -P. and {Benabed}, K. and {Cuby}, J. -G. and {Elbaz}, D. and {Fosalba}, P. and {Gavazzi}, G. and {Helmi}, A. and {Hook}, I. and {Irwin}, M. and {Kneib}, J. -P. and {Kunz}, M. and {Mannucci}, F. and {Moscardini}, L. and {Tao}, C. and {Teyssier}, R. and {Weller}, J. and {Zamorani}, G. and {Zapatero Osorio}, M.~R. and {Boulade}, O. and {Foumond}, J.~J. and {Di Giorgio}, A. and {Guttridge}, P. and {James}, A. and {Kemp}, M. and {Martignac}, J. and {Spencer}, A. and {Walton}, D. and {Bl{\"u}mchen}, T. and {Bonoli}, C. and {Bortoletto}, F. and {Cerna}, C. and {Corcione}, L. and {Fabron}, C. and {Jahnke}, K. and {Ligori}, S. and {Madrid}, F. and {Martin}, L. and {Morgante}, G. and {Pamplona}, T. and {Prieto}, E. and {Riva}, M. and {Toledo}, R. and {Trifoglio}, M. and {Zerbi}, F. and {Abdalla}, F. and {Douspis}, M. and {Grenet}, C. and {Borgani}, S. and {Bouwens}, R. and {Courbin}, F. and {Delouis}, J. -M. and {Dubath}, P. and {Fontana}, A. and {Frailis}, M. and {Grazian}, A. and {Koppenh{\"o}fer}, J. and {Mansutti}, O. and {Melchior}, M. and {Mignoli}, M. and {Mohr}, J. and {Neissner}, C. and {Noddle}, K. and {Poncet}, M. and {Scodeggio}, M. and {Serrano}, S. and {Shane}, N. and {Starck}, J. -L. and {Surace}, C. and {Taylor}, A. and {Verdoes-Kleijn}, G. and {Vuerli}, C. and {Williams}, O.~R. and {Zacchei}, A. and {Altieri}, B. and {Escudero Sanz}, I. and {Kohley}, R. and {Oosterbroek}, T. and {Astier}, P. and {Bacon}, D. and {Bardelli}, S. and {Baugh}, C. and {Bellagamba}, F. and {Benoist}, C. and {Bianchi}, D. and {Biviano}, A. and {Branchini}, E. and {Carbone}, C. and {Cardone}, V. and {Clements}, D. and {Colombi}, S. and {Conselice}, C. and {Cresci}, G. and {Deacon}, N. and {Dunlop}, J. and {Fedeli}, C. and {Fontanot}, F. and {Franzetti}, P. and {Giocoli}, C. and {Garcia-Bellido}, J. and {Gow}, J. and {Heavens}, A. and {Hewett}, P. and {Heymans}, C. and {Holland}, A. and {Huang}, Z. and {Ilbert}, O. and {Joachimi}, B. and {Jennins}, E. and {Kerins}, E. and {Kiessling}, A. and {Kirk}, D. and {Kotak}, R. and {Krause}, O. and {Lahav}, O. and {van Leeuwen}, F. and {Lesgourgues}, J. and {Lombardi}, M. and {Magliocchetti}, M. and {Maguire}, K. and {Majerotto}, E. and {Maoli}, R. and {Marulli}, F. and {Maurogordato}, S. and {McCracken}, H. and {McLure}, R. and {Melchiorri}, A. and {Merson}, A. and {Moresco}, M. and {Nonino}, M. and {Norberg}, P. and {Peacock}, J. and {Pello}, R. and {Penny}, M. and {Pettorino}, V. and {Di Porto}, C. and {Pozzetti}, L. and {Quercellini}, C. and {Radovich}, M. and {Rassat}, A. and {Roche}, N. and {Ronayette}, S. and {Rossetti}, E. and {Sartoris}, B. and {Schneider}, P. and {Semboloni}, E. and {Serjeant}, S. and {Simpson}, F. and {Skordis}, C. and {Smadja}, G. and {Smartt}, S. and {Spano}, P. and {Spiro}, S. and {Sullivan}, M. and {Tilquin}, A. and {Trotta}, R. and {Verde}, L. and {Wang}, Y. and {Williger}, G. and {Zhao}, G. and {Zoubian}, J. and {Zucca}, E.},
        title = "{Euclid Definition Study Report}",
      journal = {arXiv e-prints},
         year = 2011,
        month = oct,
          eid = {arXiv:1110.3193},
        pages = {arXiv:1110.3193},
          doi = {10.48550/arXiv.1110.3193},
archivePrefix = {arXiv},
       eprint = {1110.3193},
 primaryClass = {astro-ph.CO},
       adsurl = {https://ui.adsabs.harvard.edu/abs/2011arXiv1110.3193L}
}

@ARTICLE{Mellier2024,
       author = {{Euclid Collaboration} and {Mellier}, Y. and {Abdurro'uf} and {Acevedo Barroso}, J.~A. and {Ach{\'u}carro}, A. and {Adamek}, J. and {Adam}, R. and {Addison}, G.~E. and {Aghanim}, N. and {Aguena}, M. and {Ajani}, V. and {Akrami}, Y. and {Al-Bahlawan}, A. and {Alavi}, A. and {Albuquerque}, I.~S. and {Alestas}, G. and {Alguero}, G. and {Allaoui}, A. and {Allen}, S.~W. and {Allevato}, V. and {Alonso-Tetilla}, A.~V. and {Altieri}, B. and {Alvarez-Candal}, A. and {Alvi}, S. and {Amara}, A. and {Amendola}, L. and {Amiaux}, J. and {Andika}, I.~T. and {Andreon}, S. and {Andrews}, A. and {Angora}, G. and {Angulo}, R.~E. and {Annibali}, F. and {Anselmi}, A. and {Anselmi}, S. and {Arcari}, S. and {Archidiacono}, M. and {Aric{\`o}}, G. and {Arnaud}, M. and {Arnouts}, S. and {Asgari}, M. and {Asorey}, J. and {Atayde}, L. and {Atek}, H. and {Atrio-Barandela}, F. and {Aubert}, M. and {Aubourg}, E. and {Auphan}, T. and {Auricchio}, N. and {Aussel}, B. and {Aussel}, H. and {Avelino}, P.~P. and {Avgoustidis}, A. and {Avila}, S. and {Awan}, S. and {Azzollini}, R. and {Baccigalupi}, C. and {Bachelet}, E. and {Bacon}, D. and {Baes}, M. and {Bagley}, M.~B. and {Bahr-Kalus}, B. and {Balaguera-Antolinez}, A. and {Balbinot}, E. and {Balcells}, M. and {Baldi}, M. and {Baldry}, I. and {Balestra}, A. and {Ballardini}, M. and {Ballester}, O. and {Balogh}, M. and {Ba{\~n}ados}, E. and {Barbier}, R. and {Bardelli}, S. and {Baron}, M. and {Barreiro}, T. and {Barrena}, R. and {Barriere}, J. -C. and {Barros}, B.~J. and {Barthelemy}, A. and {Bartolo}, N. and {Basset}, A. and {Battaglia}, P. and {Battisti}, A.~J. and {Baugh}, C.~M. and {Baumont}, L. and {Bazzanini}, L. and {Beaulieu}, J. -P. and {Beckmann}, V. and {Belikov}, A.~N. and {Bel}, J. and {Bellagamba}, F. and {Bella}, M. and {Bellini}, E. and {Benabed}, K. and {Bender}, R. and {Benevento}, G. and {Bennett}, C.~L. and {Benson}, K. and {Bergamini}, P. and {Bermejo-Climent}, J.~R. and {Bernardeau}, F. and {Bertacca}, D. and {Berthe}, M. and {Berthier}, J. and {Bethermin}, M. and {Beutler}, F. and {Bevillon}, C. and {Bhargava}, S. and {Bhatawdekar}, R. and {Bianchi}, D. and {Bisigello}, L. and {Biviano}, A. and {Blake}, R.~P. and {Blanchard}, A. and {Blazek}, J. and {Blot}, L. and {Bosco}, A. and {Bodendorf}, C. and {Boenke}, T. and {B{\"o}hringer}, H. and {Boldrini}, P. and {Bolzonella}, M. and {Bonchi}, A. and {Bonici}, M. and {Bonino}, D. and {Bonino}, L. and {Bonvin}, C. and {Bon}, W. and {Booth}, J.~T. and {Borgani}, S. and {Borlaff}, A.~S. and {Borsato}, E. and {Bose}, B. and {Botticella}, M.~T. and {Boucaud}, A. and {Bouche}, F. and {Boucher}, J.~S. and {Boutigny}, D. and {Bouvard}, T. and {Bouwens}, R. and {Bouy}, H. and {Bowler}, R.~A.~A. and {Bozza}, V. and {Bozzo}, E. and {Branchini}, E. and {Brando}, G. and {Brau-Nogue}, S. and {Brekke}, P. and {Bremer}, M.~N. and {Brescia}, M. and {Breton}, M. -A. and {Brinchmann}, J. and {Brinckmann}, T. and {Brockley-Blatt}, C. and {Brodwin}, M. and {Brouard}, L. and {Brown}, M.~L. and {Bruton}, S. and {Bucko}, J. and {Buddelmeijer}, H. and {Buenadicha}, G. and {Buitrago}, F. and {Burger}, P. and {Burigana}, C. and {Busillo}, V. and {Busonero}, D. and {Cabanac}, R. and {Cabayol-Garcia}, L. and {Cagliari}, M.~S. and {Caillat}, A. and {Caillat}, L. and {Calabrese}, M. and {Calabro}, A. and {Calderone}, G. and {Calura}, F. and {Camacho Quevedo}, B. and {Camera}, S. and {Campos}, L. and {Ca{\~n}as-Herrera}, G. and {Candini}, G.~P. and {Cantiello}, M. and {Capobianco}, V. and {Cappellaro}, E. and {Cappelluti}, N. and {Cappi}, A. and {Caputi}, K.~I. and {Cara}, C. and {Carbone}, C. and {Cardone}, V.~F. and {Carella}, E. and {Carlberg}, R.~G. and {Carle}, M. and {Carminati}, L. and {Caro}, F. and {Carrasco}, J.~M. and {Carretero}, J. and {Carrilho}, P. and {Carron Duque}, J. and {Carry}, B.},
        title = "{Euclid: I. Overview of the Euclid mission}",
      journal = {\aap},
         year = 2025,
        month = may,
       volume = {697},
          eid = {A1},
        pages = {A1},
          doi = {10.1051/0004-6361/202450810},
archivePrefix = {arXiv},
       eprint = {2405.13491},
 primaryClass = {astro-ph.CO},
       adsurl = {https://ui.adsabs.harvard.edu/abs/2025A&A...697A...1E}
}

@ARTICLE{Akeson2019,
       author = {{Akeson}, Rachel and {Armus}, Lee and {Bachelet}, Etienne and {Bailey}, Vanessa and {Bartusek}, Lisa and {Bellini}, Andrea and {Benford}, Dominic and {Bennett}, David and {Bhattacharya}, Aparna and {Bohlin}, Ralph and {Boyer}, Martha and {Bozza}, Valerio and {Bryden}, Geoffrey and {Calchi Novati}, Sebastiano and {Carpenter}, Kenneth and {Casertano}, Stefano and {Choi}, Ami and {Content}, David and {Dayal}, Pratika and {Dressler}, Alan and {Dor{\'e}}, Olivier and {Fall}, S. Michael and {Fan}, Xiaohui and {Fang}, Xiao and {Filippenko}, Alexei and {Finkelstein}, Steven and {Foley}, Ryan and {Furlanetto}, Steven and {Kalirai}, Jason and {Gaudi}, B. Scott and {Gilbert}, Karoline and {Girard}, Julien and {Grady}, Kevin and {Greene}, Jenny and {Guhathakurta}, Puragra and {Heinrich}, Chen and {Hemmati}, Shoubaneh and {Hendel}, David and {Henderson}, Calen and {Henning}, Thomas and {Hirata}, Christopher and {Ho}, Shirley and {Huff}, Eric and {Hutter}, Anne and {Jansen}, Rolf and {Jha}, Saurabh and {Johnson}, Samson and {Jones}, David and {Kasdin}, Jeremy and {Kelly}, Patrick and {Kirshner}, Robert and {Koekemoer}, Anton and {Kruk}, Jeffrey and {Lewis}, Nikole and {Macintosh}, Bruce and {Madau}, Piero and {Malhotra}, Sangeeta and {Mandel}, Kaisey and {Massara}, Elena and {Masters}, Daniel and {McEnery}, Julie and {McQuinn}, Kristen and {Melchior}, Peter and {Melton}, Mark and {Mennesson}, Bertrand and {Peeples}, Molly and {Penny}, Matthew and {Perlmutter}, Saul and {Pisani}, Alice and {Plazas}, Andr{\'e}s and {Poleski}, Radek and {Postman}, Marc and {Ranc}, Cl{\'e}ment and {Rauscher}, Bernard and {Rest}, Armin and {Roberge}, Aki and {Robertson}, Brant and {Rodney}, Steven and {Rhoads}, James and {Rhodes}, Jason and {Ryan}, Russell, Jr. and {Sahu}, Kailash and {Sand}, David and {Scolnic}, Dan and {Seth}, Anil and {Shvartzvald}, Yossi and {Siellez}, Karelle and {Smith}, Arfon and {Spergel}, David and {Stassun}, Keivan and {Street}, Rachel and {Strolger}, Louis-Gregory and {Szalay}, Alexander and {Trauger}, John and {Troxel}, M.~A. and {Turnbull}, Margaret and {van der Marel}, Roeland and {von der Linden}, Anja and {Wang}, Yun and {Weinberg}, David and {Williams}, Benjamin and {Windhorst}, Rogier and {Wollack}, Edward and {Wu}, Hao-Yi and {Yee}, Jennifer and {Zimmerman}, Neil},
        title = "{The Wide Field Infrared Survey Telescope: 100 Hubbles for the 2020s}",
      journal = {arXiv e-prints},
         year = 2019,
        month = feb,
          eid = {arXiv:1902.05569},
        pages = {arXiv:1902.05569},
          doi = {10.48550/arXiv.1902.05569},
archivePrefix = {arXiv},
       eprint = {1902.05569},
 primaryClass = {astro-ph.IM},
       adsurl = {https://ui.adsabs.harvard.edu/abs/2019arXiv190205569A}
}

@ARTICLE{LSST_2019,
       author = {{Ivezi{\'c}}, {\v{Z}}eljko and {Kahn}, Steven M. and {Tyson}, J. Anthony and {Abel}, Bob and {Acosta}, Emily and {Allsman}, Robyn and {Alonso}, David and {AlSayyad}, Yusra and {Anderson}, Scott F. and {Andrew}, John and {Angel}, James Roger P. and {Angeli}, George Z. and {Ansari}, Reza and {Antilogus}, Pierre and {Araujo}, Constanza and {Armstrong}, Robert and {Arndt}, Kirk T. and {Astier}, Pierre and {Aubourg}, {\'E}ric and {Auza}, Nicole and {Axelrod}, Tim S. and {Bard}, Deborah J. and {Barr}, Jeff D. and {Barrau}, Aurelian and {Bartlett}, James G. and {Bauer}, Amanda E. and {Bauman}, Brian J. and {Baumont}, Sylvain and {Bechtol}, Ellen and {Bechtol}, Keith and {Becker}, Andrew C. and {Becla}, Jacek and {Beldica}, Cristina and {Bellavia}, Steve and {Bianco}, Federica B. and {Biswas}, Rahul and {Blanc}, Guillaume and {Blazek}, Jonathan and {Blandford}, Roger D. and {Bloom}, Josh S. and {Bogart}, Joanne and {Bond}, Tim W. and {Booth}, Michael T. and {Borgland}, Anders W. and {Borne}, Kirk and {Bosch}, James F. and {Boutigny}, Dominique and {Brackett}, Craig A. and {Bradshaw}, Andrew and {Brandt}, William Nielsen and {Brown}, Michael E. and {Bullock}, James S. and {Burchat}, Patricia and {Burke}, David L. and {Cagnoli}, Gianpietro and {Calabrese}, Daniel and {Callahan}, Shawn and {Callen}, Alice L. and {Carlin}, Jeffrey L. and {Carlson}, Erin L. and {Chandrasekharan}, Srinivasan and {Charles-Emerson}, Glenaver and {Chesley}, Steve and {Cheu}, Elliott C. and {Chiang}, Hsin-Fang and {Chiang}, James and {Chirino}, Carol and {Chow}, Derek and {Ciardi}, David R. and {Claver}, Charles F. and {Cohen-Tanugi}, Johann and {Cockrum}, Joseph J. and {Coles}, Rebecca and {Connolly}, Andrew J. and {Cook}, Kem H. and {Cooray}, Asantha and {Covey}, Kevin R. and {Cribbs}, Chris and {Cui}, Wei and {Cutri}, Roc and {Daly}, Philip N. and {Daniel}, Scott F. and {Daruich}, Felipe and {Daubard}, Guillaume and {Daues}, Greg and {Dawson}, William and {Delgado}, Francisco and {Dellapenna}, Alfred and {de Peyster}, Robert and {de Val-Borro}, Miguel and {Digel}, Seth W. and {Doherty}, Peter and {Dubois}, Richard and {Dubois-Felsmann}, Gregory P. and {Durech}, Josef and {Economou}, Frossie and {Eifler}, Tim and {Eracleous}, Michael and {Emmons}, Benjamin L. and {Fausti Neto}, Angelo and {Ferguson}, Henry and {Figueroa}, Enrique and {Fisher-Levine}, Merlin and {Focke}, Warren and {Foss}, Michael D. and {Frank}, James and {Freemon}, Michael D. and {Gangler}, Emmanuel and {Gawiser}, Eric and {Geary}, John C. and {Gee}, Perry and {Geha}, Marla and {Gessner}, Charles J.~B. and {Gibson}, Robert R. and {Gilmore}, D. Kirk and {Glanzman}, Thomas and {Glick}, William and {Goldina}, Tatiana and {Goldstein}, Daniel A. and {Goodenow}, Iain and {Graham}, Melissa L. and {Gressler}, William J. and {Gris}, Philippe and {Guy}, Leanne P. and {Guyonnet}, Augustin and {Haller}, Gunther and {Harris}, Ron and {Hascall}, Patrick A. and {Haupt}, Justine and {Hernandez}, Fabio and {Herrmann}, Sven and {Hileman}, Edward and {Hoblitt}, Joshua and {Hodgson}, John A. and {Hogan}, Craig and {Howard}, James D. and {Huang}, Dajun and {Huffer}, Michael E. and {Ingraham}, Patrick and {Innes}, Walter R. and {Jacoby}, Suzanne H. and {Jain}, Bhuvnesh and {Jammes}, Fabrice and {Jee}, M. James and {Jenness}, Tim and {Jernigan}, Garrett and {Jevremovi{\'c}}, Darko and {Johns}, Kenneth and {Johnson}, Anthony S. and {Johnson}, Margaret W.~G. and {Jones}, R. Lynne and {Juramy-Gilles}, Claire and {Juri{\'c}}, Mario and {Kalirai}, Jason S. and {Kallivayalil}, Nitya J. and {Kalmbach}, Bryce and {Kantor}, Jeffrey P. and {Karst}, Pierre and {Kasliwal}, Mansi M. and {Kelly}, Heather and {Kessler}, Richard and {Kinnison}, Veronica and {Kirkby}, David and {Knox}, Lloyd and {Kotov}, Ivan V. and {Krabbendam}, Victor L. and {Krughoff}, K. Simon and {Kub{\'a}nek}, Petr and {Kuczewski}, John and {Kulkarni}, Shri and {Ku}, John and {Kurita}, Nadine R. and {Lage}, Craig S. and {Lambert}, Ron and {Lange}, Travis and {Langton}, J. Brian and {Le Guillou}, Laurent and {Levine}, Deborah and {Liang}, Ming and {Lim}, Kian-Tat and {Lintott}, Chris J. and {Long}, Kevin E. and {Lopez}, Margaux and {Lotz}, Paul J. and {Lupton}, Robert H. and {Lust}, Nate B. and {MacArthur}, Lauren A. and {Mahabal}, Ashish and {Mandelbaum}, Rachel and {Markiewicz}, Thomas W. and {Marsh}, Darren S. and {Marshall}, Philip J. and {Marshall}, Stuart and {May}, Morgan and {McKercher}, Robert and {McQueen}, Michelle and {Meyers}, Joshua and {Migliore}, Myriam and {Miller}, Michelle and {Mills}, David J. and {Miraval}, Connor and {Moeyens}, Joachim and {Moolekamp}, Fred E. and {Monet}, David G. and {Moniez}, Marc and {Monkewitz}, Serge and {Montgomery}, Christopher and {Morrison}, Christopher B. and {Mueller}, Fritz and {Muller}, Gary P. and {Mu{\~n}oz Arancibia}, Freddy and {Neill}, Douglas R. and {Newbry}, Scott P. and {Nief}, Jean-Yves and {Nomerotski}, Andrei and {Nordby}, Martin and {O'Connor}, Paul and {Oliver}, John and {Olivier}, Scot S. and {Olsen}, Knut and {O'Mullane}, William and {Ortiz}, Sandra and {Osier}, Shawn and {Owen}, Russell E. and {Pain}, Reynald and {Palecek}, Paul E. and {Parejko}, John K. and {Parsons}, James B. and {Pease}, Nathan M. and {Peterson}, J. Matt and {Peterson}, John R. and {Petravick}, Donald L. and {Libby Petrick}, M.~E. and {Petry}, Cathy E. and {Pierfederici}, Francesco and {Pietrowicz}, Stephen and {Pike}, Rob and {Pinto}, Philip A. and {Plante}, Raymond and {Plate}, Stephen and {Plutchak}, Joel P. and {Price}, Paul A. and {Prouza}, Michael and {Radeka}, Veljko and {Rajagopal}, Jayadev and {Rasmussen}, Andrew P. and {Regnault}, Nicolas and {Reil}, Kevin A. and {Reiss}, David J. and {Reuter}, Michael A. and {Ridgway}, Stephen T. and {Riot}, Vincent J. and {Ritz}, Steve and {Robinson}, Sean and {Roby}, William and {Roodman}, Aaron and {Rosing}, Wayne and {Roucelle}, Cecille and {Rumore}, Matthew R. and {Russo}, Stefano and {Saha}, Abhijit and {Sassolas}, Benoit and {Schalk}, Terry L. and {Schellart}, Pim and {Schindler}, Rafe H. and {Schmidt}, Samuel and {Schneider}, Donald P. and {Schneider}, Michael D. and {Schoening}, William and {Schumacher}, German and {Schwamb}, Megan E. and {Sebag}, Jacques and {Selvy}, Brian and {Sembroski}, Glenn H. and {Seppala}, Lynn G. and {Serio}, Andrew and {Serrano}, Eduardo and {Shaw}, Richard A. and {Shipsey}, Ian and {Sick}, Jonathan and {Silvestri}, Nicole and {Slater}, Colin T. and {Smith}, J. Allyn and {Smith}, R. Chris and {Sobhani}, Shahram and {Soldahl}, Christine and {Storrie-Lombardi}, Lisa and {Stover}, Edward and {Strauss}, Michael A. and {Street}, Rachel A. and {Stubbs}, Christopher W. and {Sullivan}, Ian S. and {Sweeney}, Donald and {Swinbank}, John D. and {Szalay}, Alexander and {Takacs}, Peter and {Tether}, Stephen A. and {Thaler}, Jon J. and {Thayer}, John Gregg and {Thomas}, Sandrine and {Thornton}, Adam J. and {Thukral}, Vaikunth and {Tice}, Jeffrey and {Trilling}, David E. and {Turri}, Max and {Van Berg}, Richard and {Vanden Berk}, Daniel and {Vetter}, Kurt and {Virieux}, Francoise and {Vucina}, Tomislav and {Wahl}, William and {Walkowicz}, Lucianne and {Walsh}, Brian and {Walter}, Christopher W. and {Wang}, Daniel L. and {Wang}, Shin-Yawn and {Warner}, Michael and {Wiecha}, Oliver and {Willman}, Beth and {Winters}, Scott E. and {Wittman}, David and {Wolff}, Sidney C. and {Wood-Vasey}, W. Michael and {Wu}, Xiuqin and {Xin}, Bo and {Yoachim}, Peter and {Zhan}, Hu},
        title = "{LSST: From Science Drivers to Reference Design and Anticipated Data Products}",
      journal = {\apj},
         year = 2019,
        month = mar,
       volume = {873},
       number = {2},
          eid = {111},
        pages = {111},
          doi = {10.3847/1538-4357/ab042c},
archivePrefix = {arXiv},
       eprint = {0805.2366},
 primaryClass = {astro-ph},
       adsurl = {https://ui.adsabs.harvard.edu/abs/2019ApJ...873..111I}
}

@ARTICLE{Mortlock_2012_bayes,
       author = {{Mortlock}, Daniel J. and {Patel}, Mitesh and {Warren}, Stephen J. and {Hewett}, Paul C. and {Venemans}, Bram P. and {McMahon}, Richard G. and {Simpson}, Chris},
        title = "{Probabilistic selection of high-redshift quasars}",
      journal = {\mnras},
         year = 2012,
        month = jan,
       volume = {419},
       number = {1},
        pages = {390-410},
          doi = {10.1111/j.1365-2966.2011.19710.x},
archivePrefix = {arXiv},
       eprint = {1101.4965},
 primaryClass = {astro-ph.IM},
       adsurl = {https://ui.adsabs.harvard.edu/abs/2012MNRAS.419..390M}
}

@ARTICLE{Barnett2019,
       author = {{Euclid Collaboration: Barnett}, R. and {Warren}, S.~J. and {Mortlock}, D.~J. and {Cuby}, J. -G. and {Conselice}, C. and {Hewett}, P.~C. and {Willott}, C.~J. and {Auricchio}, N. and {Balaguera-Antol{\'\i}nez}, A. and {Baldi}, M. and {Bardelli}, S. and {Bellagamba}, F. and {Bender}, R. and {Biviano}, A. and {Bonino}, D. and {Bozzo}, E. and {Branchini}, E. and {Brescia}, M. and {Brinchmann}, J. and {Burigana}, C. and {Camera}, S. and {Capobianco}, V. and {Carbone}, C. and {Carretero}, J. and {Carvalho}, C.~S. and {Castander}, F.~J. and {Castellano}, M. and {Cavuoti}, S. and {Cimatti}, A. and {Cl{\'e}dassou}, R. and {Congedo}, G. and {Conversi}, L. and {Copin}, Y. and {Corcione}, L. and {Coupon}, J. and {Courtois}, H.~M. and {Cropper}, M. and {Da Silva}, A. and {Duncan}, C.~A.~J. and {Dusini}, S. and {Ealet}, A. and {Farrens}, S. and {Fosalba}, P. and {Fotopoulou}, S. and {Fourmanoit}, N. and {Frailis}, M. and {Fumana}, M. and {Galeotta}, S. and {Garilli}, B. and {Gillard}, W. and {Gillis}, B.~R. and {Graci{\'a}-Carpio}, J. and {Grupp}, F. and {Hoekstra}, H. and {Hormuth}, F. and {Israel}, H. and {Jahnke}, K. and {Kermiche}, S. and {Kilbinger}, M. and {Kirkpatrick}, C.~C. and {Kitching}, T. and {Kohley}, R. and {Kubik}, B. and {Kunz}, M. and {Kurki-Suonio}, H. and {Laureijs}, R. and {Ligori}, S. and {Lilje}, P.~B. and {Lloro}, I. and {Maiorano}, E. and {Mansutti}, O. and {Marggraf}, O. and {Martinet}, N. and {Marulli}, F. and {Massey}, R. and {Mauri}, N. and {Medinaceli}, E. and {Mei}, S. and {Mellier}, Y. and {Metcalf}, R.~B. and {Metge}, J.~J. and {Meylan}, G. and {Moresco}, M. and {Moscardini}, L. and {Munari}, E. and {Neissner}, C. and {Niemi}, S.~M. and {Nutma}, T. and {Padilla}, C. and {Paltani}, S. and {Pasian}, F. and {Paykari}, P. and {Percival}, W.~J. and {Pettorino}, V. and {Polenta}, G. and {Poncet}, M. and {Pozzetti}, L. and {Raison}, F. and {Renzi}, A. and {Rhodes}, J. and {Rix}, H. -W. and {Romelli}, E. and {Roncarelli}, M. and {Rossetti}, E. and {Saglia}, R. and {Sapone}, D. and {Scaramella}, R. and {Schneider}, P. and {Scottez}, V. and {Secroun}, A. and {Serrano}, S. and {Sirri}, G. and {Stanco}, L. and {Sureau}, F. and {Tallada-Cresp{\'\i}}, P. and {Tavagnacco}, D. and {Taylor}, A.~N. and {Tenti}, M. and {Tereno}, I. and {Toledo-Moreo}, R. and {Torradeflot}, F. and {Valenziano}, L. and {Vassallo}, T. and {Wang}, Y. and {Zacchei}, A. and {Zamorani}, G. and {Zoubian}, J. and {Zucca}, E.},
        title = "{Euclid preparation. V. Predicted yield of redshift 7 < z < 9 quasars from the wide survey}",
      journal = {\aap},
         year = 2019,
        month = nov,
       volume = {631},
          eid = {A85},
        pages = {A85},
          doi = {10.1051/0004-6361/201936427},
archivePrefix = {arXiv},
       eprint = {1908.04310},
 primaryClass = {astro-ph.GA},
       adsurl = {https://ui.adsabs.harvard.edu/abs/2019A&A...631A..85E}
}

@ARTICLE{Nanni2023,
       author = {{Nanni}, Riccardo and {Hennawi}, Joseph F. and {Wang}, Feige and {Yang}, Jinyi and {Schindler}, Jan-Torge and {Fan}, Xiaohui},
        title = "{Paving the way for Euclid and JWST via probabilistic selection of high-redshift quasars}",
      journal = {\mnras},
         year = 2022,
        month = sep,
       volume = {515},
       number = {3},
        pages = {3224-3248},
          doi = {10.1093/mnras/stac1944},
archivePrefix = {arXiv},
       eprint = {2111.03073},
 primaryClass = {astro-ph.GA},
       adsurl = {https://ui.adsabs.harvard.edu/abs/2022MNRAS.515.3224N}
}

@ARTICLE{Kang2024,
       author = {{Kang}, Yi and {Hennawi}, Joseph F. and {Schindler}, Jan-Torge and {Tamanas}, John and {Nanni}, Riccardo},
        title = "{Extreme Deconvolution Reimagined: Conditional Densities via Neural Networks and an Application in Quasar Classification}",
      journal = {arXiv e-prints},
         year = 2024,
        month = dec,
          eid = {arXiv:2412.03029},
        pages = {arXiv:2412.03029},
          doi = {10.48550/arXiv.2412.03029},
archivePrefix = {arXiv},
       eprint = {2412.03029},
 primaryClass = {astro-ph.IM},
       adsurl = {https://ui.adsabs.harvard.edu/abs/2024arXiv241203029K}
}

@ARTICLE{Bovy2009,
       author = {{Bovy}, Jo and {Hogg}, David W. and {Roweis}, Sam T.},
        title = "{Extreme deconvolution: Inferring complete distribution functions from noisy, heterogeneous and incomplete observations}",
      journal = {Annals of Applied Statistics},
         year = 2011,
        month = jun,
       volume = {5},
       number = {2},
        pages = {1657-1677},
          doi = {10.1214/10-AOAS439},
archivePrefix = {arXiv},
       eprint = {0905.2979},
 primaryClass = {stat.ME},
       adsurl = {https://ui.adsabs.harvard.edu/abs/2011AnApS...5.1657B}
}

@ARTICLE{Belladitta2020,
       author = {{Belladitta}, S. and {Moretti}, A. and {Caccianiga}, A. and {Spingola}, C. and {Severgnini}, P. and {Della Ceca}, R. and {Ghisellini}, G. and {Dallacasa}, D. and {Sbarrato}, T. and {Cicone}, C. and {Cassar{\`a}}, L.~P. and {Pedani}, M.},
        title = "{The first blazar observed at z > 6}",
      journal = {\aap},
         year = 2020,
        month = mar,
       volume = {635},
          eid = {L7},
        pages = {L7},
          doi = {10.1051/0004-6361/201937395},
archivePrefix = {arXiv},
       eprint = {2002.05178},
 primaryClass = {astro-ph.CO},
       adsurl = {https://ui.adsabs.harvard.edu/abs/2020A&A...635L...7B}
}

@ARTICLE{Banados2014,
       author = {{Ba{\~n}ados}, E. and {Venemans}, B.~P. and {Morganson}, E. and {Decarli}, R. and {Walter}, F. and {Chambers}, K.~C. and {Rix}, H. -W. and {Farina}, E.~P. and {Fan}, X. and {Jiang}, L. and {McGreer}, I. and {De Rosa}, G. and {Simcoe}, R. and {Wei{\ss}}, A. and {Price}, P.~A. and {Morgan}, J.~S. and {Burgett}, W.~S. and {Greiner}, J. and {Kaiser}, N. and {Kudritzki}, R. -P. and {Magnier}, E.~A. and {Metcalfe}, N. and {Stubbs}, C.~W. and {Sweeney}, W. and {Tonry}, J.~L. and {Wainscoat}, R.~J. and {Waters}, C.},
        title = "{Discovery of Eight z \raisebox{-0.5ex}\textasciitilde 6 Quasars from Pan-STARRS1}",
      journal = {\aj},
         year = 2014,
        month = jul,
       volume = {148},
       number = {1},
          eid = {14},
        pages = {14},
          doi = {10.1088/0004-6256/148/1/14},
archivePrefix = {arXiv},
       eprint = {1405.3986},
 primaryClass = {astro-ph.GA},
       adsurl = {https://ui.adsabs.harvard.edu/abs/2014AJ....148...14B}
}

@ARTICLE{Banados2016,
       author = {{Ba{\~n}ados}, E. and {Venemans}, B.~P. and {Decarli}, R. and {Farina}, E.~P. and {Mazzucchelli}, C. and {Walter}, F. and {Fan}, X. and {Stern}, D. and {Schlafly}, E. and {Chambers}, K.~C. and {Rix}, H. -W. and {Jiang}, L. and {McGreer}, I. and {Simcoe}, R. and {Wang}, F. and {Yang}, J. and {Morganson}, E. and {De Rosa}, G. and {Greiner}, J. and {Balokovi{\'c}}, M. and {Burgett}, W.~S. and {Cooper}, T. and {Draper}, P.~W. and {Flewelling}, H. and {Hodapp}, K.~W. and {Jun}, H.~D. and {Kaiser}, N. and {Kudritzki}, R. -P. and {Magnier}, E.~A. and {Metcalfe}, N. and {Miller}, D. and {Schindler}, J. -T. and {Tonry}, J.~L. and {Wainscoat}, R.~J. and {Waters}, C. and {Yang}, Q.},
        title = "{The Pan-STARRS1 Distant z > 5.6 Quasar Survey: More than 100 Quasars within the First Gyr of the Universe}",
      journal = {\apjs},
         year = 2016,
        month = nov,
       volume = {227},
       number = {1},
          eid = {11},
        pages = {11},
          doi = {10.3847/0067-0049/227/1/11},
archivePrefix = {arXiv},
       eprint = {1608.03279},
 primaryClass = {astro-ph.GA},
       adsurl = {https://ui.adsabs.harvard.edu/abs/2016ApJS..227...11B}
}

@ARTICLE{Wenzl2021,
       author = {{Wenzl}, Lukas and {Schindler}, Jan-Torge and {Fan}, Xiaohui and {Andika}, Irham Taufik and {Ba{\~n}ados}, Eduardo and {Decarli}, Roberto and {Jahnke}, Knud and {Mazzucchelli}, Chiara and {Onoue}, Masafusa and {Venemans}, Bram P. and {Walter}, Fabian and {Yang}, Jinyi},
        title = "{Random Forests as a Viable Method to Select and Discover High-redshift Quasars}",
      journal = {\aj},
         year = 2021,
        month = aug,
       volume = {162},
       number = {2},
          eid = {72},
        pages = {72},
          doi = {10.3847/1538-3881/ac0254},
archivePrefix = {arXiv},
       eprint = {2105.09171},
 primaryClass = {astro-ph.GA},
       adsurl = {https://ui.adsabs.harvard.edu/abs/2021AJ....162...72W}
}

@ARTICLE{Scaramella2022,
       author = {{Euclid Collaboration: Scaramella}, R. and {Amiaux}, J. and {Mellier}, Y. and {Burigana}, C. and {Carvalho}, C.~S. and {Cuillandre}, J. -C. and {Da Silva}, A. and {Derosa}, A. and {Dinis}, J. and {Maiorano}, E. and {Maris}, M. and {Tereno}, I. and {Laureijs}, R. and {Boenke}, T. and {Buenadicha}, G. and {Dupac}, X. and {Gaspar Venancio}, L.~M. and {G{\'o}mez-{\'A}lvarez}, P. and {Hoar}, J. and {Lorenzo Alvarez}, J. and {Racca}, G.~D. and {Saavedra-Criado}, G. and {Schwartz}, J. and {Vavrek}, R. and {Schirmer}, M. and {Aussel}, H. and {Azzollini}, R. and {Cardone}, V.~F. and {Cropper}, M. and {Ealet}, A. and {Garilli}, B. and {Gillard}, W. and {Granett}, B.~R. and {Guzzo}, L. and {Hoekstra}, H. and {Jahnke}, K. and {Kitching}, T. and {Maciaszek}, T. and {Meneghetti}, M. and {Miller}, L. and {Nakajima}, R. and {Niemi}, S.~M. and {Pasian}, F. and {Percival}, W.~J. and {Pottinger}, S. and {Sauvage}, M. and {Scodeggio}, M. and {Wachter}, S. and {Zacchei}, A. and {Aghanim}, N. and {Amara}, A. and {Auphan}, T. and {Auricchio}, N. and {Awan}, S. and {Balestra}, A. and {Bender}, R. and {Bodendorf}, C. and {Bonino}, D. and {Branchini}, E. and {Brau-Nogue}, S. and {Brescia}, M. and {Candini}, G.~P. and {Capobianco}, V. and {Carbone}, C. and {Carlberg}, R.~G. and {Carretero}, J. and {Casas}, R. and {Castander}, F.~J. and {Castellano}, M. and {Cavuoti}, S. and {Cimatti}, A. and {Cledassou}, R. and {Congedo}, G. and {Conselice}, C.~J. and {Conversi}, L. and {Copin}, Y. and {Corcione}, L. and {Costille}, A. and {Courbin}, F. and {Degaudenzi}, H. and {Douspis}, M. and {Dubath}, F. and {Duncan}, C.~A.~J. and {Dusini}, S. and {Farrens}, S. and {Ferriol}, S. and {Fosalba}, P. and {Fourmanoit}, N. and {Frailis}, M. and {Franceschi}, E. and {Franzetti}, P. and {Fumana}, M. and {Gillis}, B. and {Giocoli}, C. and {Grazian}, A. and {Grupp}, F. and {Haugan}, S.~V.~H. and {Holmes}, W. and {Hormuth}, F. and {Hudelot}, P. and {Kermiche}, S. and {Kiessling}, A. and {Kilbinger}, M. and {Kohley}, R. and {Kubik}, B. and {K{\"u}mmel}, M. and {Kunz}, M. and {Kurki-Suonio}, H. and {Lahav}, O. and {Ligori}, S. and {Lilje}, P.~B. and {Lloro}, I. and {Mansutti}, O. and {Marggraf}, O. and {Markovic}, K. and {Marulli}, F. and {Massey}, R. and {Maurogordato}, S. and {Melchior}, M. and {Merlin}, E. and {Meylan}, G. and {Mohr}, J.~J. and {Moresco}, M. and {Morin}, B. and {Moscardini}, L. and {Munari}, E. and {Nichol}, R.~C. and {Padilla}, C. and {Paltani}, S. and {Peacock}, J. and {Pedersen}, K. and {Pettorino}, V. and {Pires}, S. and {Poncet}, M. and {Popa}, L. and {Pozzetti}, L. and {Raison}, F. and {Rebolo}, R. and {Rhodes}, J. and {Rix}, H. -W. and {Roncarelli}, M. and {Rossetti}, E. and {Saglia}, R. and {Schneider}, P. and {Schrabback}, T. and {Secroun}, A. and {Seidel}, G. and {Serrano}, S. and {Sirignano}, C. and {Sirri}, G. and {Skottfelt}, J. and {Stanco}, L. and {Starck}, J.~L. and {Tallada-Cresp{\'\i}}, P. and {Tavagnacco}, D. and {Taylor}, A.~N. and {Teplitz}, H.~I. and {Toledo-Moreo}, R. and {Torradeflot}, F. and {Trifoglio}, M. and {Valentijn}, E.~A. and {Valenziano}, L. and {Verdoes Kleijn}, G.~A. and {Wang}, Y. and {Welikala}, N. and {Weller}, J. and {Wetzstein}, M. and {Zamorani}, G. and {Zoubian}, J. and {Andreon}, S. and {Baldi}, M. and {Bardelli}, S. and {Boucaud}, A. and {Camera}, S. and {Di Ferdinando}, D. and {Fabbian}, G. and {Farinelli}, R. and {Galeotta}, S. and {Graci{\'a}-Carpio}, J. and {Maino}, D. and {Medinaceli}, E. and {Mei}, S. and {Neissner}, C. and {Polenta}, G. and {Renzi}, A. and {Romelli}, E. and {Rosset}, C. and {Sureau}, F. and {Tenti}, M. and {Vassallo}, T. and {Zucca}, E. and {Baccigalupi}, C. and {Balaguera-Antol{\'\i}nez}, A. and {Battaglia}, P. and {Biviano}, A. and {Borgani}, S. and {Bozzo}, E. and {Cabanac}, R. and {Cappi}, A. and {Casas}, S. and {Castignani}, G. and {Colodro-Conde}, C. and {Coupon}, J. and {Courtois}, H.~M. and {Cuby}, J. and {de la Torre}, S. and {Desai}, S. and {Dole}, H. and {Fabricius}, M. and {Farina}, M. and {Ferreira}, P.~G. and {Finelli}, F. and {Flose-Reimberg}, P. and {Fotopoulou}, S. and {Ganga}, K. and {Gozaliasl}, G. and {Hook}, I.~M. and {Keihanen}, E. and {Kirkpatrick}, C.~C. and {Liebing}, P. and {Lindholm}, V. and {Mainetti}, G. and {Martinelli}, M. and {Martinet}, N. and {Maturi}, M. and {McCracken}, H.~J. and {Metcalf}, R.~B. and {Morgante}, G. and {Nightingale}, J. and {Nucita}, A. and {Patrizii}, L. and {Potter}, D. and {Riccio}, G. and {S{\'a}nchez}, A.~G. and {Sapone}, D. and {Schewtschenko}, J.~A. and {Schultheis}, M. and {Scottez}, V. and {Teyssier}, R. and {Tutusaus}, I. and {Valiviita}, J. and {Viel}, M. and {Vriend}, W. and {Whittaker}, L.},
        title = "{Euclid preparation. I. The Euclid Wide Survey}",
      journal = {\aap},
         year = 2022,
        month = jun,
       volume = {662},
          eid = {A112},
        pages = {A112},
          doi = {10.1051/0004-6361/202141938},
archivePrefix = {arXiv},
       eprint = {2108.01201},
 primaryClass = {astro-ph.CO},
       adsurl = {https://ui.adsabs.harvard.edu/abs/2022A&A...662A.112E}
}

@ARTICLE{Oke1983,
       author = {{Oke}, J.~B. and {Gunn}, J.~E.},
        title = "{Secondary standard stars for absolute spectrophotometry.}",
      journal = {\apj},
         year = 1983,
        month = mar,
       volume = {266},
        pages = {713-717},
          doi = {10.1086/160817},
       adsurl = {https://ui.adsabs.harvard.edu/abs/1983ApJ...266..713O} 
}

@ARTICLE{Pipien2018,
       author = {{Pipien}, S. and {Cuby}, J. -G. and {Basa}, S. and {Willott}, C.~J. and {Cuillandre}, J. -C. and {Arnouts}, S. and {Hudelot}, P.},
        title = "{High-redshift quasar selection from the CFHQSIR survey}",
      journal = {\aap},
         year = 2018,
        month = sep,
       volume = {617},
          eid = {A127},
        pages = {A127},
          doi = {10.1051/0004-6361/201833488},
archivePrefix = {arXiv},
       eprint = {1808.10672},
 primaryClass = {astro-ph.GA},
       adsurl = {https://ui.adsabs.harvard.edu/abs/2018A&A...617A.127P}
}

@ARTICLE{vanMierlo2022,
       author = {{Euclid Collaboration: van Mierlo}, S.~E. and {Caputi}, K.~I. and {Ashby}, M. and {Atek}, H. and {Bolzonella}, M. and {Bowler}, R.~A.~A. and {Brammer}, G. and {Conselice}, C.~J. and {Cuby}, J. and {Dayal}, P. and {D{\'\i}az-S{\'a}nchez}, A. and {Finkelstein}, S.~L. and {Hoekstra}, H. and {Humphrey}, A. and {Ilbert}, O. and {McCracken}, H.~J. and {Milvang-Jensen}, B. and {Oesch}, P.~A. and {Pello}, R. and {Rodighiero}, G. and {Schirmer}, M. and {Toft}, S. and {Weaver}, J.~R. and {Wilkins}, S.~M. and {Willott}, C.~J. and {Zamorani}, G. and {Amara}, A. and {Auricchio}, N. and {Baldi}, M. and {Bender}, R. and {Bodendorf}, C. and {Bonino}, D. and {Branchini}, E. and {Brescia}, M. and {Brinchmann}, J. and {Camera}, S. and {Capobianco}, V. and {Carbone}, C. and {Carretero}, J. and {Castellano}, M. and {Cavuoti}, S. and {Cimatti}, A. and {Cledassou}, R. and {Congedo}, G. and {Conversi}, L. and {Copin}, Y. and {Corcione}, L. and {Courbin}, F. and {Da Silva}, A. and {Degaudenzi}, H. and {Douspis}, M. and {Dubath}, F. and {Dupac}, X. and {Dusini}, S. and {Farrens}, S. and {Ferriol}, S. and {Frailis}, M. and {Franceschi}, E. and {Franzetti}, P. and {Fumana}, M. and {Galeotta}, S. and {Garilli}, B. and {Gillard}, W. and {Gillis}, B. and {Giocoli}, C. and {Grazian}, A. and {Grupp}, F. and {Haugan}, S.~V.~H. and {Holmes}, W. and {Hormuth}, F. and {Hornstrup}, A. and {Jahnke}, K. and {K{\"u}mmel}, M. and {Kiessling}, A. and {Kilbinger}, M. and {Kitching}, T. and {Kohley}, R. and {Kunz}, M. and {Kurki-Suonio}, H. and {Laureijs}, R. and {Ligori}, S. and {Lilje}, P.~B. and {Lloro}, I. and {Maiorano}, E. and {Mansutti}, O. and {Marggraf}, O. and {Markovic}, K. and {Marulli}, F. and {Massey}, R. and {Maurogordato}, S. and {Medinaceli}, E. and {Meneghetti}, M. and {Merlin}, E. and {Meylan}, G. and {Moresco}, M. and {Moscardini}, L. and {Munari}, E. and {Niemi}, S.~M. and {Padilla}, C. and {Paltani}, S. and {Pasian}, F. and {Pedersen}, K. and {Pettorino}, V. and {Pires}, S. and {Poncet}, M. and {Popa}, L. and {Pozzetti}, L. and {Raison}, F. and {Renzi}, A. and {Rhodes}, J. and {Riccio}, G. and {Romelli}, E. and {Rossetti}, E. and {Saglia}, R. and {Sapone}, D. and {Sartoris}, B. and {Schneider}, P. and {Secroun}, A. and {Sirignano}, C. and {Sirri}, G. and {Stanco}, L. and {Starck}, J. -L. and {Surace}, C. and {Tallada-Cresp{\'\i}}, P. and {Taylor}, A.~N. and {Tereno}, I. and {Toledo-Moreo}, R. and {Torradeflot}, F. and {Tutusaus}, I. and {Valentijn}, E.~A. and {Valenziano}, L. and {Vassallo}, T. and {Wang}, Y. and {Zacchei}, A. and {Zoubian}, J. and {Andreon}, S. and {Bardelli}, S. and {Boucaud}, A. and {Graci{\'a}-Carpio}, J. and {Maino}, D. and {Mauri}, N. and {Mei}, S. and {Sureau}, F. and {Zucca}, E. and {Aussel}, H. and {Baccigalupi}, C. and {Balaguera-Antol{\'\i}nez}, A. and {Biviano}, A. and {Blanchard}, A. and {Borgani}, S. and {Bozzo}, E. and {Burigana}, C. and {Cabanac}, R. and {Calura}, F. and {Cappi}, A. and {Carvalho}, C.~S. and {Casas}, S. and {Castignani}, G. and {Colodro-Conde}, C. and {Cooray}, A.~R. and {Coupon}, J. and {Courtois}, H.~M. and {Crocce}, M. and {Cucciati}, O. and {Davini}, S. and {Dole}, H. and {Escartin}, J.~A. and {Escoffier}, S. and {Fabricius}, M. and {Farina}, M. and {Ganga}, K. and {Garc{\'\i}a-Bellido}, J. and {George}, K. and {Giacomini}, F. and {Gozaliasl}, G. and {Gwyn}, S. and {Hook}, I. and {Huertas-Company}, M. and {Kansal}, V. and {Kashlinsky}, A. and {Keihanen}, E. and {Kirkpatrick}, C.~C. and {Lindholm}, V. and {Maoli}, R. and {Martinelli}, M. and {Martinet}, N. and {Maturi}, M. and {Metcalf}, R.~B. and {Monaco}, P. and {Morgante}, G. and {Nucita}, A.~A. and {Patrizii}, L. and {Peel}, A. and {Pollack}, J. and {Popa}, V. and {Porciani}, C. and {Potter}, D. and {Reimberg}, P. and {S{\'a}nchez}, A.~G. and {Scottez}, V. and {Sefusatti}, E. and {Stadel}, J. and {Teyssier}, R. and {Valiviita}, J. and {Viel}, M.},
        title = "{Euclid preparation. XXI. Intermediate-redshift contaminants in the search for z > 6 galaxies within the Euclid Deep Survey}",
      journal = {\aap},
         year = 2022,
        month = oct,
       volume = {666},
          eid = {A200},
        pages = {A200},
          doi = {10.1051/0004-6361/202243950},
archivePrefix = {arXiv},
       eprint = {2205.02871},
 primaryClass = {astro-ph.GA},
       adsurl = {https://ui.adsabs.harvard.edu/abs/2022A&A...666A.200V}
}

@article{Stern2007,
doi = {10.1086/516833},
url = {https://dx.doi.org/10.1086/516833},
year = {2007},
month = {jul},
publisher = {},
volume = {663},
number = {1},
pages = {677},
author = {Daniel Stern and J. Davy Kirkpatrick and Lori E. Allen and Chao Bian and Andrew Blain and Kate Brand and Mark Brodwin and Michael J. I. Brown and Richard Cool and Vandana Desai and Arjun Dey and Peter Eisenhardt and Anthony Gonzalez and Buell T. Jannuzi and Karin Menendez-Delmestre and Howard A. Smith and B. T. Soifer and Glenn P. Tiede and E. Wright},
title = {Mid-Infrared Selection of Brown Dwarfs and High-Redshift Quasars},
journal = {\apj}
}

@ARTICLE{Caballero2008,
       author = {{Caballero}, J.~A. and {Burgasser}, A.~J. and {Klement}, R.},
        title = "{Contamination by field late-M, L, and T dwarfs in deep surveys}",
      journal = {\aap},
         year = 2008,
        month = sep,
       volume = {488},
       number = {1},
        pages = {181-190},
          doi = {10.1051/0004-6361:200809520},
archivePrefix = {arXiv},
       eprint = {0805.4480},
 primaryClass = {astro-ph},
       adsurl = {https://ui.adsabs.harvard.edu/abs/2008A&A...488..181C}
}

@article{Wilkins2014,
    author = {Wilkins, Stephen M. and Stanway, Elizabeth R. and Bremer, Malcolm N.},
    title = "{High-redshift galaxies and low-mass stars}",
    journal = {\mnras},
    volume = {439},
    number = {1},
    pages = {1038-1050},
    year = {2014},
    month = {02},
    issn = {0035-8711},
    doi = {10.1093/mnras/stu029},
    url = {https://doi.org/10.1093/mnras/stu029},
    eprint = {https://academic.oup.com/mnras/article-pdf/439/1/1038/5627219/stu029.pdf},
}

@ARTICLE{Hainline2024_2,
       author = {{Hainline}, Kevin N. and {D'Eugenio}, Francesco and {Sun}, Fengwu and {Helton}, Jakob M. and {Miles}, Brittany E. and {Marley}, Mark S. and {Lew}, Ben W.~P. and {Leisenring}, Jarron M. and {Bunker}, Andrew J. and {Cargile}, Phillip A. and {Carniani}, Stefano and {Eisenstein}, Daniel J. and {Juod{\v{z}}balis}, Ignas and {Johnson}, Benjamin D. and {Robertson}, Brant and {Tacchella}, Sandro and {Williams}, Christina C. and {Willmer}, Christopher N.~A.},
        title = "{JADES: Spectroscopic Confirmation and Proper Motion for a T-Dwarf at 2 kpc}",
      journal = {\apj},
         year = 2024,
        month = nov,
       volume = {975},
       number = {1},
          eid = {31},
        pages = {31},
          doi = {10.3847/1538-4357/ad76a7},
archivePrefix = {arXiv},
       eprint = {2407.08781},
 primaryClass = {astro-ph.GA},
       adsurl = {https://ui.adsabs.harvard.edu/abs/2024ApJ...975...31H}
}

@ARTICLE{Schirmer2022,
       author = {{Euclid Collaboration} and {Schirmer}, M. and {Jahnke}, K. and {Seidel}, G. and {Aussel}, H. and {Bodendorf}, C. and {Grupp}, F. and {Hormuth}, F. and {Wachter}, S. and {Appleton}, P.~N. and {Barbier}, R. and {Brinchmann}, J. and {Carrasco}, J.~M. and {Castander}, F.~J. and {Coupon}, J. and {De Paolis}, F. and {Franco}, A. and {Ganga}, K. and {Hudelot}, P. and {Jullo}, E. and {Lan{\c{c}}on}, A. and {Nucita}, A.~A. and {Paltani}, S. and {Smadja}, G. and {Strafella}, F. and {Venancio}, L.~M.~G. and {Weiler}, M. and {Amara}, A. and {Auphan}, T. and {Auricchio}, N. and {Balestra}, A. and {Bender}, R. and {Bonino}, D. and {Branchini}, E. and {Brescia}, M. and {Capobianco}, V. and {Carbone}, C. and {Carretero}, J. and {Casas}, R. and {Castellano}, M. and {Cavuoti}, S. and {Cimatti}, A. and {Cledassou}, R. and {Congedo}, G. and {Conselice}, C.~J. and {Conversi}, L. and {Copin}, Y. and {Corcione}, L. and {Costille}, A. and {Courbin}, F. and {Da Silva}, A. and {Degaudenzi}, H. and {Douspis}, M. and {Dubath}, F. and {Dupac}, X. and {Dusini}, S. and {Ealet}, A. and {Farrens}, S. and {Ferriol}, S. and {Fosalba}, P. and {Frailis}, M. and {Franceschi}, E. and {Franzetti}, P. and {Fumana}, M. and {Garilli}, B. and {Gillard}, W. and {Gillis}, B. and {Giocoli}, C. and {Grazian}, A. and {Guzzo}, L. and {Haugan}, S.~V.~H. and {Hoekstra}, H. and {Holmes}, W. and {Hornstrup}, A. and {K{\"u}mmel}, M. and {Kermiche}, S. and {Kiessling}, A. and {Kilbinger}, M. and {Kitching}, T. and {Kohley}, R. and {Kunz}, M. and {Kurki-Suonio}, H. and {Laureijs}, R. and {Ligori}, S. and {Lilje}, P.~B. and {Lloro}, I. and {Maciaszek}, T. and {Maiorano}, E. and {Mansutti}, O. and {Marggraf}, O. and {Markovic}, K. and {Marulli}, F. and {Massey}, R. and {Maurogordato}, S. and {Mellier}, Y. and {Meneghetti}, M. and {Merlin}, E. and {Meylan}, G. and {Moresco}, M. and {Moscardini}, L. and {Munari}, E. and {Nakajima}, R. and {Nichol}, R.~C. and {Niemi}, S.~M. and {Padilla}, C. and {Pasian}, F. and {Pedersen}, K. and {Percival}, W.~J. and {Pettorino}, V. and {Pires}, S. and {Poncet}, M. and {Popa}, L. and {Pozzetti}, L. and {Prieto}, E. and {Raison}, F. and {Rhodes}, J. and {Rix}, H. -W. and {Roncarelli}, M. and {Rossetti}, E. and {Saglia}, R. and {Sartoris}, B. and {Scaramella}, R. and {Schneider}, P. and {Secroun}, A. and {Serrano}, S. and {Sirignano}, C. and {Sirri}, G. and {Stanco}, L. and {Tallada-Cresp{\'\i}}, P. and {Taylor}, A.~N. and {Teplitz}, H.~I. and {Tereno}, I. and {Toledo-Moreo}, R. and {Torradeflot}, F. and {Trifoglio}, M. and {Valentijn}, E.~A. and {Valenziano}, L. and {Wang}, Y. and {Weller}, J. and {Zamorani}, G. and {Zoubian}, J. and {Andreon}, S. and {Bardelli}, S. and {Boucaud}, A. and {Camera}, S. and {Farinelli}, R. and {Graci{\'a}-Carpio}, J. and {Maino}, D. and {Medinaceli}, E. and {Mei}, S. and {Morisset}, N. and {Polenta}, G. and {Renzi}, A. and {Romelli}, E. and {Tenti}, M. and {Vassallo}, T. and {Zacchei}, A. and {Zucca}, E. and {Baccigalupi}, C. and {Balaguera-Antol{\'\i}nez}, A. and {Biviano}, A. and {Blanchard}, A. and {Borgani}, S. and {Bozzo}, E. and {Burigana}, C. and {Cabanac}, R. and {Cappi}, A. and {Carvalho}, C.~S. and {Casas}, S. and {Castignani}, G. and {Colodro-Conde}, C. and {Cooray}, A.~R. and {Courtois}, H.~M. and {Crocce}, M. and {Cuby}, J. -G. and {Davini}, S. and {de la Torre}, S. and {Di Ferdinando}, D. and {Escartin}, J.~A. and {Farina}, M. and {Ferreira}, P.~G. and {Finelli}, F. and {Fotopoulou}, S. and {Galeotta}, S. and {Garcia-Bellido}, J. and {Gaztanaga}, E. and {George}, K. and {Gozaliasl}, G. and {Hook}, I.~M. and {Ili{\'c}}, S. and {Kansal}, V. and {Kashlinsky}, A. and {Keihanen}, E. and {Kirkpatrick}, C.~C. and {Lindholm}, V. and {Mainetti}, G. and {Maoli}, R. and {Martinelli}, M. and {Martinet}, N. and {Maturi}, M.},
        title = "{Euclid preparation. XVIII. The NISP photometric system}",
      journal = {\aap},
         year = 2022,
        month = jun,
       volume = {662},
          eid = {A92},
        pages = {A92},
          doi = {10.1051/0004-6361/202142897},
archivePrefix = {arXiv},
       eprint = {2203.01650},
 primaryClass = {astro-ph.IM},
       adsurl = {https://ui.adsabs.harvard.edu/abs/2022A&A...662A..92E}
}

@book{robert2007bayesian,
  title     = {The Bayesian choice: from decision-theoretic foundations to computational implementation},
  author    = {Robert, Christian P},
  edition   = {2nd},
  year      = {2007},
  publisher = {Springer}
}

@ARTICLE{Banados2016_template,
       author = {{Ba{\~n}ados}, E. and {Venemans}, B.~P. and {Decarli}, R. and {Farina}, E.~P. and {Mazzucchelli}, C. and {Walter}, F. and {Fan}, X. and {Stern}, D. and {Schlafly}, E. and {Chambers}, K.~C. and {Rix}, H. -W. and {Jiang}, L. and {McGreer}, I. and {Simcoe}, R. and {Wang}, F. and {Yang}, J. and {Morganson}, E. and {De Rosa}, G. and {Greiner}, J. and {Balokovi{\'c}}, M. and {Burgett}, W.~S. and {Cooper}, T. and {Draper}, P.~W. and {Flewelling}, H. and {Hodapp}, K.~W. and {Jun}, H.~D. and {Kaiser}, N. and {Kudritzki}, R. -P. and {Magnier}, E.~A. and {Metcalfe}, N. and {Miller}, D. and {Schindler}, J. -T. and {Tonry}, J.~L. and {Wainscoat}, R.~J. and {Waters}, C. and {Yang}, Q.},
        title = "{The Pan-STARRS1 Distant z > 5.6 Quasar Survey: More than 100 Quasars within the First Gyr of the Universe}",
      journal = {\apjs},
         year = 2016,
        month = nov,
       volume = {227},
       number = {1},
          eid = {11},
        pages = {11},
          doi = {10.3847/0067-0049/227/1/11},
archivePrefix = {arXiv},
       eprint = {1608.03279},
 primaryClass = {astro-ph.GA},
       adsurl = {https://ui.adsabs.harvard.edu/abs/2016ApJS..227...11B}
}

@ARTICLE{Willott2010,
       author = {{Willott}, Chris J. and {Delorme}, Philippe and {Reyl{\'e}}, C{\'e}line and {Albert}, Loic and {Bergeron}, Jacqueline and {Crampton}, David and {Delfosse}, Xavier and {Forveille}, Thierry and {Hutchings}, John B. and {McLure}, Ross J. and {Omont}, Alain and {Schade}, David},
        title = "{The Canada-France High-z Quasar Survey: Nine New Quasars and the Luminosity Function at Redshift 6}",
      journal = {\aj},
         year = 2010,
        month = mar,
       volume = {139},
       number = {3},
        pages = {906-918},
          doi = {10.1088/0004-6256/139/3/906},
archivePrefix = {arXiv},
       eprint = {0912.0281},
 primaryClass = {astro-ph.CO},
       adsurl = {https://ui.adsabs.harvard.edu/abs/2010AJ....139..906W}
}

@ARTICLE{Matsuoka2023,
       author = {{Matsuoka}, Yoshiki and {Onoue}, Masafusa and {Iwasawa}, Kazushi and {Strauss}, Michael A. and {Kashikawa}, Nobunari and {Izumi}, Takuma and {Nagao}, Tohru and {Imanishi}, Masatoshi and {Akiyama}, Masayuki and {Silverman}, John D. and {Asami}, Naoko and {Bosch}, James and {Furusawa}, Hisanori and {Goto}, Tomotsugu and {Gunn}, James E. and {Harikane}, Yuichi and {Ikeda}, Hiroyuki and {Inayoshi}, Kohei and {Ishimoto}, Rikako and {Kawaguchi}, Toshihiro and {Kikuta}, Satoshi and {Kohno}, Kotaro and {Komiyama}, Yutaka and {Lee}, Chien-Hsiu and {Lupton}, Robert H. and {Minezaki}, Takeo and {Miyazaki}, Satoshi and {Murayama}, Hitoshi and {Nishizawa}, Atsushi J. and {Oguri}, Masamune and {Ono}, Yoshiaki and {Oogi}, Taira and {Ouchi}, Masami and {Price}, Paul A. and {Sameshima}, Hiroaki and {Sugiyama}, Naoshi and {Tait}, Philip J. and {Takada}, Masahiro and {Takahashi}, Ayumi and {Takata}, Tadafumi and {Tanaka}, Masayuki and {Toba}, Yoshiki and {Wang}, Shiang-Yu and {Yamashita}, Takuji},
        title = "{Quasar Luminosity Function at z = 7}",
      journal = {\apjl},
         year = 2023,
        month = jun,
       volume = {949},
       number = {2},
          eid = {L42},
        pages = {L42},
          doi = {10.3847/2041-8213/acd69f},
archivePrefix = {arXiv},
       eprint = {2305.11225},
 primaryClass = {astro-ph.GA},
       adsurl = {https://ui.adsabs.harvard.edu/abs/2023ApJ...949L..42M}
}

@ARTICLE{Bakos2002,
       author = {{Bakos}, G{\'a}sp{\'a}r {\'A}. and {Sahu}, Kailash C. and {N{\'e}meth}, P{\'e}ter},
        title = "{Revised Coordinates and Proper Motions of the Stars in the Luyten Half-Second Catalog}",
      journal = {\apjs},
         year = 2002,
        month = jul,
       volume = {141},
       number = {1},
        pages = {187-193},
          doi = {10.1086/340115},
archivePrefix = {arXiv},
       eprint = {astro-ph/0202164},
 primaryClass = {astro-ph},
       adsurl = {https://ui.adsabs.harvard.edu/abs/2002ApJS..141..187B}
}

@INPROCEEDINGS{Burgasser2014,
       author = {{Burgasser}, Adam J.},
        title = "{The SpeX Prism Library: 1000+ low-resolution, near-infrared spectra of ultracool M, L, T and Y dwarfs}",
    booktitle = {Astronomical Society of India Conference Series},
         year = 2014,
       series = {Astronomical Society of India Conference Series},
       volume = {11},
        month = jan,
        pages = {7-16},
          doi = {10.48550/arXiv.1406.4887},
archivePrefix = {arXiv},
       eprint = {1406.4887},
 primaryClass = {astro-ph.SR},
       adsurl = {https://ui.adsabs.harvard.edu/abs/2014ASInC..11....7B}
}

@INPROCEEDINGS{Burgasser2017,
       author = {{Burgasser}, A.~J. and {Splat Development Team}},
        title = "{The SpeX Prism Library Analysis Toolkit (SPLAT): A Data Curation Model}",
    booktitle = {Astronomical Society of India Conference Series},
         year = 2017,
       series = {Astronomical Society of India Conference Series},
       volume = {14},
        month = jan,
        pages = {7-12},
          doi = {10.48550/arXiv.1707.00062},
archivePrefix = {arXiv},
       eprint = {1707.00062},
 primaryClass = {astro-ph.SR},
       adsurl = {https://ui.adsabs.harvard.edu/abs/2017ASInC..14....7B}
}

@ARTICLE{Best2017,
       author = {{Best}, William M.~J. and {Liu}, Michael C. and {Dupuy}, Trent J. and {Magnier}, Eugene A.},
        title = "{The Young L Dwarf 2MASS J11193254-1137466 Is a Planetary-mass Binary}",
      journal = {\apjl},
         year = 2017,
        month = jul,
       volume = {843},
       number = {1},
          eid = {L4},
        pages = {L4},
          doi = {10.3847/2041-8213/aa76df},
archivePrefix = {arXiv},
       eprint = {1706.01883},
 primaryClass = {astro-ph.SR},
       adsurl = {https://ui.adsabs.harvard.edu/abs/2017ApJ...843L...4B}
}

@article{Sysoliatina2022,
	author = {{Sysoliatina, K.} and {Just, A.}},
	title = {Towards a fully consistent Milky Way disk model - V. The disk model for 4–14 kpc},
	DOI= "10.1051/0004-6361/202243780",
	url= "https://doi.org/10.1051/0004-6361/202243780",
	journal = {A\&A},
	year = 2022,
	volume = 666,
	pages = "A130",
}

@ARTICLE{Recio-Blanco2023,
       author = {{Gaia Collaboration} and {Recio-Blanco}, A. and {Kordopatis}, G. and {de Laverny}, P. and {Palicio}, P.~A. and {Spagna}, A. and {Spina}, L. and {Katz}, D. and {Re Fiorentin}, P. and {Poggio}, E. and {McMillan}, P.~J. and {Vallenari}, A. and {Lattanzi}, M.~G. and {Seabroke}, G.~M. and {Casamiquela}, L. and {Bragaglia}, A. and {Antoja}, T. and {Bailer-Jones}, C.~A.~L. and {Schultheis}, M. and {Andrae}, R. and {Fouesneau}, M. and {Cropper}, M. and {Cantat-Gaudin}, T. and {Bijaoui}, A. and {Heiter}, U. and {Brown}, A.~G.~A. and {Prusti}, T. and {de Bruijne}, J.~H.~J. and {Arenou}, F. and {Babusiaux}, C. and {Biermann}, M. and {Creevey}, O.~L. and {Ducourant}, C. and {Evans}, D.~W. and {Eyer}, L. and {Guerra}, R. and {Hutton}, A. and {Jordi}, C. and {Klioner}, S.~A. and {Lammers}, U.~L. and {Lindegren}, L. and {Luri}, X. and {Mignard}, F. and {Panem}, C. and {Pourbaix}, D. and {Randich}, S. and {Sartoretti}, P. and {Soubiran}, C. and {Tanga}, P. and {Walton}, N.~A. and {Bastian}, U. and {Drimmel}, R. and {Jansen}, F. and {van Leeuwen}, F. and {Bakker}, J. and {Cacciari}, C. and {Casta{\~n}eda}, J. and {De Angeli}, F. and {Fabricius}, C. and {Fr{\'e}mat}, Y. and {Galluccio}, L. and {Guerrier}, A. and {Masana}, E. and {Messineo}, R. and {Mowlavi}, N. and {Nicolas}, C. and {Nienartowicz}, K. and {Pailler}, F. and {Panuzzo}, P. and {Riclet}, F. and {Roux}, W. and {Sordo}, R. and {Th{\'e}venin}, F. and {Gracia-Abril}, G. and {Portell}, J. and {Teyssier}, D. and {Altmann}, M. and {Audard}, M. and {Bellas-Velidis}, I. and {Benson}, K. and {Berthier}, J. and {Blomme}, R. and {Burgess}, P.~W. and {Busonero}, D. and {Busso}, G. and {C{\'a}novas}, H. and {Carry}, B. and {Cellino}, A. and {Cheek}, N. and {Clementini}, G. and {Damerdji}, Y. and {Davidson}, M. and {de Teodoro}, P. and {Nu{\~n}ez Campos}, M. and {Delchambre}, L. and {Dell'Oro}, A. and {Esquej}, P. and {Fern{\'a}ndez-Hern{\'a}ndez}, J. and {Fraile}, E. and {Garabato}, D. and {Garc{\'\i}a-Lario}, P. and {Gosset}, E. and {Haigron}, R. and {Halbwachs}, J. -L. and {Hambly}, N.~C. and {Harrison}, D.~L. and {Hern{\'a}ndez}, J. and {Hestroffer}, D. and {Hodgkin}, S.~T. and {Holl}, B. and {Jan{\ss}en}, K. and {Jevardat de Fombelle}, G. and {Jordan}, S. and {Krone-Martins}, A. and {Lanzafame}, A.~C. and {L{\"o}ffler}, W. and {Marchal}, O. and {Marrese}, P.~M. and {Moitinho}, A. and {Muinonen}, K. and {Osborne}, P. and {Pancino}, E. and {Pauwels}, T. and {Reyl{\'e}}, C. and {Riello}, M. and {Rimoldini}, L. and {Roegiers}, T. and {Rybizki}, J. and {Sarro}, L.~M. and {Siopis}, C. and {Smith}, M. and {Sozzetti}, A. and {Utrilla}, E. and {van Leeuwen}, M. and {Abbas}, U. and {{\'A}brah{\'a}m}, P. and {Abreu Aramburu}, A. and {Aerts}, C. and {Aguado}, J.~J. and {Ajaj}, M. and {Aldea-Montero}, F. and {Altavilla}, G. and {{\'A}lvarez}, M.~A. and {Alves}, J. and {Anders}, F. and {Anderson}, R.~I. and {Anglada Varela}, E. and {Baines}, D. and {Baker}, S.~G. and {Balaguer-N{\'u}{\~n}ez}, L. and {Balbinot}, E. and {Balog}, Z. and {Barache}, C. and {Barbato}, D. and {Barros}, M. and {Barstow}, M.~A. and {Bartolom{\'e}}, S. and {Bassilana}, J. -L. and {Bauchet}, N. and {Becciani}, U. and {Bellazzini}, M. and {Berihuete}, A. and {Bernet}, M. and {Bertone}, S. and {Bianchi}, L. and {Binnenfeld}, A. and {Blanco-Cuaresma}, S. and {Boch}, T. and {Bombrun}, A. and {Bossini}, D. and {Bouquillon}, S. and {Bramante}, L. and {Breedt}, E. and {Bressan}, A. and {Brouillet}, N. and {Brugaletta}, E. and {Bucciarelli}, B. and {Burlacu}, A. and {Butkevich}, A.~G. and {Buzzi}, R. and {Caffau}, E. and {Cancelliere}, R. and {Carballo}, R. and {Carlucci}, T. and {Carnerero}, M.~I. and {Carrasco}, J.~M. and {Castellani}, M. and {Castro-Ginard}, A. and {Chaoul}, L. and {Charlot}, P. and {Chemin}, L. and {Chiaramida}, V. and {Chiavassa}, A. and {Chornay}, N. and {Comoretto}, G. and {Contursi}, G. and {Cooper}, W.~J. and {Cornez}, T. and {Cowell}, S. and {Crifo}, F. and {Crosta}, M. and {Crowley}, C. and {Dafonte}, C. and {Dapergolas}, A. and {David}, P. and {De Luise}, F. and {De March}, R. and {De Ridder}, J. and {de Souza}, R. and {de Torres}, A. and {del Peloso}, E.~F. and {del Pozo}, E. and {Delbo}, M. and {Delgado}, A. and {Delisle}, J. -B. and {Demouchy}, C. and {Dharmawardena}, T.~E. and {Di Matteo}, P. and {Diakite}, S. and {Diener}, C. and {Distefano}, E. and {Dolding}, C. and {Edvardsson}, B. and {Enke}, H. and {Fabre}, C. and {Fabrizio}, M. and {Faigler}, S. and {Fedorets}, G. and {Fernique}, P. and {Figueras}, F. and {Fournier}, Y. and {Fouron}, C. and {Fragkoudi}, F. and {Gai}, M. and {Garcia-Gutierrez}, A. and {Garcia-Reinaldos}, M. and {Garc{\'\i}a-Torres}, M. and {Garofalo}, A. and {Gavel}, A. and {Gavras}, P. and {Gerlach}, E. and {Geyer}, R. and {Giacobbe}, P. and {Gilmore}, G. and {Girona}, S. and {Giuffrida}, G. and {Gomel}, R. and {Gomez}, A. and {Gonz{\'a}lez-N{\'u}{\~n}ez}, J. and {Gonz{\'a}lez-Santamar{\'\i}a}, I. and {Gonz{\'a}lez-Vidal}, J.~J. and {Granvik}, M. and {Guillout}, P. and {Guiraud}, J. and {Guti{\'e}rrez-S{\'a}nchez}, R. and {Guy}, L.~P. and {Hatzidimitriou}, D. and {Hauser}, M. and {Haywood}, M. and {Helmer}, A. and {Helmi}, A. and {Sarmiento}, M.~H. and {Hidalgo}, S.~L. and {H{\l}adczuk}, N. and {Hobbs}, D. and {Holland}, G. and {Huckle}, H.~E. and {Jardine}, K. and {Jasniewicz}, G. and {Jean-Antoine Piccolo}, A. and {Jim{\'e}nez-Arranz}, {\'O}. and {Juaristi Campillo}, J. and {Julbe}, F. and {Karbevska}, L. and {Kervella}, P. and {Khanna}, S. and {Korn}, A.~J. and {K{\'o}sp{\'a}l}, {\'A}. and {Kostrzewa-Rutkowska}, Z. and {Kruszy{\'n}ska}, K. and {Kun}, M. and {Laizeau}, P. and {Lambert}, S. and {Lanza}, A.~F. and {Lasne}, Y. and {Le Campion}, J. -F. and {Lebreton}, Y. and {Lebzelter}, T. and {Leccia}, S. and {Leclerc}, N. and {Lecoeur-Taibi}, I. and {Liao}, S. and {Licata}, E.~L. and {Lindstr{\o}m}, H.~E.~P. and {Lister}, T.~A. and {Livanou}, E. and {Lobel}, A. and {Lorca}, A. and {Loup}, C. and {Madrero Pardo}, P. and {Magdaleno Romeo}, A. and {Managau}, S. and {Mann}, R.~G. and {Manteiga}, M. and {Marchant}, J.~M. and {Marconi}, M. and {Marcos}, J. and {Marcos Santos}, M.~M.~S. and {Mar{\'\i}n Pina}, D. and {Marinoni}, S. and {Marocco}, F. and {Marshall}, D.~J. and {Martin Polo}, L. and {Mart{\'\i}n-Fleitas}, J.~M. and {Marton}, G. and {Mary}, N. and {Masip}, A. and {Massari}, D. and {Mastrobuono-Battisti}, A. and {Mazeh}, T. and {Messina}, S. and {Michalik}, D. and {Millar}, N.~R. and {Mints}, A. and {Molina}, D. and {Molinaro}, R. and {Moln{\'a}r}, L. and {Monari}, G. and {Mongui{\'o}}, M. and {Montegriffo}, P. and {Montero}, A. and {Mor}, R. and {Mora}, A. and {Morbidelli}, R. and {Morel}, T. and {Morris}, D. and {Muraveva}, T. and {Murphy}, C.~P. and {Musella}, I. and {Nagy}, Z. and {Noval}, L. and {Oca{\~n}a}, F. and {Ogden}, A. and {Ordenovic}, C. and {Osinde}, J.~O. and {Pagani}, C. and {Pagano}, I. and {Palaversa}, L. and {Pallas-Quintela}, L. and {Panahi}, A. and {Payne-Wardenaar}, S. and {Pe{\~n}alosa Esteller}, X. and {Penttil{\"a}}, A. and {Pichon}, B. and {Piersimoni}, A.~M. and {Pineau}, F. -X. and {Plachy}, E. and {Plum}, G. and {Pr{\v{s}}a}, A. and {Pulone}, L. and {Racero}, E. and {Ragaini}, S. and {Rainer}, M. and {Raiteri}, C.~M. and {Ramos}, P. and {Ramos-Lerate}, M. and {Regibo}, S. and {Richards}, P.~J. and {Rios Diaz}, C. and {Ripepi}, V. and {Riva}, A. and {Rix}, H. -W. and {Rixon}, G. and {Robichon}, N. and {Robin}, A.~C. and {Robin}, C. and {Roelens}, M. and {Rogues}, H.~R.~O. and {Rohrbasser}, L. and {Romero-G{\'o}mez}, M. and {Rowell}, N. and {Royer}, F. and {Ruz Mieres}, D. and {Rybicki}, K.~A. and {Sadowski}, G. and {S{\'a}ez N{\'u}{\~n}ez}, A. and {Sagrist{\`a} Sell{\'e}s}, A. and {Sahlmann}, J. and {Salguero}, E. and {Samaras}, N. and {Sanchez Gimenez}, V. and {Sanna}, N. and {Santove{\~n}a}, R. and {Sarasso}, M. and {Sciacca}, E. and {Segol}, M. and {Segovia}, J.~C. and {S{\'e}gransan}, D. and {Semeux}, D. and {Shahaf}, S. and {Siddiqui}, H.~I. and {Siebert}, A. and {Siltala}, L. and {Silvelo}, A. and {Slezak}, E. and {Slezak}, I. and {Smart}, R.~L. and {Snaith}, O.~N. and {Solano}, E. and {Solitro}, F. and {Souami}, D. and {Souchay}, J. and {Spoto}, F. and {Steele}, I.~A. and {Steidelm{\"u}ller}, H. and {Stephenson}, C.~A. and {S{\"u}veges}, M. and {Surdej}, J. and {Szabados}, L. and {Szegedi-Elek}, E. and {Taris}, F. and {Taylor}, M.~B. and {Teixeira}, R. and {Tolomei}, L. and {Tonello}, N. and {Torra}, F. and {Torra}, J. and {Torralba Elipe}, G. and {Trabucchi}, M. and {Tsounis}, A.~T. and {Turon}, C. and {Ulla}, A. and {Unger}, N. and {Vaillant}, M.~V. and {van Dillen}, E. and {van Reeven}, W. and {Vanel}, O. and {Vecchiato}, A. and {Viala}, Y. and {Vicente}, D. and {Voutsinas}, S. and {Weiler}, M. and {Wevers}, T. and {Wyrzykowski}, {\L}. and {Yoldas}, A. and {Yvard}, P. and {Zhao}, H. and {Zorec}, J. and {Zucker}, S. and {Zwitter}, T.},
        title = "{Gaia Data Release 3. Chemical cartography of the Milky Way}",
      journal = {\aap},
         year = 2023,
        month = jun,
       volume = {674},
          eid = {A38},
        pages = {A38},
          doi = {10.1051/0004-6361/202243511},
archivePrefix = {arXiv},
       eprint = {2206.05534},
 primaryClass = {astro-ph.GA},
       adsurl = {https://ui.adsabs.harvard.edu/abs/2023A&A...674A..38G}
}

@article{Vieira2022,
doi = {10.3847/1538-4357/ac6b9b},
url = {https://dx.doi.org/10.3847/1538-4357/ac6b9b},
year = {2022},
month = {jun},
publisher = {The American Astronomical Society},
volume = {932},
number = {1},
pages = {28},
author = {Katherine Vieira and Giovanni Carraro and Vladimir Korchagin and Artem Lutsenko and Terrence M. Girard and William van Altena},
title = {Milky Way Thin and Thick Disk Kinematics with Gaia EDR3 and RAVE DR5},
journal = {\apj}
}

@article{Burgasser2024,
doi = {10.3847/1538-4357/ad206f},
url = {https://dx.doi.org/10.3847/1538-4357/ad206f},
year = {2024},
month = {feb},
publisher = {The American Astronomical Society},
volume = {962},
number = {2},
pages = {177},
author = {Adam J. Burgasser and Rachel Bezanson and Ivo Labbe and Gabriel Brammer and Sam E. Cutler and Lukas J. Furtak and Jenny E. Greene and Roman Gerasimov and Joel Leja and Richard Pan and Sedona H. Price and Bingjie Wang and John R. Weaver and Katherine E. Whitaker and Seiji Fujimoto and Vasily Kokorev and Pratika Dayal and Themiya Nanayakkara and Christina C. Williams and Danilo Marchesini and Adi Zitrin and Pieter van Dokkum},
title = {UNCOVER: JWST Spectroscopy of Three Cold Brown Dwarfs at Kiloparsec-scale Distances},
journal = {\apj}
}

@ARTICLE{Ryan2016,
       author = {{Ryan}, R.~E., Jr. and {Reid}, I.~N.},
        title = "{The Surface Densities of Disk Brown Dwarfs in JWST Surveys}",
      journal = {\aj},
         year = 2016,
        month = apr,
       volume = {151},
       number = {4},
          eid = {92},
        pages = {92},
          doi = {10.3847/0004-6256/151/4/92},
archivePrefix = {arXiv},
       eprint = {1510.05019},
 primaryClass = {astro-ph.GA},
       adsurl = {https://ui.adsabs.harvard.edu/abs/2016AJ....151...92R}
}

@ARTICLE{Vieira2023,
       author = {{Vieira}, Katherine and {Korchagin}, Vladimir and {Carraro}, Giovanni and {Lutsenko}, Artem},
        title = "{Vertical Structure of the Milky Way Disk with Gaia DR3}",
      journal = {Galaxies},
         year = 2023,
        month = jun,
       volume = {11},
       number = {3},
          eid = {77},
        pages = {77},
          doi = {10.3390/galaxies11030077},
       adsurl = {https://ui.adsabs.harvard.edu/abs/2023Galax..11...77V}
}

@ARTICLE{Dupuy2012,
       author = {{Dupuy}, Trent J. and {Liu}, Michael C.},
        title = "{The Hawaii Infrared Parallax Program. I. Ultracool Binaries and the L/T Transition}",
      journal = {\apjs},
         year = 2012,
        month = aug,
       volume = {201},
       number = {2},
          eid = {19},
        pages = {19},
          doi = {10.1088/0067-0049/201/2/19},
archivePrefix = {arXiv},
       eprint = {1201.2465},
 primaryClass = {astro-ph.SR},
       adsurl = {https://ui.adsabs.harvard.edu/abs/2012ApJS..201...19D}
}

@ARTICLE{Skrzypek2016,
       author = {{Skrzypek}, N. and {Warren}, S.~J. and {Faherty}, J.~K.},
        title = "{Photometric brown-dwarf classification. II. A homogeneous sample of 1361 L and T dwarfs brighter than J = 17.5 with accurate spectral types}",
      journal = {\aap},
         year = 2016,
        month = may,
       volume = {589},
          eid = {A49},
        pages = {A49},
          doi = {10.1051/0004-6361/201527359},
archivePrefix = {arXiv},
       eprint = {1602.08582},
 primaryClass = {astro-ph.IM},
       adsurl = {https://ui.adsabs.harvard.edu/abs/2016A&A...589A..49S}
}

@ARTICLE{Bochanski2010,
       author = {{Bochanski}, John J. and {Hawley}, Suzanne L. and {Covey}, Kevin R. and {West}, Andrew A. and {Reid}, I. Neill and {Golimowski}, David A. and {Ivezi{\'c}}, {\v{Z}}eljko},
        title = "{The Luminosity and Mass Functions of Low-mass Stars in the Galactic Disk. II. The Field}",
      journal = {\aj},
         year = 2010,
        month = jun,
       volume = {139},
       number = {6},
        pages = {2679-2699},
          doi = {10.1088/0004-6256/139/6/2679},
archivePrefix = {arXiv},
       eprint = {1004.4002},
 primaryClass = {astro-ph.SR},
       adsurl = {https://ui.adsabs.harvard.edu/abs/2010AJ....139.2679B}
}

@Inbook{Reid2013,
author="Reid, I. Neill",
editor="Oswalt, Terry D.
and Barstow, Martin A.",
title="Brown Dwarfs",
bookTitle="Planets, Stars and Stellar Systems: Volume 4: Stellar Structure and Evolution",
year="2013",
publisher="Springer Netherlands",
address="Dordrecht",
pages="337--395",
isbn="978-94-007-5615-1",
doi="10.1007/978-94-007-5615-1_7",
url="https://doi.org/10.1007/978-94-007-5615-1_7"
}

@ARTICLE{Hainline2024_1,
       author = {{Hainline}, Kevin N. and {Helton}, Jakob M. and {Johnson}, Benjamin D. and {Sun}, Fengwu and {Topping}, Michael W. and {Leisenring}, Jarron M. and {Baker}, William M. and {Eisenstein}, Daniel J. and {Hausen}, Ryan and {Hviding}, Raphael E. and {Lyu}, Jianwei and {Robertson}, Brant and {Tacchella}, Sandro and {Williams}, Christina C. and {Willmer}, Christopher N.~A. and {Roellig}, Thomas L.},
        title = "{Brown Dwarf Candidates in the JADES and CEERS Extragalactic Surveys}",
      journal = {\apj},
         year = 2024,
        month = mar,
       volume = {964},
       number = {1},
          eid = {66},
        pages = {66},
          doi = {10.3847/1538-4357/ad20d1},
archivePrefix = {arXiv},
       eprint = {2309.03250},
 primaryClass = {astro-ph.SR},
       adsurl = {https://ui.adsabs.harvard.edu/abs/2024ApJ...964...66H}
}

@article{Weaver_2022,
   title={COSMOS2020: A Panchromatic View of the Universe to z ∼ 10 from Two Complementary Catalogs},
   volume={258},
   ISSN={1538-4365},
   url={http://dx.doi.org/10.3847/1538-4365/ac3078},
   DOI={10.3847/1538-4365/ac3078},
   number={1},
   journal={ApJ Supplement Series},
   publisher={American Astronomical Society},
   author={Weaver, J. R. and Kauffmann, O. B. and Ilbert, O. and McCracken, H. J. and Moneti, A. and Toft, S. and Brammer, G. and Shuntov, M. and Davidzon, I. and Hsieh, B. C. and Laigle, C. and Anastasiou, A. and Jespersen, C. K. and Vinther, J. and Capak, P. and Casey, C. M. and McPartland, C. J. R. and Milvang-Jensen, B. and Mobasher, B. and Sanders, D. B. and Zalesky, L. and Arnouts, S. and Aussel, H. and Dunlop, J. S. and Faisst, A. and Franx, M. and Furtak, L. J. and Fynbo, J. P. U. and Gould, K. M. L. and Greve, T. R. and Gwyn, S. and Kartaltepe, J. S. and Kashino, D. and Koekemoer, A. M. and Kokorev, V. and Le Fèvre, O. and Lilly, S. and Masters, D. and Magdis, G. and Mehta, V. and Peng, Y. and Riechers, D. A. and Salvato, M. and Sawicki, M. and Scarlata, C. and Scoville, N. and Shirley, R. and Silverman, J. D. and Sneppen, A. and Smolc̆ić, V. and Steinhardt, C. and Stern, D. and Tanaka, M. and Taniguchi, Y. and Teplitz, H. I. and Vaccari, M. and Wang, W.-H. and Zamorani, G.},
   year={2022},
   month=jan, pages={11} }

@ARTICLE{Calzetti,
       author = {{Calzetti}, Daniela and {Armus}, Lee and {Bohlin}, Ralph C. and {Kinney}, Anne L. and {Koornneef}, Jan and {Storchi-Bergmann}, Thaisa},
        title = "{The Dust Content and Opacity of Actively Star-forming Galaxies}",
      journal = {\apj},
         year = 2000,
        month = apr,
       volume = {533},
       number = {2},
        pages = {682-695},
          doi = {10.1086/308692},
archivePrefix = {arXiv},
       eprint = {astro-ph/9911459},
 primaryClass = {astro-ph},
       adsurl = {https://ui.adsabs.harvard.edu/abs/2000ApJ...533..682C}
}

@ARTICLE{Bolzonella2000,
       author = {{Bolzonella}, M. and {Miralles}, J. -M. and {Pell{\'o}}, R.},
        title = "{Photometric redshifts based on standard SED fitting procedures}",
      journal = {\aap},
         year = 2000,
        month = nov,
       volume = {363},
        pages = {476-492},
          doi = {10.48550/arXiv.astro-ph/0003380},
archivePrefix = {arXiv},
       eprint = {astro-ph/0003380},
 primaryClass = {astro-ph},
       adsurl = {https://ui.adsabs.harvard.edu/abs/2000A&A...363..476B}
}

@ARTICLE{BC2003,
   author = {{Bruzual}, G. and {Charlot}, S.},
    title = "{Stellar population synthesis at the resolution of 2003}",
  journal = {\mnras},
   eprint = {arXiv:astro-ph/0309134},
     year = 2003,
    month = oct,
   volume = 344,
    pages = {1000-1028},
      doi = {10.1046/j.1365-8711.2003.06897.x}, 
   adsurl = {http://cdsads.u-strasbg.fr/abs/2003MNRAS.344.1000B}
}

@ARTICLE{Chabrier2003,
   author = {{Chabrier}, G.},
    title = "{Galactic Stellar and Substellar Initial Mass Function}",
  journal = {\pasp},
   eprint = {arXiv:astro-ph/0304382},
     year = 2003,
    month = jul,
   volume = 115,
    pages = {763-795},
      doi = {10.1086/376392},
   adsurl = {http://cdsads.u-strasbg.fr/abs/2003PASP..115..763C}
}

@ARTICLE{STARBURST99,
       author = {{Leitherer}, Claus and {Schaerer}, Daniel and {Goldader}, Jeffrey D. and {Delgado}, Rosa M. Gonz{\'a}lez and {Robert}, Carmelle and {Kune}, Denis Foo and {de Mello}, Du{\'\i}lia F. and {Devost}, Daniel and {Heckman}, Timothy M.},
        title = "{Starburst99: Synthesis Models for Galaxies with Active Star Formation}",
      journal = {\apjs},
         year = 1999,
        month = jul,
       volume = {123},
       number = {1},
        pages = {3-40},
          doi = {10.1086/313233},
archivePrefix = {arXiv},
       eprint = {astro-ph/9902334},
 primaryClass = {astro-ph},
       adsurl = {https://ui.adsabs.harvard.edu/abs/1999ApJS..123....3L}
}

@ARTICLE{Zucca2006,
       author = {{Zucca}, E. and {Ilbert}, O. and {Bardelli}, S. and {Tresse}, L. and {Zamorani}, G. and {Arnouts}, S. and {Pozzetti}, L. and {Bolzonella}, M. and {McCracken}, H.~J. and {Bottini}, D. and {Garilli}, B. and {Le Brun}, V. and {Le F{\`e}vre}, O. and {Maccagni}, D. and {Picat}, J.~P. and {Scaramella}, R. and {Scodeggio}, M. and {Vettolani}, G. and {Zanichelli}, A. and {Adami}, C. and {Arnaboldi}, M. and {Cappi}, A. and {Charlot}, S. and {Ciliegi}, P. and {Contini}, T. and {Foucaud}, S. and {Franzetti}, P. and {Gavignaud}, I. and {Guzzo}, L. and {Iovino}, A. and {Marano}, B. and {Marinoni}, C. and {Mazure}, A. and {Meneux}, B. and {Merighi}, R. and {Paltani}, S. and {Pell{\`o}}, R. and {Pollo}, A. and {Radovich}, M. and {Bondi}, M. and {Bongiorno}, A. and {Busarello}, G. and {Cucciati}, O. and {Gregorini}, L. and {Lamareille}, F. and {Mathez}, G. and {Mellier}, Y. and {Merluzzi}, P. and {Ripepi}, V. and {Rizzo}, D.},
        title = "{The VIMOS VLT Deep Survey. Evolution of the luminosity functions by galaxy type up to z = 1.5 from first epoch data}",
      journal = {\aap},
         year = 2006,
        month = sep,
       volume = {455},
       number = {3},
        pages = {879-890},
          doi = {10.1051/0004-6361:20053645},
archivePrefix = {arXiv},
       eprint = {astro-ph/0506393},
 primaryClass = {astro-ph},
       adsurl = {https://ui.adsabs.harvard.edu/abs/2006A&A...455..879Z}
}

@ARTICLE{Trenti-Stiavelli08,
       author = {{Trenti}, M. and {Stiavelli}, M.},
        title = "{Cosmic Variance and Its Effect on the Luminosity Function Determination in Deep High-z Surveys}",
      journal = {\apj},
         year = 2008,
        month = apr,
       volume = {676},
       number = {2},
        pages = {767-780},
          doi = {10.1086/528674},
archivePrefix = {arXiv},
       eprint = {0712.0398},
 primaryClass = {astro-ph},
       adsurl = {https://ui.adsabs.harvard.edu/abs/2008ApJ...676..767T}
}

@ARTICLE{DECALS,
       author = {{Dey}, Arjun and {Schlegel}, David J. and {Lang}, Dustin and {Blum}, Robert and {Burleigh}, Kaylan and {Fan}, Xiaohui and {Findlay}, Joseph R. and {Finkbeiner}, Doug and {Herrera}, David and {Juneau}, St{\'e}phanie and {Landriau}, Martin and {Levi}, Michael and {McGreer}, Ian and {Meisner}, Aaron and {Myers}, Adam D. and {Moustakas}, John and {Nugent}, Peter and {Patej}, Anna and {Schlafly}, Edward F. and {Walker}, Alistair R. and {Valdes}, Francisco and {Weaver}, Benjamin A. and {Y{\`e}che}, Christophe and {Zou}, Hu and {Zhou}, Xu and {Abareshi}, Behzad and {Abbott}, T.~M.~C. and {Abolfathi}, Bela and {Aguilera}, C. and {Alam}, Shadab and {Allen}, Lori and {Alvarez}, A. and {Annis}, James and {Ansarinejad}, Behzad and {Aubert}, Marie and {Beechert}, Jacqueline and {Bell}, Eric F. and {BenZvi}, Segev Y. and {Beutler}, Florian and {Bielby}, Richard M. and {Bolton}, Adam S. and {Brice{\~n}o}, C{\'e}sar and {Buckley-Geer}, Elizabeth J. and {Butler}, Karen and {Calamida}, Annalisa and {Carlberg}, Raymond G. and {Carter}, Paul and {Casas}, Ricard and {Castander}, Francisco J. and {Choi}, Yumi and {Comparat}, Johan and {Cukanovaite}, Elena and {Delubac}, Timoth{\'e}e and {DeVries}, Kaitlin and {Dey}, Sharmila and {Dhungana}, Govinda and {Dickinson}, Mark and {Ding}, Zhejie and {Donaldson}, John B. and {Duan}, Yutong and {Duckworth}, Christopher J. and {Eftekharzadeh}, Sarah and {Eisenstein}, Daniel J. and {Etourneau}, Thomas and {Fagrelius}, Parker A. and {Farihi}, Jay and {Fitzpatrick}, Mike and {Font-Ribera}, Andreu and {Fulmer}, Leah and {G{\"a}nsicke}, Boris T. and {Gaztanaga}, Enrique and {George}, Koshy and {Gerdes}, David W. and {Gontcho}, Satya Gontcho A. and {Gorgoni}, Claudio and {Green}, Gregory and {Guy}, Julien and {Harmer}, Diane and {Hernandez}, M. and {Honscheid}, Klaus and {Huang}, Lijuan Wendy and {James}, David J. and {Jannuzi}, Buell T. and {Jiang}, Linhua and {Joyce}, Richard and {Karcher}, Armin and {Karkar}, Sonia and {Kehoe}, Robert and {Kneib}, Jean-Paul and {Kueter-Young}, Andrea and {Lan}, Ting-Wen and {Lauer}, Tod R. and {Le Guillou}, Laurent and {Le Van Suu}, Auguste and {Lee}, Jae Hyeon and {Lesser}, Michael and {Perreault Levasseur}, Laurence and {Li}, Ting S. and {Mann}, Justin L. and {Marshall}, Robert and {Mart{\'\i}nez-V{\'a}zquez}, C.~E. and {Martini}, Paul and {du Mas des Bourboux}, H{\'e}lion and {McManus}, Sean and {Meier}, Tobias Gabriel and {M{\'e}nard}, Brice and {Metcalfe}, Nigel and {Mu{\~n}oz-Guti{\'e}rrez}, Andrea and {Najita}, Joan and {Napier}, Kevin and {Narayan}, Gautham and {Newman}, Jeffrey A. and {Nie}, Jundan and {Nord}, Brian and {Norman}, Dara J. and {Olsen}, Knut A.~G. and {Paat}, Anthony and {Palanque-Delabrouille}, Nathalie and {Peng}, Xiyan and {Poppett}, Claire L. and {Poremba}, Megan R. and {Prakash}, Abhishek and {Rabinowitz}, David and {Raichoor}, Anand and {Rezaie}, Mehdi and {Robertson}, A.~N. and {Roe}, Natalie A. and {Ross}, Ashley J. and {Ross}, Nicholas P. and {Rudnick}, Gregory and {Safonova}, Sasha and {Saha}, Abhijit and {S{\'a}nchez}, F. Javier and {Savary}, Elodie and {Schweiker}, Heidi and {Scott}, Adam and {Seo}, Hee-Jong and {Shan}, Huanyuan and {Silva}, David R. and {Slepian}, Zachary and {Soto}, Christian and {Sprayberry}, David and {Staten}, Ryan and {Stillman}, Coley M. and {Stupak}, Robert J. and {Summers}, David L. and {Sien Tie}, Suk and {Tirado}, H. and {Vargas-Maga{\~n}a}, Mariana and {Vivas}, A. Katherina and {Wechsler}, Risa H. and {Williams}, Doug and {Yang}, Jinyi and {Yang}, Qian and {Yapici}, Tolga and {Zaritsky}, Dennis and {Zenteno}, A. and {Zhang}, Kai and {Zhang}, Tianmeng and {Zhou}, Rongpu and {Zhou}, Zhimin},
        title = "{Overview of the DESI Legacy Imaging Surveys}",
      journal = {\aj},
         year = 2019,
        month = may,
       volume = {157},
       number = {5},
          eid = {168},
        pages = {168},
          doi = {10.3847/1538-3881/ab089d},
archivePrefix = {arXiv},
       eprint = {1804.08657},
 primaryClass = {astro-ph.IM},
       adsurl = {https://ui.adsabs.harvard.edu/abs/2019AJ....157..168D}
}

@ARTICLE{PS1,
       author = {{Chambers}, K.~C. and {Magnier}, E.~A. and {Metcalfe}, N. and {Flewelling}, H.~A. and {Huber}, M.~E. and {Waters}, C.~Z. and {Denneau}, L. and {Draper}, P.~W. and {Farrow}, D. and {Finkbeiner}, D.~P. and {Holmberg}, C. and {Koppenhoefer}, J. and {Price}, P.~A. and {Rest}, A. and {Saglia}, R.~P. and {Schlafly}, E.~F. and {Smartt}, S.~J. and {Sweeney}, W. and {Wainscoat}, R.~J. and {Burgett}, W.~S. and {Chastel}, S. and {Grav}, T. and {Heasley}, J.~N. and {Hodapp}, K.~W. and {Jedicke}, R. and {Kaiser}, N. and {Kudritzki}, R. -P. and {Luppino}, G.~A. and {Lupton}, R.~H. and {Monet}, D.~G. and {Morgan}, J.~S. and {Onaka}, P.~M. and {Shiao}, B. and {Stubbs}, C.~W. and {Tonry}, J.~L. and {White}, R. and {Ba{\~n}ados}, E. and {Bell}, E.~F. and {Bender}, R. and {Bernard}, E.~J. and {Boegner}, M. and {Boffi}, F. and {Botticella}, M.~T. and {Calamida}, A. and {Casertano}, S. and {Chen}, W. -P. and {Chen}, X. and {Cole}, S. and {Deacon}, N. and {Frenk}, C. and {Fitzsimmons}, A. and {Gezari}, S. and {Gibbs}, V. and {Goessl}, C. and {Goggia}, T. and {Gourgue}, R. and {Goldman}, B. and {Grant}, P. and {Grebel}, E.~K. and {Hambly}, N.~C. and {Hasinger}, G. and {Heavens}, A.~F. and {Heckman}, T.~M. and {Henderson}, R. and {Henning}, T. and {Holman}, M. and {Hopp}, U. and {Ip}, W. -H. and {Isani}, S. and {Jackson}, M. and {Keyes}, C.~D. and {Koekemoer}, A.~M. and {Kotak}, R. and {Le}, D. and {Liska}, D. and {Long}, K.~S. and {Lucey}, J.~R. and {Liu}, M. and {Martin}, N.~F. and {Masci}, G. and {McLean}, B. and {Mindel}, E. and {Misra}, P. and {Morganson}, E. and {Murphy}, D.~N.~A. and {Obaika}, A. and {Narayan}, G. and {Nieto-Santisteban}, M.~A. and {Norberg}, P. and {Peacock}, J.~A. and {Pier}, E.~A. and {Postman}, M. and {Primak}, N. and {Rae}, C. and {Rai}, A. and {Riess}, A. and {Riffeser}, A. and {Rix}, H.~W. and {R{\"o}ser}, S. and {Russel}, R. and {Rutz}, L. and {Schilbach}, E. and {Schultz}, A.~S.~B. and {Scolnic}, D. and {Strolger}, L. and {Szalay}, A. and {Seitz}, S. and {Small}, E. and {Smith}, K.~W. and {Soderblom}, D.~R. and {Taylor}, P. and {Thomson}, R. and {Taylor}, A.~N. and {Thakar}, A.~R. and {Thiel}, J. and {Thilker}, D. and {Unger}, D. and {Urata}, Y. and {Valenti}, J. and {Wagner}, J. and {Walder}, T. and {Walter}, F. and {Watters}, S.~P. and {Werner}, S. and {Wood-Vasey}, W.~M. and {Wyse}, R.},
        title = "{The Pan-STARRS1 Surveys}",
      journal = {arXiv e-prints},
         year = 2016,
        month = dec,
          eid = {arXiv:1612.05560},
        pages = {arXiv:1612.05560},
          doi = {10.48550/arXiv.1612.05560},
archivePrefix = {arXiv},
       eprint = {1612.05560},
 primaryClass = {astro-ph.IM},
       adsurl = {https://ui.adsabs.harvard.edu/abs/2016arXiv161205560C}
}

@ARTICLE{UKIDSS,
       author = {{Warren}, S.~J. and {Cross}, N.~J.~G. and {Dye}, S. and {Hambly}, N.~C. and {Almaini}, O. and {Edge}, A.~C. and {Hewett}, P.~C. and {Hodgkin}, S.~T. and {Irwin}, M.~J. and {Jameson}, R.~F. and {Lawrence}, A. and {Lucas}, P.~W. and {Mortlock}, D.~J. and {Adamson}, A.~J. and {Bryant}, J. and {Collins}, R.~S. and {Davis}, C.~J. and {Emerson}, J.~P. and {Evans}, D.~W. and {Gonzales-Solares}, E.~A. and {Hirst}, P. and {Kerr}, T.~H. and {Lewis}, J.~R. and {Mann}, R.~G. and {Rawlings}, M.~G. and {Read}, M.~A. and {Riello}, M. and {Sutorius}, E.~T.~W. and {Varricatt}, W.~P.},
        title = "{The UKIRT Infrared Deep Sky Survey Second Data Release}",
      journal = {arXiv e-prints},
         year = 2007,
        month = mar,
          eid = {astro-ph/0703037},
        pages = {astro-ph/0703037},
          doi = {10.48550/arXiv.astro-ph/0703037},
archivePrefix = {arXiv},
       eprint = {astro-ph/0703037},
 primaryClass = {astro-ph},
       adsurl = {https://ui.adsabs.harvard.edu/abs/2007astro.ph..3037W}
}

@ARTICLE{Wang2021,
       author = {{Wang}, Feige and {Yang}, Jinyi and {Fan}, Xiaohui and {Hennawi}, Joseph F. and {Barth}, Aaron J. and {Banados}, Eduardo and {Bian}, Fuyan and {Boutsia}, Konstantina and {Connor}, Thomas and {Davies}, Frederick B. and {Decarli}, Roberto and {Eilers}, Anna-Christina and {Farina}, Emanuele Paolo and {Green}, Richard and {Jiang}, Linhua and {Li}, Jiang-Tao and {Mazzucchelli}, Chiara and {Nanni}, Riccardo and {Schindler}, Jan-Torge and {Venemans}, Bram and {Walter}, Fabian and {Wu}, Xue-Bing and {Yue}, Minghao},
        title = "{A Luminous Quasar at Redshift 7.642}",
      journal = {\apjl},
         year = 2021,
        month = jan,
       volume = {907},
       number = {1},
          eid = {L1},
        pages = {L1},
          doi = {10.3847/2041-8213/abd8c6},
archivePrefix = {arXiv},
       eprint = {2101.03179},
 primaryClass = {astro-ph.GA},
       adsurl = {https://ui.adsabs.harvard.edu/abs/2021ApJ...907L...1W}
}

@ARTICLE{Wang2018,
       author = {{Wang}, Feige and {Yang}, Jinyi and {Fan}, Xiaohui and {Yue}, Minghao and {Wu}, Xue-Bing and {Schindler}, Jan-Torge and {Bian}, Fuyan and {Li}, Jiang-Tao and {Farina}, Emanuele P. and {Ba{\~n}ados}, Eduardo and {Davies}, Frederick B. and {Decarli}, Roberto and {Green}, Richard and {Jiang}, Linhua and {Hennawi}, Joseph F. and {Huang}, Yun-Hsin and {Mazzucchelli}, Chiara and {McGreer}, Ian D. and {Venemans}, Bram and {Walter}, Fabian and {Beletsky}, Yuri},
        title = "{The Discovery of a Luminous Broad Absorption Line Quasar at a Redshift of 7.02}",
      journal = {\apjl},
         year = 2018,
        month = dec,
       volume = {869},
       number = {1},
          eid = {L9},
        pages = {L9},
          doi = {10.3847/2041-8213/aaf1d2},
archivePrefix = {arXiv},
       eprint = {1810.11925},
 primaryClass = {astro-ph.GA},
       adsurl = {https://ui.adsabs.harvard.edu/abs/2018ApJ...869L...9W}
}

@ARTICLE{Matsuoka2019J124353.93+010038.5,
       author = {{Matsuoka}, Yoshiki and {Onoue}, Masafusa and {Kashikawa}, Nobunari and {Strauss}, Michael A. and {Iwasawa}, Kazushi and {Lee}, Chien-Hsiu and {Imanishi}, Masatoshi and {Nagao}, Tohru and {Akiyama}, Masayuki and {Asami}, Naoko and {Bosch}, James and {Furusawa}, Hisanori and {Goto}, Tomotsugu and {Gunn}, James E. and {Harikane}, Yuichi and {Ikeda}, Hiroyuki and {Izumi}, Takuma and {Kawaguchi}, Toshihiro and {Kato}, Nanako and {Kikuta}, Satoshi and {Kohno}, Kotaro and {Komiyama}, Yutaka and {Koyama}, Shuhei and {Lupton}, Robert H. and {Minezaki}, Takeo and {Miyazaki}, Satoshi and {Murayama}, Hitoshi and {Niida}, Mana and {Nishizawa}, Atsushi J. and {Noboriguchi}, Akatoki and {Oguri}, Masamune and {Ono}, Yoshiaki and {Ouchi}, Masami and {Price}, Paul A. and {Sameshima}, Hiroaki and {Schulze}, Andreas and {Shirakata}, Hikari and {Silverman}, John D. and {Sugiyama}, Naoshi and {Tait}, Philip J. and {Takada}, Masahiro and {Takata}, Tadafumi and {Tanaka}, Masayuki and {Tang}, Ji-Jia and {Toba}, Yoshiki and {Utsumi}, Yousuke and {Wang}, Shiang-Yu and {Yamashita}, Takuji},
        title = "{Discovery of the First Low-luminosity Quasar at z > 7}",
      journal = {\apjl},
         year = 2019,
        month = feb,
       volume = {872},
       number = {1},
          eid = {L2},
        pages = {L2},
          doi = {10.3847/2041-8213/ab0216},
archivePrefix = {arXiv},
       eprint = {1901.10487},
 primaryClass = {astro-ph.GA},
       adsurl = {https://ui.adsabs.harvard.edu/abs/2019ApJ...872L...2M}
}

@ARTICLE{Matsuoka2019,
       author = {{Matsuoka}, Yoshiki and {Iwasawa}, Kazushi and {Onoue}, Masafusa and {Kashikawa}, Nobunari and {Strauss}, Michael A. and {Lee}, Chien-Hsiu and {Imanishi}, Masatoshi and {Nagao}, Tohru and {Akiyama}, Masayuki and {Asami}, Naoko and {Bosch}, James and {Furusawa}, Hisanori and {Goto}, Tomotsugu and {Gunn}, James E. and {Harikane}, Yuichi and {Ikeda}, Hiroyuki and {Izumi}, Takuma and {Kawaguchi}, Toshihiro and {Kato}, Nanako and {Kikuta}, Satoshi and {Kohno}, Kotaro and {Komiyama}, Yutaka and {Koyama}, Shuhei and {Lupton}, Robert H. and {Minezaki}, Takeo and {Miyazaki}, Satoshi and {Murayama}, Hitoshi and {Niida}, Mana and {Nishizawa}, Atsushi J. and {Noboriguchi}, Akatoki and {Oguri}, Masamune and {Ono}, Yoshiaki and {Ouchi}, Masami and {Price}, Paul A. and {Sameshima}, Hiroaki and {Schulze}, Andreas and {Silverman}, John D. and {Sugiyama}, Naoshi and {Tait}, Philip J. and {Takada}, Masahiro and {Takata}, Tadafumi and {Tanaka}, Masayuki and {Tang}, Ji-Jia and {Toba}, Yoshiki and {Utsumi}, Yousuke and {Wang}, Shiang-Yu and {Yamashita}, Takuji},
        title = "{Subaru High-z Exploration of Low-luminosity Quasars (SHELLQs). X. Discovery of 35 Quasars and Luminous Galaxies at 5.7 {\ensuremath{\leq}} z {\ensuremath{\leq}} 7.0}",
      journal = {\apj},
         year = 2019,
        month = oct,
       volume = {883},
       number = {2},
          eid = {183},
        pages = {183},
          doi = {10.3847/1538-4357/ab3c60},
archivePrefix = {arXiv},
       eprint = {1908.07910},
 primaryClass = {astro-ph.GA},
       adsurl = {https://ui.adsabs.harvard.edu/abs/2019ApJ...883..183M}
}

@ARTICLE{Yang2019,
       author = {{Yang}, Jinyi and {Wang}, Feige and {Fan}, Xiaohui and {Yue}, Minghao and {Wu}, Xue-Bing and {Li}, Jiang-Tao and {Bian}, Fuyan and {Jiang}, Linhua and {Ba{\~n}ados}, Eduardo and {Beletsky}, Yuri},
        title = "{Exploring Reionization-era Quasars. IV. Discovery of Six New z {\ensuremath{\gtrsim}} 6.5 Quasars with DES, VHS, and unWISE Photometry}",
      journal = {\aj},
         year = 2019,
        month = jun,
       volume = {157},
       number = {6},
          eid = {236},
        pages = {236},
          doi = {10.3847/1538-3881/ab1be1},
archivePrefix = {arXiv},
       eprint = {1811.11915},
 primaryClass = {astro-ph.GA},
       adsurl = {https://ui.adsabs.harvard.edu/abs/2019AJ....157..236Y}
}

@ARTICLE{Yang2020,
       author = {{Yang}, Jinyi and {Wang}, Feige and {Fan}, Xiaohui and {Hennawi}, Joseph F. and {Davies}, Frederick B. and {Yue}, Minghao and {Banados}, Eduardo and {Wu}, Xue-Bing and {Venemans}, Bram and {Barth}, Aaron J. and {Bian}, Fuyan and {Boutsia}, Konstantina and {Decarli}, Roberto and {Farina}, Emanuele Paolo and {Green}, Richard and {Jiang}, Linhua and {Li}, Jiang-Tao and {Mazzucchelli}, Chiara and {Walter}, Fabian},
        title = "{P{\={o}}niu{\={a}}'ena: A Luminous z = 7.5 Quasar Hosting a 1.5 Billion Solar Mass Black Hole}",
      journal = {\apjl},
         year = 2020,
        month = jul,
       volume = {897},
       number = {1},
          eid = {L14},
        pages = {L14},
          doi = {10.3847/2041-8213/ab9c26},
archivePrefix = {arXiv},
       eprint = {2006.13452},
 primaryClass = {astro-ph.GA},
       adsurl = {https://ui.adsabs.harvard.edu/abs/2020ApJ...897L..14Y}
}

@ARTICLE{Mortlock2011,
       author = {{Mortlock}, Daniel J. and {Warren}, Stephen J. and {Venemans}, Bram P. and {Patel}, Mitesh and {Hewett}, Paul C. and {McMahon}, Richard G. and {Simpson}, Chris and {Theuns}, Tom and {Gonz{\'a}les-Solares}, Eduardo A. and {Adamson}, Andy and {Dye}, Simon and {Hambly}, Nigel C. and {Hirst}, Paul and {Irwin}, Mike J. and {Kuiper}, Ernst and {Lawrence}, Andy and {R{\"o}ttgering}, Huub J.~A.},
        title = "{A luminous quasar at a redshift of z = 7.085}",
      journal = {\nat},
         year = 2011,
        month = jun,
       volume = {474},
       number = {7353},
        pages = {616-619},
          doi = {10.1038/nature10159},
archivePrefix = {arXiv},
       eprint = {1106.6088},
 primaryClass = {astro-ph.CO},
       adsurl = {https://ui.adsabs.harvard.edu/abs/2011Natur.474..616M}
}

@software{LEPHARE,
       author = {{Arnouts}, S. and {Ilbert}, O.},
        title = "{LePHARE: Photometric Analysis for Redshift Estimate}",
 howpublished = {Astrophysics Source Code Library, record ascl:1108.009},
         year = 2011,
        month = aug,
          eid = {ascl:1108.009},
       adsurl = {https://ui.adsabs.harvard.edu/abs/2011ascl.soft08009A}
}

@book{fmeasure,
author = {Baeza-Yates, Ricardo and Ribeiro-Neto, Berthier},
year = {2011},
month = {01},
pages = {},
title = {Modern Information Retrieval the Concepts and Technology Behind Search},
isbn = {978-0-321-41691-9}
}

@ARTICLE{CFHQSIR,
       author = {{Pipien}, S. and {Basa}, S. and {Cuby}, J. -G. and {Cuillandre}, J. -C. and {Willott}, C. and {Moutard}, T. and {Chatron}, J. and {Arnouts}, S. and {Hudelot}, P.},
        title = "{The CFHQSIR survey: a Y-band extension of the CFHTLS-Wide survey}",
      journal = {\aap},
         year = 2018,
        month = aug,
       volume = {616},
          eid = {A55},
        pages = {A55},
          doi = {10.1051/0004-6361/201731944},
archivePrefix = {arXiv},
       eprint = {1709.04308},
 primaryClass = {astro-ph.IM},
       adsurl = {https://ui.adsabs.harvard.edu/abs/2018A&A...616A..55P}
}

@ARTICLE{Temple,
       author = {{Temple}, Matthew J. and {Hewett}, Paul C. and {Banerji}, Manda},
        title = "{Modelling type 1 quasar colours in the era of Rubin and Euclid}",
      journal = {\mnras},
         year = 2021,
        month = nov,
       volume = {508},
       number = {1},
        pages = {737-754},
          doi = {10.1093/mnras/stab2586},
archivePrefix = {arXiv},
       eprint = {2109.04472},
 primaryClass = {astro-ph.GA},
       adsurl = {https://ui.adsabs.harvard.edu/abs/2021MNRAS.508..737T}
}

@ARTICLE{hogg,
       author = {{Hogg}, David W. and {Baldry}, Ivan K. and {Blanton}, Michael R. and {Eisenstein}, Daniel J.},
        title = "{The K correction}",
      journal = {arXiv e-prints},
         year = 2002,
        month = oct,
          eid = {astro-ph/0210394},
        pages = {astro-ph/0210394},
          doi = {10.48550/arXiv.astro-ph/0210394},
archivePrefix = {arXiv},
       eprint = {astro-ph/0210394},
 primaryClass = {astro-ph},
       adsurl = {https://ui.adsabs.harvard.edu/abs/2002astro.ph.10394H}
}

@ARTICLE{Sadman,
       author = {{Ali}, Sadman S. and {De Propris}, Roberto and {Chung}, Chul and {Phillipps}, Steven and {Bremer}, Malcolm N. and {Onodera}, Masato and {Sawicki}, Marcin and {Desprez}, Guillaume and {Gwyn}, Stephen},
        title = "{Probing the Stellar Populations and Star Formation History of Early-type Galaxies at 0 < z < 1.1 in the Rest-frame Ultraviolet}",
      journal = {\apj},
         year = 2024,
        month = may,
       volume = {966},
       number = {1},
          eid = {50},
        pages = {50},
          doi = {10.3847/1538-4357/ad3209},
archivePrefix = {arXiv},
       eprint = {2403.08301},
 primaryClass = {astro-ph.GA},
       adsurl = {https://ui.adsabs.harvard.edu/abs/2024ApJ...966...50A}
}

@ARTICLE{WISE,
       author = {{Wright}, Edward L. and {Eisenhardt}, Peter R.~M. and {Mainzer}, Amy K. and {Ressler}, Michael E. and {Cutri}, Roc M. and {Jarrett}, Thomas and {Kirkpatrick}, J. Davy and {Padgett}, Deborah and {McMillan}, Robert S. and {Skrutskie}, Michael and {Stanford}, S.~A. and {Cohen}, Martin and {Walker}, Russell G. and {Mather}, John C. and {Leisawitz}, David and {Gautier}, Thomas N., III and {McLean}, Ian and {Benford}, Dominic and {Lonsdale}, Carol J. and {Blain}, Andrew and {Mendez}, Bryan and {Irace}, William R. and {Duval}, Valerie and {Liu}, Fengchuan and {Royer}, Don and {Heinrichsen}, Ingolf and {Howard}, Joan and {Shannon}, Mark and {Kendall}, Martha and {Walsh}, Amy L. and {Larsen}, Mark and {Cardon}, Joel G. and {Schick}, Scott and {Schwalm}, Mark and {Abid}, Mohamed and {Fabinsky}, Beth and {Naes}, Larry and {Tsai}, Chao-Wei},
        title = "{The Wide-field Infrared Survey Explorer (WISE): Mission Description and Initial On-orbit Performance}",
      journal = {\aj},
         year = 2010,
        month = dec,
       volume = {140},
       number = {6},
        pages = {1868-1881},
          doi = {10.1088/0004-6256/140/6/1868},
archivePrefix = {arXiv},
       eprint = {1008.0031},
 primaryClass = {astro-ph.IM},
       adsurl = {https://ui.adsabs.harvard.edu/abs/2010AJ....140.1868W}
}

@BOOK{LF,
       author = {{Schechter}, Paul},
        title = "{The luminosity function for galaxies and the clustering of galaxies}",
         year = 1978,
       adsurl = {https://ui.adsabs.harvard.edu/abs/1978lfgc.book.....S}
}
\end{document}